\documentclass[%
reprint,
jmp,
superscriptaddress,
nofootinbib,
amsmath,amssymb,
table
]{revtex4-2}

\usepackage[english]{babel}
\usepackage{graphicx}
\usepackage{dcolumn}
\usepackage{bm}
\usepackage[version=4]{mhchem}
\usepackage{svg}
\usepackage{multirow}
\usepackage{comment}
\usepackage{tabularx}
\usepackage{array}
\usepackage{makecell}
\usepackage{amssymb}
\usepackage{float}
\usepackage{afterpage}
\usepackage[utf8]{inputenc}
\usepackage{braket}
\usepackage{amsmath}
\usepackage{amsfonts}
\usepackage{booktabs} 
\usepackage{hyperref}
\usepackage{cleveref}
\usepackage{titlesec}
\usepackage{dsfont}
\usepackage{pifont}
\usepackage{adjustbox}
\usepackage[normalem]{ulem}
\usepackage{tabularray}
\usepackage[left=0.8in,right=0.8in,top=0.8in,bottom=0.8in]{geometry}
\usepackage{upgreek}
\usepackage{textgreek}
\usepackage[super]{nth}
\usepackage{csquotes}
\usepackage{xcolor}
\usepackage{hhline}
\usepackage{enumitem}
\usepackage{xspace}

\usepackage{graphicx}
\usepackage{longtable}
\usepackage{array}
\usepackage{booktabs}
\usepackage{xurl}

\newcolumntype{Y}{>{\centering\arraybackslash}X}
\renewcommand{\arraystretch}{1.1}  

\newcommand{\lambdamace}{$\lambda$\text{-MACE }}

\definecolor{myblue}{RGB}{71,120,207}
\definecolor{mygreen}{RGB}{106,204,100}
\definecolor{myred}{RGB}{213,95,95}

\date{\today}

\makeatletter
\def\@email#1#2{%
 \endgroup
 \patchcmd{\titleblock@produce}
  {\frontmatter@RRAPformat}
  {\frontmatter@RRAPformat{\produce@RRAP{*#1\href{mailto:#2}{#2}}}\frontmatter@RRAPformat}
  {}{}
}%
\makeatother

\begin{document}

\preprint{AIP/123-QED}

\title{Computing hydration free energies and crystalline solubility of drug-like molecules with MACE}

\author{J. Harry Moore}
\affiliation{Ångström AI, San Francisco, California, USA}
\affiliation{Engineering Laboratory, University of Cambridge, Cambridge, CB2 1PZ, UK}
\author{Paul Mortenson}
\affiliation{Astex Pharmaceuticals, Cambridge, CB4 0QA, UK}
\author{Davide Branduardi}
\affiliation{Astex Pharmaceuticals, Cambridge, CB4 0QA, UK}
\author{Pavel Buslaev}
\affiliation{Astex Pharmaceuticals, Cambridge, CB4 0QA, UK}
\author{Javier Antoran}
\affiliation{Ångström AI, San Francisco, California, USA}
\affiliation{Engineering Laboratory, University of Cambridge, Cambridge, CB2 1PZ, UK}
\author{Laurence Midgley}
\affiliation{Ångström AI, San Francisco, California, USA}
\affiliation{Engineering Laboratory, University of Cambridge, Cambridge, CB2 1PZ, UK}
\author{Nicholas J. Browning}
\affiliation{Ångström AI, San Francisco, California, USA}
\author{G\'abor Cs\'anyi}
\affiliation{Ångström AI, San Francisco, California, USA}
\affiliation{Engineering Laboratory, University of Cambridge, Cambridge, CB2 1PZ, UK}

\begin{abstract}
Accurate prediction of thermodynamic solubility is critical in drug discovery, where it influences oral bioavailability and drives decisions in formulation chemistry. 
First-principles calculations of intrinsic solubility offer the potential to achieve superior generalization to existing direct, data-driven predictors, which often suffer significant performance degradation outside of their training distribution. 
Whilst free energy perturbation calculations provide a rigorous framework to estimate these quantities, their accuracy is typically limited by the quality of the empirical force field used.

In this work, we present a computational workflow driven by the MACE family of machine learning interatomic potentials to compute the solubility of neutral drug-like molecules. 
We introduce \lambdamace, a potential optimised for condensed-phase free energy calculations, which enables hydration free energy calculations at experimental accuracy on an evaluation set of 70 compounds from the FreeSolv database selected for their relevance to medicinal chemistry. 
We further combine these hydration free energy calculations with sublimation free energy calculations to predict crystalline solubility given a known experimental crystal structure, on an additional set of 10 compounds. 
We achieve state-of-the-art performance for each leg of the calculation, with experimental accuracy for the hydration leg, and accuracy competitive with industry-standard DFT for the sublimation leg. 
However, the overall errors after combining both legs lead to accuracy of 1.53 log units. 
This is significantly above the experimental accuracy (0.17-0.6 log units). 
Closing this gap will require substantially tighter accuracy in each leg, on the order of half of chemical accuracy (0.5 kcal/mol), with crystalline thermodynamics the hardest remaining challenge.

\end{abstract}

\maketitle
\footnotetext{Correspondence to laurence@angstrom-ai.com}

\section{Introduction}

In drug discovery, solubility is a fundamental physicochemical property.
For oral dosing, the solubility of a drug determines the maximum concentration that will be available for absorption through the gastrointestinal tract, in turn limiting the maximum systemic exposure that is achievable~\cite{diProfilingDruglikeProperties2003, lipinskiExperimentalComputationalApproaches1997}.  Poorly soluble compounds may also show increased variability in exposure between patients if dosed close to the limit of their solubility.  Its importance to oral bioavailability is reflected in solubility being one of the two parameters (along with gastrointestinal permeability) that form the widely used Biopharmaceutics Classification System \citep{Amidon1995biopharmaceutic}.  While poor drug solubility (for both oral and IV dosing) can sometimes be mitigated via alternative formulations, this can be costly and time-consuming to investigate.

Poor solubility also presents problems at early stages of drug discovery, where compounds are typically less efficacious and need to be tested at higher concentrations~\cite{harnerFragmentbasedDrugDiscovery2013}.  For example, in the case of fragment-based drug discovery (FBDD) it is not unusual to test early compounds in biophysical or biochemical assays at millimolar concentrations.  Inadequate solubility in these experiments can lead to incomplete or even misleading concentration response curves, which makes interpretation of the results (and hence optimisation of molecules) significantly more challenging.

The most common computational method of solubility estimation is via direct prediction, whereby a model is fit via regression to predict solubility given a  molecular structure and/or descriptors \cite{llinas2020findings}. 
However, these models suffer from poor generalisation outside of the training distribution and systematic improvement is hampered by the limited availability of high quality data \cite{second_challenge_post_hoc}.

Free energy perturbation (FEP) calculations provide a rigorous approach to computationally estimating thermodynamic properties, like solubility or binding free energy. Its use has become commonplace in pharmaceutical R\&D, where they see broad application in structure-based drug discovery programs, augmenting rational design approaches  \citep{SchindlerFEP2020,MooreAutomatedRBFE2023}.

Among free energy methods, alchemical transformations are perhaps the most popular approach for constructing sample-efficient estimators of free energies. These methods integrate the free energy difference along a non-physical path that connects the physical end-states of interest.
For thermodynamic solubility, the end states correspond to the molecule of interest in solid phase, and the same molecule in aqueous solution. The alchemical pathway connecting these end-states combines calculation of both hydration and sublimation free energy.

FEP relies on Molecular Dynamics (MD) simulations to sample the conformational phase space of the molecular systems under consideration. Thus, the accuracy of free energy calculations depends crucially on the force field being used.
Empirical force fields, e.g. the General Amber force field (GAFF) \citep{WangGaff2004} or OpenFF \citep{BoothroydOpeff2023}, typically have their nonbonded parameters fitted to experimental condensed phase data. 
This often includes measurements of density, hydration and transfer free energies, and heats of vaporisation, covering commonly occurring functional groups. 
Directly fitting to experimental data raises the concern that the force field may reproduce the fitted observables without adequately capturing the underlying interactions in a transferable manner. Consequently, applying the resulting nonbonded parameters to functional group interactions, or combinations thereof, not represented in the fitting data may lead to large errors in free energy calculations.

Considerable work has gone into using machine learning interatomic potentials (MLIP) to ameliorate the deficiencies of empirical force fields~\cite{unkeMachineLearningForce2021a}. MLIPs are typically trained to predict energies and forces using datasets of quantum mechanical calculations, most commonly Density Functional Theory (DFT). 
This avoids the risk of relying on cancellation of errors to accurately estimate observables. 
As a result, it is hypothesized that MLIPs may more accurately capture nonbonded interactions.
Of particular relevance to this paper is the MACE organic force field (MACE-OFF)\citep{kovacs2023maceoff23transferablemachinelearning}. This MLIP is trained on a subset of the SPICE dataset, consisting of energies and forces for ~1.1 million neutral organic molecules and peptides, computed at the ωB97M-D3 level of theory \citep{EastmanSPICE2023}.

Although MLIPs are fast compared to DFT, they are significantly slower than classical mechanical force fields. Thus, MLIPs are often applied to model only intramolecular ligand energetics, while remaining interactions are treated with an empirical force field.
Whilst there is limited evidence that such a corrective scheme improves performance~\cite{rufaChemicalAccuracyAlchemical2020}, a growing consensus suggests that embedding MLIPs within classical force fields can introduce inconsistencies between the modified and retained interaction terms, making it difficult to achieve accuracy gains across a broad range of molecular systems~\cite{karwounopoulosInsightsChallengesCorrecting2024}.
The alternative setting, and the one considered in this work, is to use an MLIP to model the full molecular systems under consideration. Examples of this include \citet{moore2024computinghydrationfreeenergies}, where the authors use MACE to compute hydration free energies on a limited set of organic molecules, and \citet{namInterpolationDifferentiationAlchemical2024} who use the same force field to compute the free energy of vacancy formation and element changes in crystals. 

Hydration free energy calculations are commonly used to benchmark small molecule force field accuracy. This is due to their relatively smaller sampling requirement compared to other free energy calculations, e.g. protein-ligand binding free energies. 
To this end, the FreeSolv database \cite{mobleyFreeSolvDatabaseExperimental2014} provides a collection of around 600 experimental hydration free energy measurements against which to compare computational predictions. These values have been manually curated from the existing literature, and compounds are typically closer to fragment-like, rather than drug-like chemical space, but it nonetheless provides a more exhaustive benchmark of what has been offered in \citet{moore2024computinghydrationfreeenergies} and an essential step towards general applicability of this technology

Given that classical force fields are parametrised directly against experimental data such as FreeSolv, one would expect that a well parametrised empirical FF would be highly accurate. 
However, it has previously been shown that empirical force fields' error tends to increase with the size and complexity of the molecule~\cite{mobleyBlindPredictionSolvation2014}. \\

Given the improved accuracy promised by MLIPs, an intriguing application is the accurate calculation of free energies. However, without modification, standard machine learned force fields cannot rigorously compute free energies by alchemical paths in the liquid phase, since the DFT they are fit to contains a singularity as atoms overlap. 

Recently, \citet{moore2024computinghydrationfreeenergies} demonstrated that one can train a `softcore' MACE potential that avoids the singularity at r=0 without modifying the equilibrium behaviour of the potential~\cite{moore2024computinghydrationfreeenergies}, enabling rigorous alchemical free energy calculations. The authors introduced an alchemical free energy pathway that modifies the MACE MLIP by smoothly and selectively scaling two-body interactions, as well as introducing `softcore` repulsion by fitting on synthetic training data. 
This resulted in highly accurate free energies predictions ($\sim$0.5 kcal/mol) across the set of 36 molecules. 

In addition to alchemical approaches to liquid phase free energy calculations, force fields, and more recently MLIPs have been applied to the calculation of the thermodynamic properties of crystalline organic molecules~\cite{taylorPredictiveCrystallographyScale2024a}.

Accurate modelling of crystal structures, including sublimation free energies, requires extremely precise prediction of nonbonded interactions. These calculations have traditionally been performed using dispersion-corrected DFT, incurring large computational costs.  
Indeed, the best performing method on the most recent blind crystal structure prediction (CSP) challenge required more than 9 million CPU hours~\cite{hunnisettSeventhBlindTest2024} for ranking putative crystal structures. 
One of the most successful approaches for ameliorating the costs of accurately modelling organic crystals is training compound-specific molecular mechanics force fields on dispersion-corrected DFT \citep{brooTransferableForceField2016,firahaPredictingCrystalForm2023}. However, the need for bespoke DFT calculation makes the cost of this approach high. Very recently, it has also been shown that MACE-MP0, a variant of the MLIP trained on a large dataset of inorganic crystals \citep{batatia2024foundationmodelatomisticmaterials}, can be fine tuned by transfer learning to predict highly accurate crystal sublimation enthalpies ~\cite{kaurDataefficientFinetuningFoundational2024}. 
\\

Previous state-of-the-art work on physics based calculation of solubility combined FEP using a classical force field (GAFF) for hydration, and DFT for the sublimation leg to achieve a promising accuracy of 0.72 log units, albeit on a limited set of three molecules \cite{physics-solubility}.
Motivated by the initial success of MACE variants in computing hydration free energies and crystal sublimation enthalpies, this paper pursues the design of an MLIP-driven workflow capable of accurately estimating the thermodynamic solubility of neutral drug-like molecules. 
To this end, we build upon the alchemical MACE model of \citet{moore2024computinghydrationfreeenergies}. 
We augment its training data with additional DFT calculations, at the ωB97M-D3/def2-TZVPPD level of theory, targetting molecular systems relevant to medicinal chemistry. 
The resulting MLIP, which we call \lambdamace, predicts hydration free energies with accuracy comparable to experiment on a set of 70 molecules curated from the FreeSolv database based on their relevance to medicinal chemistry. 
We further use \lambdamace to compute sublimation free energies from experimental crystal structures, achieving performance competitive with dispersion-corrected DFT~\cite{brooTransferableForceField2016}.
Combining the two legs yields crystalline solubility predictions with an RMSE of 1.53 log units on a set of 10 compounds. 
Although each leg individually reaches state-of-the-art accuracy, we find that their errors compound, leaving the overall prediction short of experimental accuracy.
The size of this gap tells us how far each leg must improve: to better than half of chemical accuracy (0.5 kcal/mol), with crystalline thermodynamics the hardest remaining challenge.

\section{Results \& Discussion}

\subsection{Drug-like hydration benchmark dataset}\label{subsec:results-dataset}

We curate a subset of the FreeSolv dataset focusing on both functional group diversity and pharmacological relevance. 

Freesolv contains experimental hydration free energy values for 643 compounds. Of these, we only consider compounds that are expected to exist in only the neutral form under physiological conditions, as predicted by Schr\"odinger's \texttt{epik} tool \cite{epikx}. 
This ensures compatibility with MACE, which is trained only on neutral compounds. 
We then identify a subset of 70 unique compounds considering a number of features that suggest relevance to medicinal chemistry.
The complete list of compounds is provided in \autoref{appendix}. 

In particular, we prioritise molecules containing only the elements H, B, C, N, O, F, S, Cl and Br. 
We include a small proportion of molecules with I and P for completeness.
We exclude aliphatic nitro groups, nitrates, and sulfonates.
We prioritise molecules that have at least one aromatic ring and another species other than carbon.
We de-prioritise molecules that have more than two rings (fused and non-fused) or that appear excessively lipophilic, large or flexible.
Although severely reduced with respect to original Freesolv set, we believe the tight selection provides a balanced view in a chemical space of pharmaceutical interest.

\subsection{Improving the MACE organic potential for intermolecular interactions}\label{subsec:mace-hyperparams}

Building on recent applications of the MACE architecture to accurate modelling of condensed phase systems, we adopted the hyperparameters of the most recent MACE-OFF-24 model~\cite{kovacs2023maceoff23transferablemachinelearning, moore2024computinghydrationfreeenergies}. Notably this resulted in a predicted water density within  2\% of experiment, which is indicative of more accurate modelling of the intermolecular interactions which are crucial for application to solution-phase free energy calculations.

\subsubsection{Generating additional DFT data}

The chemical space covered by the training data is the primary determinant of hydration
free energy accuracy: predictions degrade for compounds whose solute--water interactions
are not represented in the training set.
Early versions of the \lambdamace{} potentials, fitted purely on SPICE~1, illustrate this
clearly: with solvated amino acid fragments as the only strongly relevant configurations,
predictions were accurate within that chemical neighbourhood but deteriorated for
out-of-distribution functional groups, particularly halogen-containing moieties.
We therefore broadened the coverage of solute--water interactions using two additional sources.

\textbf{SPICE 2.} Relative to SPICE~1, this adds 14,000 solvated PubChem configurations
spanning 1,400 unique molecules, 1,000 further water clusters, and $\sim$200,000
amino acid--ligand configurations covering 80,000 unique amino acid--ligand pairs
\cite{Eastman2024Spice2}.

\textbf{Solvated Enamine fragments.} Because  thio-containing and phenolic compounds interactions remained
under-represented even with SPICE~2, we generated a further 73,000 configurations in house
at the SPICE level of theory. 
These derive from 2,240 fragment-like compounds drawn from the
Enamine `essential' (320 compounds) and `high fidelity' (1,920 compounds)
libraries, the
latter contributing finer-grained pharmacophores \cite{enamine_fragment_libraries}. 
Both libraries were filtered to neutral
compounds bearing at most two functional groups, so that individual functional
group--water interactions could be sampled in isolation; the resulting set spans 825 unique
Bemis--Murcko extended scaffolds. Configurations were subsampled from OpenFF2/TIP3P molecular
dynamics, retaining the solute and its primary hydration shell
\cite{BoothroydOpeff2023,tip3p}.

\begin{figure}
    \centering
    \includegraphics[width=0.9\linewidth]{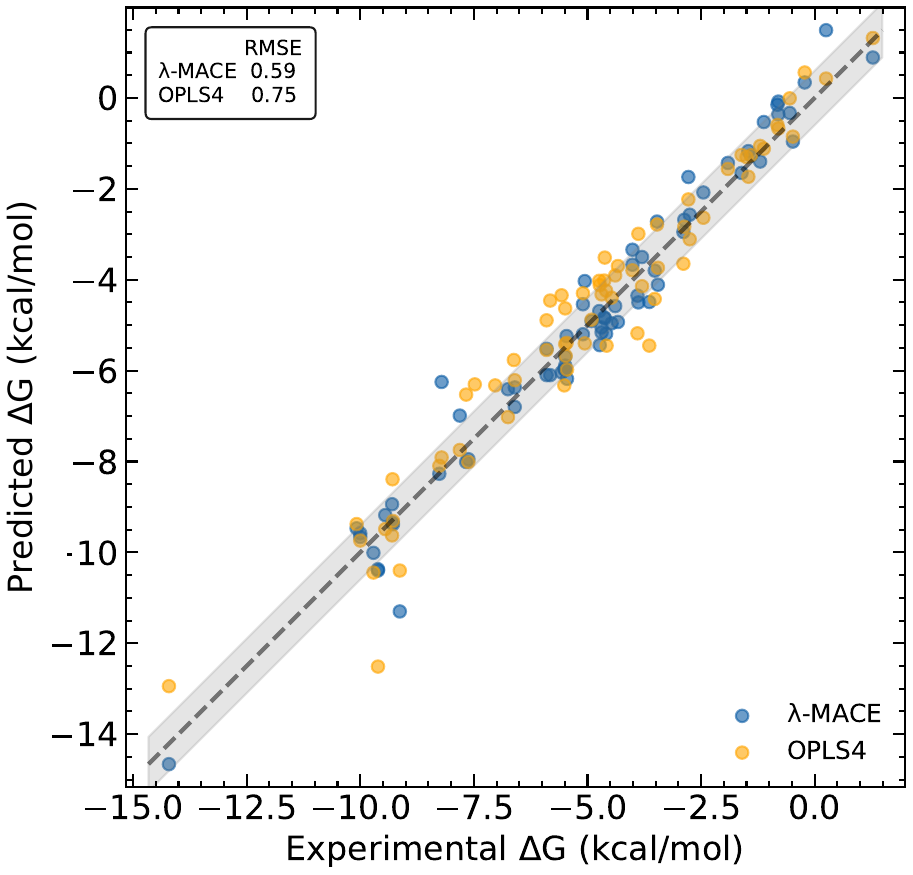}
    \caption{Performance of \lambdamace vs OPLS4 on hydration free energy of 70 medicinal chemistry relevant molecules selected from FreeSolv. Error within 1 kcal/mol of experiment indicated in grey.}
    \label{fig:astx_freesolv}
\end{figure}

\subsection{Performance on FreeSolv}

Compared to the empirical force field OPLS4, \lambdamace performs favourably across our test set of 70 FreeSolv compounds, with a modest RMSE decrease from 0.75 to 0.59 (Figure~\ref{fig:astx_freesolv}). The outliers in the \lambdamace predictions are accounted for by the large uncertainty of the corresponding experimental values. Given that the accuracy of \lambdamace matches the experimental uncertainty of the dataset, one would not expect to be able to improve significantly on this performance without benchmarking on a more accurate experimental dataset. However, as mentioned previously, it is encouraging that \lambdamace is capable of achieving superior performance on these benchmarks compared to an empirical force field, given that the model has learnt the required interactions from the ground up, without requiring explicit parametrisation to predict these values.

\subsection{Generalisation to larger compounds}
\label{sec:generalise-large}

Whilst hydration free energies are routinely used to parametrise empirical force fields, performance on this task is not expected to correlate directly with accuracy on more complex free energy calculations, since, as previously discussed, this depends on the accuracy of the atom-type assignment.

Hydration free energies are not routinely collected for drug-like molecules, and those values that do exist are obtained from disparate experimental literature, where the experimental uncertainty can be large or entirely unknown.  This makes any direct parametrisation against such values a challenging prospect, given the inherent amount of noise in the signal. 
However, there are limited examples of some larger or more highly functionalised compounds in FreeSolv, which correlate with the largest prediction errors.

It is therefore natural to ask how the performance of empirical force fields scales with larger and more heavily functionalised compounds.  Analysing the trends in GAFF errors in the original FreeSolv database compared to experiment reveals that the highest prediction errors indeed correspond to large, polyfunctionalised compounds. This indicates that, even within the relatively small FreeSolv dataset, the limitations of atom-typing and the functional form of the force field become apparent, since, presumably, these compounds have been included in the parametrisation process.

We therefore selected a set of large compounds on which to confirm the ability of \lambdamace to generalise in chemical space. 
We identified compounds where the GAFF hydration free energy error is largest, but where the compounds are still of interest as defined by the selection approach outlined in section~\ref{subsec:results-dataset} and directly compare these predictions with \lambdamace.
\autoref{appendix} provides further details on the selected molecules for the high error set. 

\begin{figure*}
    \centering
    \includegraphics[width=1.0\textwidth]{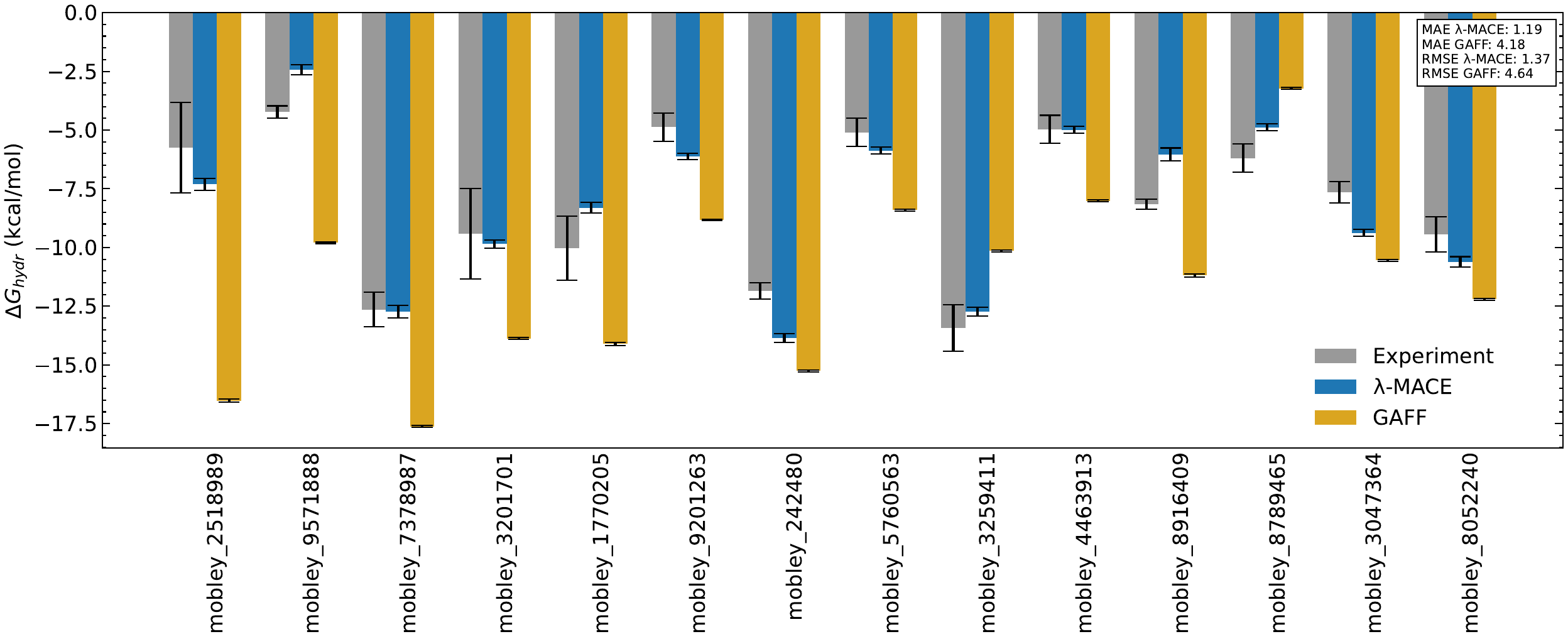}
    \caption{High error test set. Comparison of \lambdamace hydration free energies with GAFF and experimental values for the high error test set.}
    \label{fig:high-gaff-errors}
\end{figure*}

Although the \lambdamace free energy errors on this test set are indeed higher than for the set of smaller fragments (Figure~\ref{fig:high-gaff-errors}), it is also worth noting that the experimental uncertainty for several of these measurements is significantly higher than those typically found for smaller molecules. The RMSE is significantly reduced by a factor of 3.5 compared to GAFF, indicating that \lambdamace is better able to generalise to larger or polyfunctionalised compounds.

\subsection{Crystalline Solubility Calculations}\label{subsec:results-crystaline}

Having confirmed the ability of \lambdamace to accurately predict hydration free energies, we turned next to consider the question of crystalline solubilities. 

For this, we leverage the existing benchmark dataset from~\cite{brooTransferableForceField2016,palmer_2012}. As noted in the original benchmark, sublimation free energies for organic crystals are not usually reported in the experimental literature, hence benchmarking data is sparse. 

Thermodynamic crystalline solubility computes the free energy difference of taking a molecule from the crystal to the aqueous phase. That is 
\begin{gather}
    \Delta G_\text{solubility} = G_\text{solvated} - G_\text{crystal}
\end{gather}
for $G_\text{solvated}$ the Gibbs free energy of the solute in aqueous state and $G_\text{crystal}$ that of the solid crystal state.
This is directly relevant to late state drug discovery, in which the solubility of the active pharmaceutical ingredient is key to bioavailability.  

Practically, we compute the free energy of dissolution, $\Delta G_\text{solubility}$, via an alchemical path that combines hydration and crystal sublimation free energies as
\begin{gather}
    \Delta G_\text{solubility} = \Delta G_\text{hydration} + \Delta  G_\text{sublimation}, \\
    \text{where} \quad \Delta G_\text{hydration} = G_\text{solvated} - G_\text{vacuum} \\ 
    \text{and} \quad \Delta G_\text{sublimation} = G_\text{vacuum} - F_\text{crystal}/Z.
\end{gather}
where we use the Helmhotz free energy of the crystal $F_\text{crystal}$ and Z is the number of formula units in the unit cell.

The intrinsic solubility is then
\[
S_0
=
\frac{P_0}{RT}
\exp\left[
-\frac{\Delta G_\mathrm{sub}
+
\Delta G_\mathrm{hyd}}{RT}
\right],
\]
where \(P_0\) is the standard pressure. 

\subsubsection{Hydration free energy results}

\begin{figure}[H]
    \centering
    \includegraphics[width=0.9\linewidth]{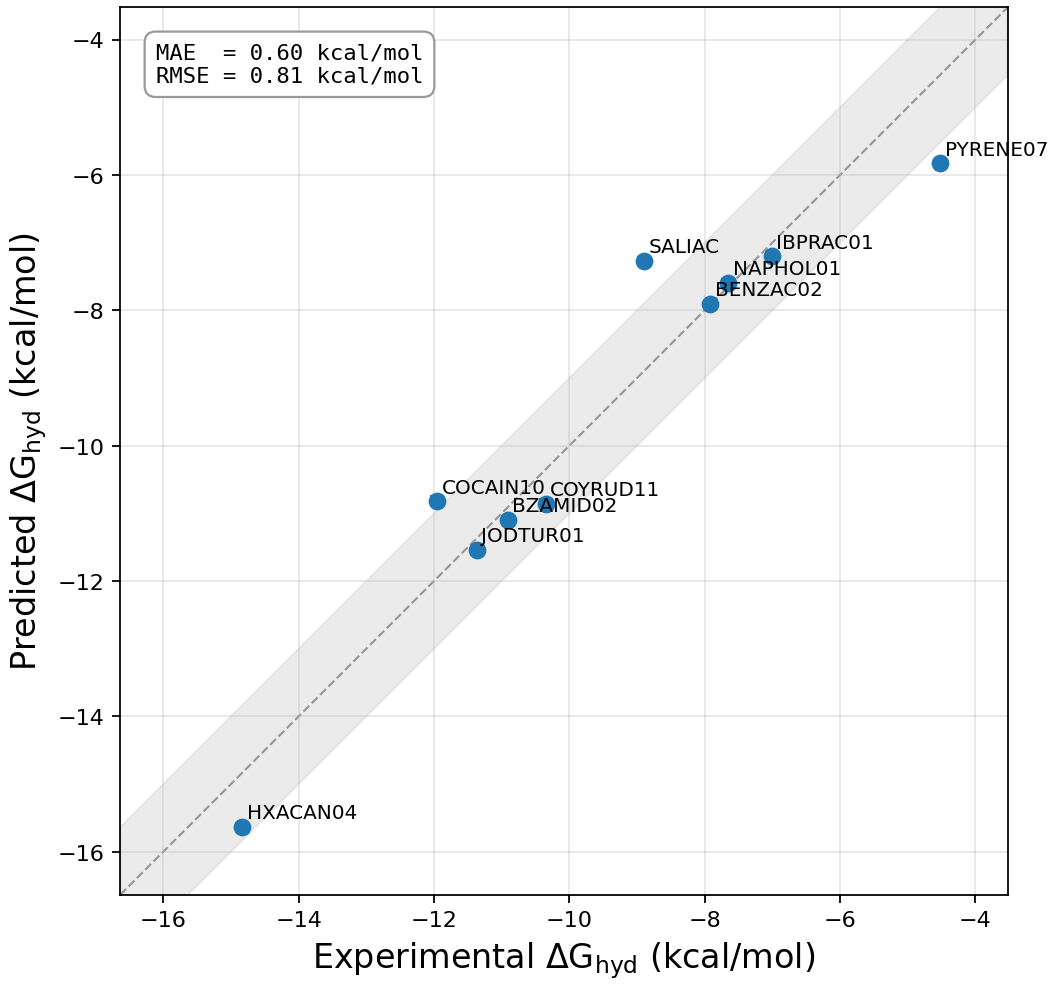}
    \caption{Hydration free energies of the molecules from the organic crystal sublimation dataset. Error within 1 kcal/mol of experiment indicated in grey.}
    \label{fig:lambda_mace_crystals_hydration}
\end{figure}

Figure~\ref{fig:lambda_mace_crystals_hydration} shows the correlation of our predicted hydration free energies for the set of molecular crystals compared to experiment. 
We find that, in agreement with earlier results on the FreeSolv dataset, our predicted hydration free energies are within the typically experimental error across the set of compounds. 
The RMSE of \lambdamace predictions is in line with those of the FreeSolv set, further indicating that the model successfully generalises to larger, more complex chemistries than those in the training set.

\subsubsection{Sublimation free energy}

For the sublimation free energy, we include comparison to PBE DFT with the Grimme dispersion
correction scheme D2 \cite{grimme2006semiempirical} and norm-conserving (NC) core potentials \cite{hamann1979norm}, taken from \cite{brooTransferableForceField2016}, which we refer to as PBE-G NC. 

\begin{figure}
    \centering
    \includegraphics[width=0.9\linewidth]{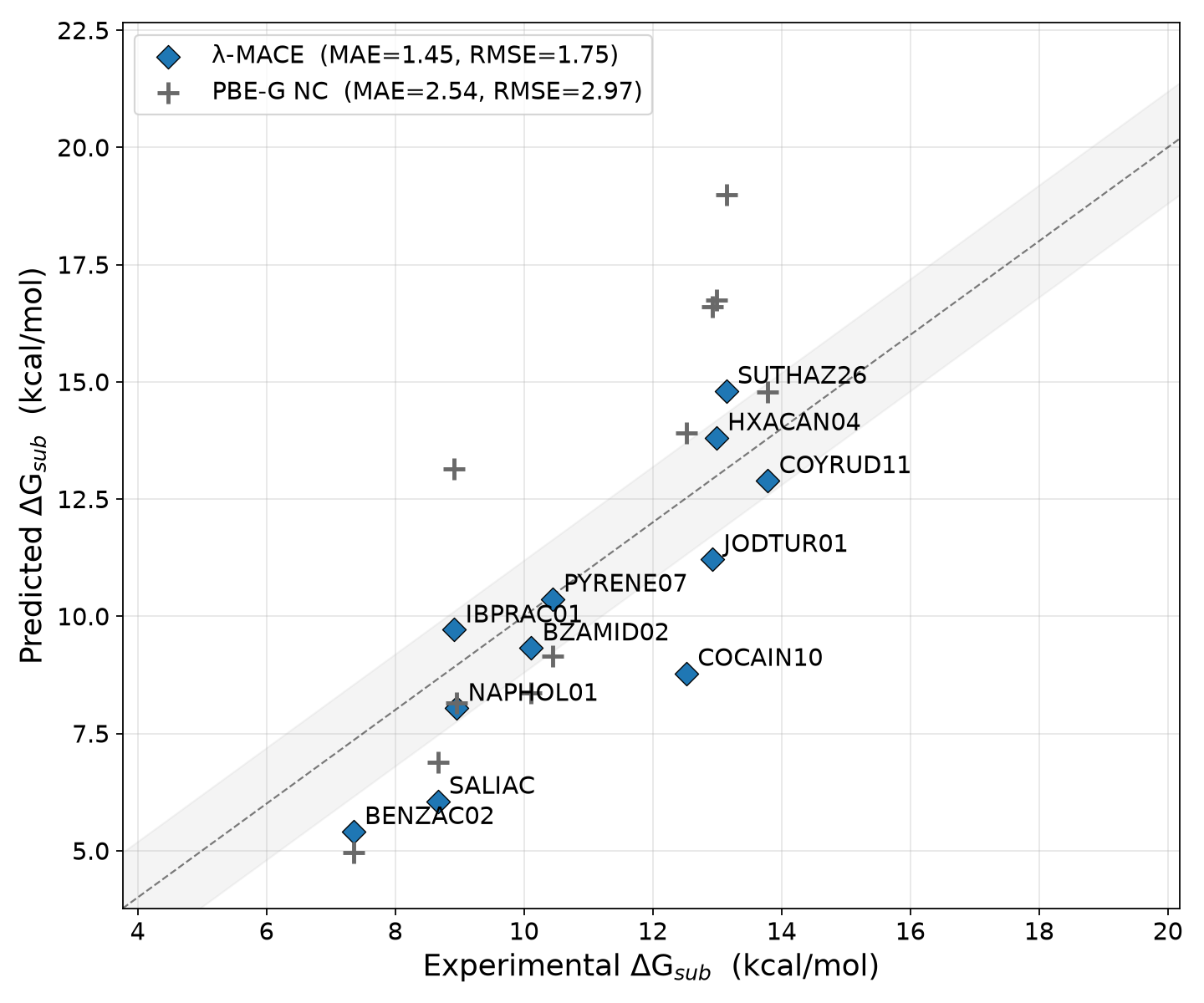}
    \caption{Sublimation free energies predicted by \lambdamace vs disperson corrected DFT. Error within 1 kcal/mol of experiment indicated in grey.}
    \label{fig:sublimation}
\end{figure}

\autoref{fig:sublimation} shows that \lambdamace achieves superior performance to PBE-G NC DFT.
The most likely source of the higher accuracy of \lambdamace is the higher level of theory (ωB97M-D3/def2-TZVPPD) of the DFT dataset that \lambdamace was trained on. 

\subsubsection{Crystalline Solubility}

\begin{figure}
    \centering
    \includegraphics[width=0.9\linewidth]{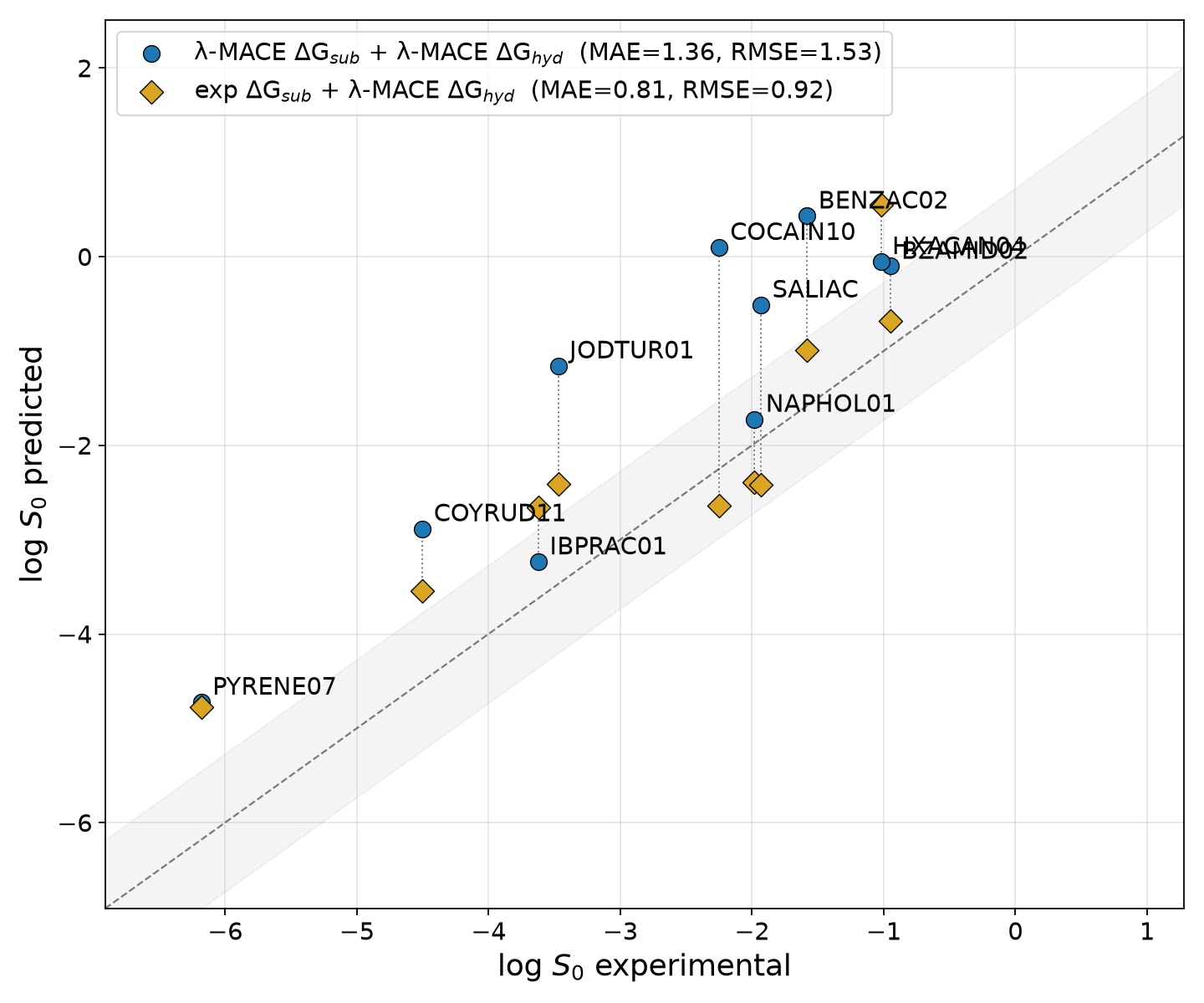}
    \caption{Predicted crystalline solubility with sublimation free energies and hydration free energies calculated with \lambdamace. Additionally, calculating the crystalline solubility using the experimental sublimation with \lambdamace hydration is shown. Error within 1 kcal/mol=0.73 log units of experiment indicated in grey.}
    \label{fig:logS}
\end{figure}

Finally, we combine the predictions from the component free energy calculations to predict the overall crystalline solubility for this set of 10 compounds. 

\autoref{fig:logS} shows reasonable accuracy in the intrinsic solubility prediction (RMSE=1.53 log units), although the compounding of errors between the hydration and sublimation leg provide worse than experimental accuracy (0.17 to 0.6 log units \cite{llinas2020findings}) and chemical accuracy (1 kcal/mol=0.73 log units). 
\autoref{fig:logS} also shows that replacing the predicted sublimation free energies, which is the larger source of error, with the experimental sublimation free energies significantly improves the overall accuracy. 
This is consistent with the individual errors for each leg shown in \autoref{fig:lambda_mace_crystals_hydration} and \autoref{fig:sublimation}. 
Although \citet{brooTransferableForceField2016} provide values for each compound's sublimation free energy, the hydration and log S values reported use large bias corrections and therefore do not permit direct comparison. 
\citet{palmer_2012} report an RMSE of 1.45 on a 25 molecule dataset (a superset of the 10 molecules here), combining molecular integral-equation theory of liquids (3D-RSIM) for the hydration leg, and a classical force field for the sublimation leg. 
Although their performance provides similar accuracy to the \lambdamace based method, the first-principles grounding of \lambdamace would most likely ensure more reliable generalisation as shown in \autoref{sec:generalise-large}. 
\\

Although each leg of our calculations has state-of-the-art accuracy, the compounding errors across each leg mean that the challenge in achieving experimental accuracy is greater, and further improvements to each are needed. 
For both hydration and sublimation legs, coupled cluster fine tuning is a promising avenue for further accuracy \cite{cc-finetuning-1}. 
Additionally, architectural improvements may be beneficial, specifically a promising avenue is using the recently released MACE-POLAR architecture \cite{batatia2026macepolar} which includes principled treatment of long-range electrostatics. 
The current MACE-POLAR-1 family of models is trained on the OMol25 dataset, and is documented to overestimate the water density by 8 \% due to overbinding of the VV-10 dispersion \cite{batatia2026macepolar}. 
MACE-POLAR-1 is therefore unsuitable for hydration free energies and this is an outstanding issue.  
For the sublimation leg, using more accurate approximations to the free energy than the harmonic approximation may provide greater accuracy, specifically to capture finite temperature effects and anharmonicity. 
We leave exploration of these improvements to future work. 

\section{Conclusion}

We have introduced a computational workflow, built on the MACE family of MLIPs, that delivers highly accurate free energy calculations, and we have used it to probe how far a single machine-learned potential can be transferred across distinct thermodynamic problems.
Building on the methodology of \citet{moore2024computinghydrationfreeenergies}, we improve the performance of the existing MACE-OFF models for solvation free energies by expanding their training set, with sampling targeted at the key organic–water interactions that govern this property. 
The resulting potential outperforms classical force fields on industry-standard hydration free energy benchmarks, and, crucially, does so without sacrificing the generality needed to describe the condensed phase. 
We therefore extend its use beyond solvation, combining accurate hydration and sublimation free energies within a thermodynamic cycle for the crystalline solubility of compounds with known experimental crystal structures, and benchmark this combined cycle against experimental values for drug-like molecules. Within that cycle, the hydration leg retains the accuracy established in the original benchmark, demonstrating that performance transfers intact from the isolated solvation problem to the composite one, while the sublimation leg is competitive with dispersion-corrected DFT at a cost that is orders of magnitude lower. 
Taken together, these results map out a domain of applicability in which a single MLIP can be relied upon for both solvation and crystalline thermodynamics — and, equally, they show where that domain ends: despite the promising accuracy reported here, a significant gap to experiment remains, particularly for crystalline thermodynamics, where errors in the lattice contribution dominate the solubility prediction. Closing this gap, and replacing the experimental crystal structures used here with structures obtained from crystal structure prediction, are the two developments required to make de novo solubility prediction viable in drug discovery.

\section{Methods}

\subsection{Solvated cluster generation}

Clusters containing the organic molecule and surrounding solvent were extracted from classical MD simulations. Small molecules were parametrised with OpenFF-2.0 and the water was described by TIP3P. For each system, 2 ns of MD simulation was performed, and 50 equally spaced snapshots were extracted from the trajectory. The central cluster was extracted with a maximum cluster size of 150 atoms in order to make the DFT calculations tractable. Energies and forces were computed with DFT according to the settings used in SPICE directly from the MD configurations. 

\subsection{Hydration Free Energy Calculation}

Hydration free energies were computed using the alchemical pathway introduced by \citet{moore2024computinghydrationfreeenergies}. This procedure exploits the graph neural network architecture of  MACE to define a smooth sequence of alchemical potentials indexed by $\lambda \in [0,1]$. In particular, MACE's first layer messages along edges communicating the solvent and solute are scaled by $\lambda$. This results in the solute-solvent interactions disappearing as $\lambda$ goes to 0. 

For each compound, we simulated a single  molecule of solute in a box of solvent. Boxes were seeded with 1.3~nm padding between the solvent and the box edge.  Systems were equilibrated for 50 ps under the NPT ensemble. Production simulations were performed in NPT ensemble, with the temperature set to 297 Kelvin and pressure of 1 atmosphere. 
Temperature was maintained with Langevin thermostat using BAOAB integration scheme with a step size of 1 fs.

Replica exchange calculations employed 16 equidistantly spaced $\lambda$ windows. A single replica was equilibrated with $\lambda$ fixed to 1 for 50 ps. The resulting state was used to initialise the replicas for all other $\lambda$ values. At least 2 ns of MD was performed per replica, recording states every picosecond. The Multi Bennet Acceptance Ratio (MBAR) was used to estimate free energy differences \cite{mbar}.

\subsection{Crystal Sublimation Free Energy Calculation}
\label{subsec:methods-sub}

For the estimation of sublimation free energy, we use experimental crystal structures from the CCDC database. 
For de novo sublimation free energy prediction, one would need to replace the use of experimental structures with structures output from crystal structure prediction \cite{hunnisett2024seventh_gen,hunnisett2024seventh_rank,gharakhanyan2025fastcsp}.

The Helmholtz free energy of the periodic crystal was computed via the harmonic approximation as
\begin{gather}
  F_\text{crystal}(T) = E_\text{pot} + \sum_{i=1}^{3N} \frac{\hbar\omega_i}{2} + k_B T\ln \left(1 - \exp(\frac{-\hbar\omega_i}{k_B T}) \right)
\end{gather}
for $N$ the number of atoms in a periodic cell and $\omega_i$ the frequencies. 
$E_\text{pot}$ is the zero K lattice potential contribution to the free energy and the remaining terms are due to the vibrational energy and system entropy. 
We assumed room temperature of 298 K. 

These calculations were performed using the Atomic Simulation Environment (ASE) \cite{hjorth2017atomic} and Phononpy \cite{TOGOphononpy}. 
For this, the \lambdamace potential was used to find the atom positions and box vectors that minimize the energy. 
\lambdamace was used to evaluate the zero temperature energy and also to compute the density of states, via finite difference approximation.

The gas state free energy was obtained under the ideal gas rigid-rotator/harmonic-oscillator approximation. 
The Hessian was computed using \lambdamace, via finite difference approximation. 
This was done using the \texttt{Vibrations} and \texttt{IdealGasThermo} ASE modules. 
For each compound, CREST \cite{pracht2024crest} was used to generate candidate gas-phase conformers and tautomers, and the lowest computed gas-phase free
energy across this set was used in the sublimation cycle.

\subsection{Computational performance}

Calculation of sublimation free energies required approximately 1 GPU hour per compound, using an atoms budget of 512 to create the supercell. 
Hydration free energy calculations were run on 8 NVIDIA A100s and required around 12 hours of wall time (\~100 GPU hours) per compound.

\section{Acknowledgments}
The authors gratefully acknowledge financial support from Astex Pharmaceuticals.
We also thank Graeme Day for the helpful comments.

\bibliography{refs}

\appendix
\section{Molecular structures}
\label{appendix}

\begingroup
\scriptsize
\setlength{\tabcolsep}{1.5pt}
\renewcommand{\arraystretch}{1.0}

\refstepcounter{table}
\label{tab:freesolv-molecule-appendix}
\makeatletter
\@makecaption{\fnum@table}{Molecular structures for the selected 70 pharmaceutically relevant FreeSolv compounds}
\makeatother
\vspace{2pt}
\noindent\begin{tabular}{@{}lll@{}}
\toprule
\parbox[t]{0.47\columnwidth}{\raggedright\bfseries SMILES} &
\parbox[t]{0.27\columnwidth}{\raggedright\bfseries freesolv\_id} &
\parbox[t]{0.18\columnwidth}{\centering\bfseries Molecule} \\
\midrule
\parbox[c][1.48cm][c]{0.47\columnwidth}{\raggedright\tiny\ttfamily c\allowbreak{}1\allowbreak{}c\allowbreak{}c\allowbreak{}c\allowbreak{}2\allowbreak{}c\allowbreak{}(\allowbreak{}c\allowbreak{}1\allowbreak{})\allowbreak{}C\allowbreak{}(\allowbreak{}=\allowbreak{}O\allowbreak{})\allowbreak{}c\allowbreak{}3\allowbreak{}c\allowbreak{}c\allowbreak{}c\allowbreak{}c\allowbreak{}(\allowbreak{}c\allowbreak{}3\allowbreak{}C\allowbreak{}2\allowbreak{}=\allowbreak{}O\allowbreak{})\allowbreak{}N\allowbreak{}C\allowbreak{}C\allowbreak{}O}
 & \parbox[c][1.48cm][c]{0.27\columnwidth}{\raggedright\scriptsize\ttfamily mobley\_\allowbreak{}2316618}
 & \parbox[c][1.48cm][c]{0.18\columnwidth}{\centering\includegraphics[width=\linewidth,height=1.32cm,keepaspectratio]{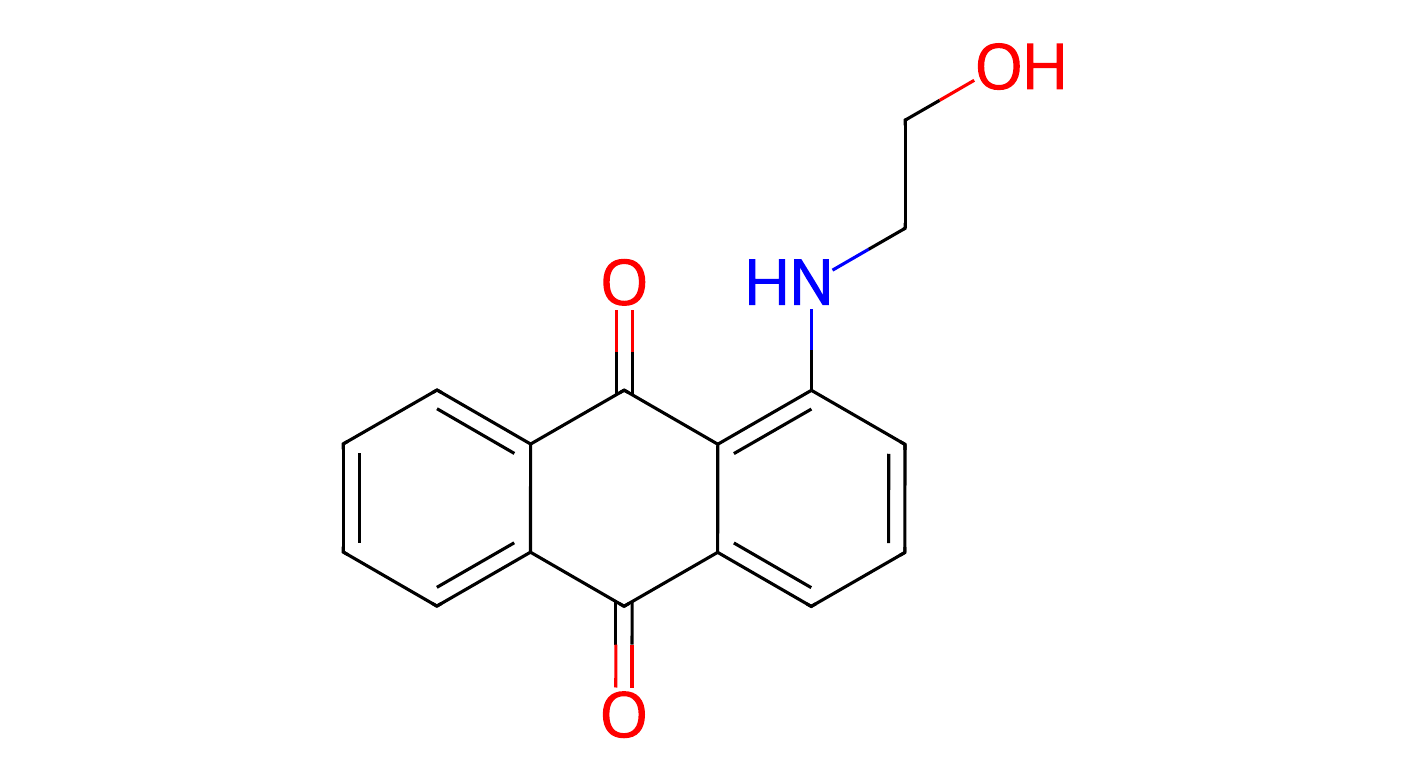}} \\
\parbox[c][1.48cm][c]{0.47\columnwidth}{\raggedright\scriptsize\ttfamily C=\allowbreak{}C(\allowbreak{}Cl)\allowbreak{}Cl}
 & \parbox[c][1.48cm][c]{0.27\columnwidth}{\raggedright\scriptsize\ttfamily mobley\_\allowbreak{}2493732}
 & \parbox[c][1.48cm][c]{0.18\columnwidth}{\centering\includegraphics[width=\linewidth,height=1.32cm,keepaspectratio]{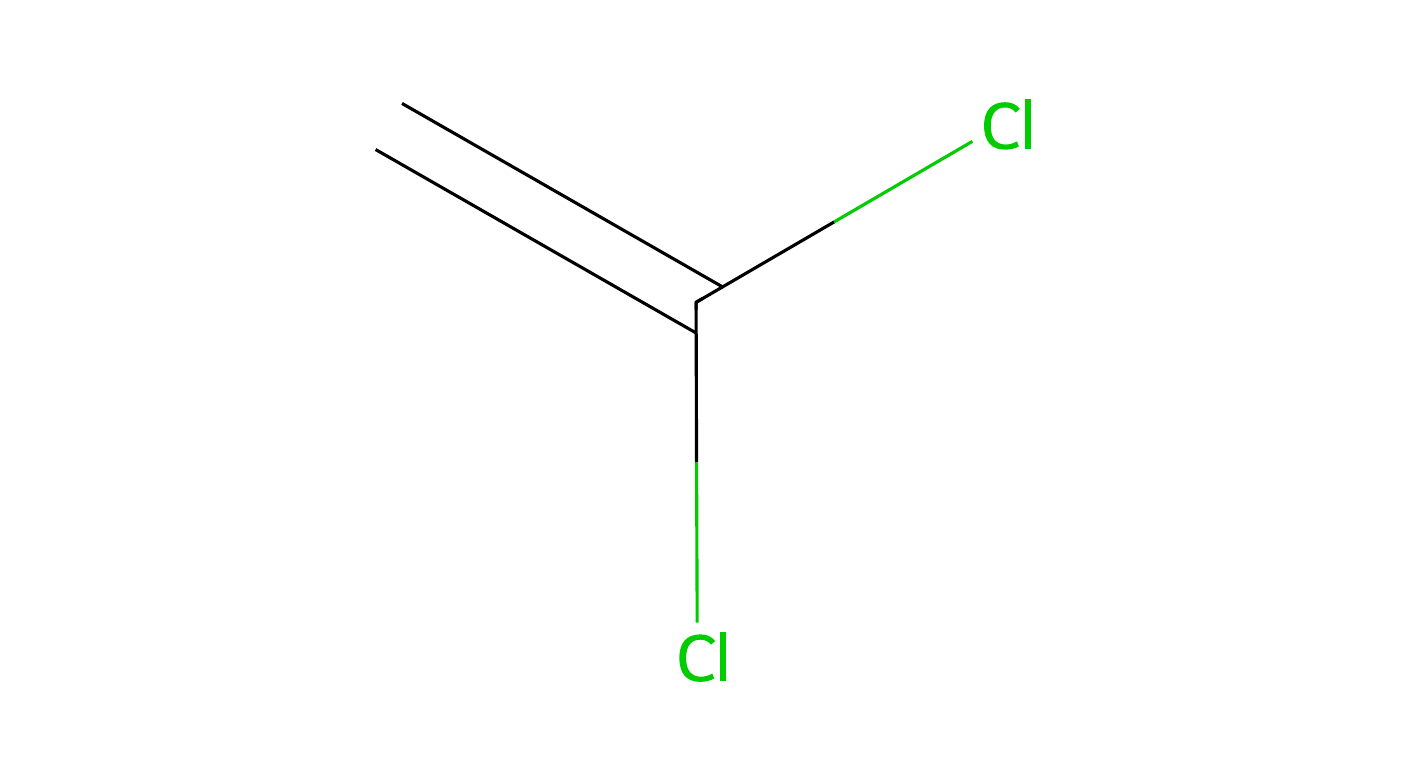}} \\
\parbox[c][1.48cm][c]{0.47\columnwidth}{\raggedright\scriptsize\ttfamily CC(\allowbreak{}C)\allowbreak{}(\allowbreak{}C)\allowbreak{}O}
 & \parbox[c][1.48cm][c]{0.27\columnwidth}{\raggedright\scriptsize\ttfamily mobley\_\allowbreak{}1650157}
 & \parbox[c][1.48cm][c]{0.18\columnwidth}{\centering\includegraphics[width=\linewidth,height=1.32cm,keepaspectratio]{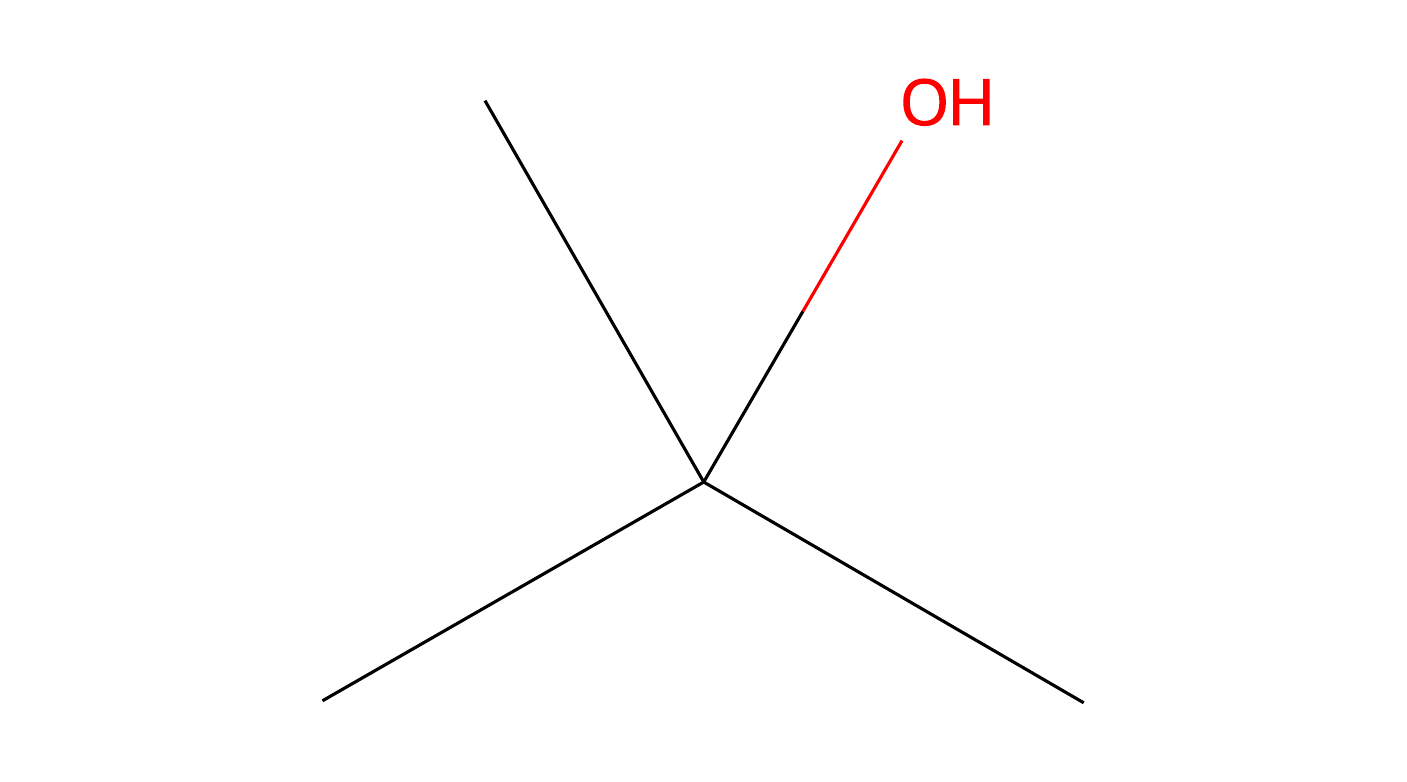}} \\
\parbox[c][1.48cm][c]{0.47\columnwidth}{\raggedright\scriptsize\ttfamily c1cc(\allowbreak{}cnc1)\allowbreak{}Cl}
 & \parbox[c][1.48cm][c]{0.27\columnwidth}{\raggedright\scriptsize\ttfamily mobley\_\allowbreak{}2789243}
 & \parbox[c][1.48cm][c]{0.18\columnwidth}{\centering\includegraphics[width=\linewidth,height=1.32cm,keepaspectratio]{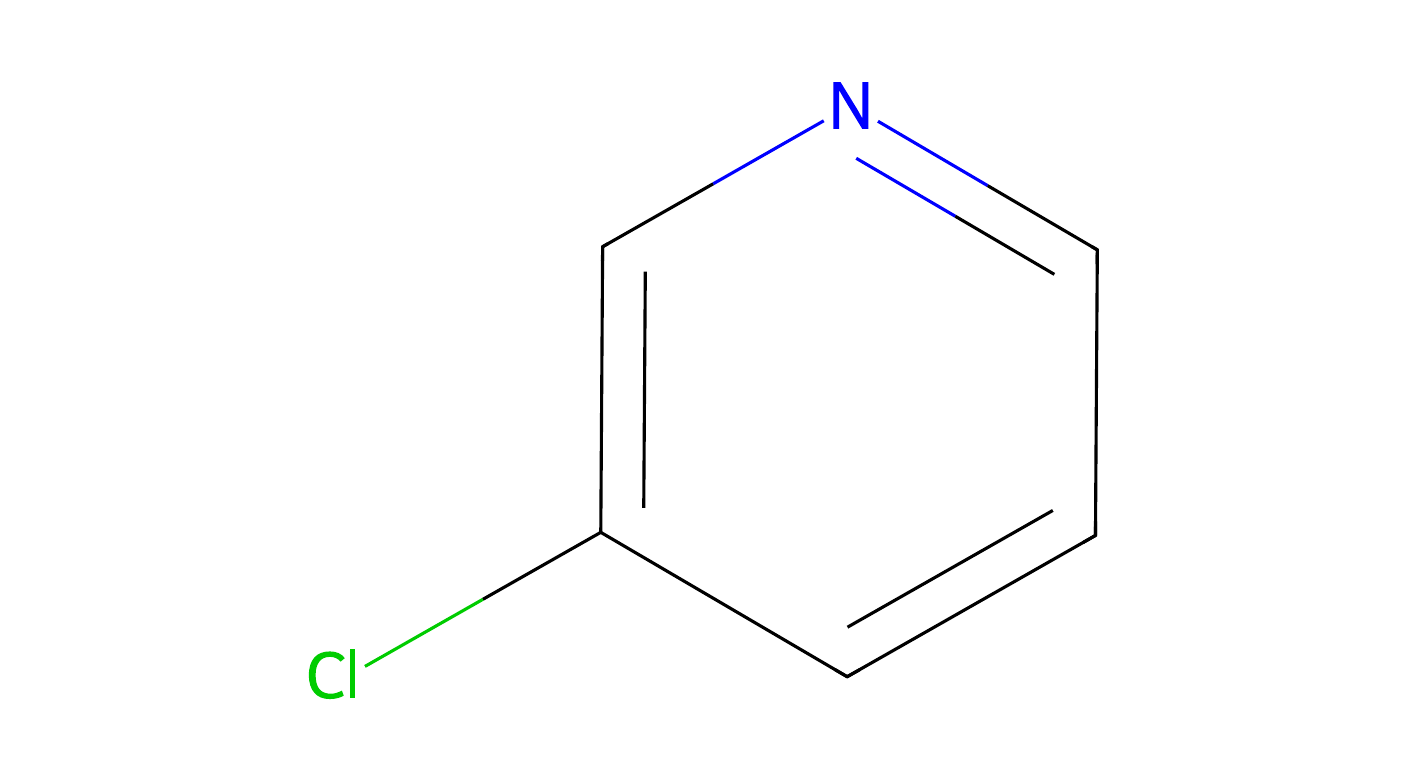}} \\
\parbox[c][1.48cm][c]{0.47\columnwidth}{\raggedright\scriptsize\ttfamily CC(\allowbreak{}=\allowbreak{}O)\allowbreak{}N}
 & \parbox[c][1.48cm][c]{0.27\columnwidth}{\raggedright\scriptsize\ttfamily mobley\_\allowbreak{}8048190}
 & \parbox[c][1.48cm][c]{0.18\columnwidth}{\centering\includegraphics[width=\linewidth,height=1.32cm,keepaspectratio]{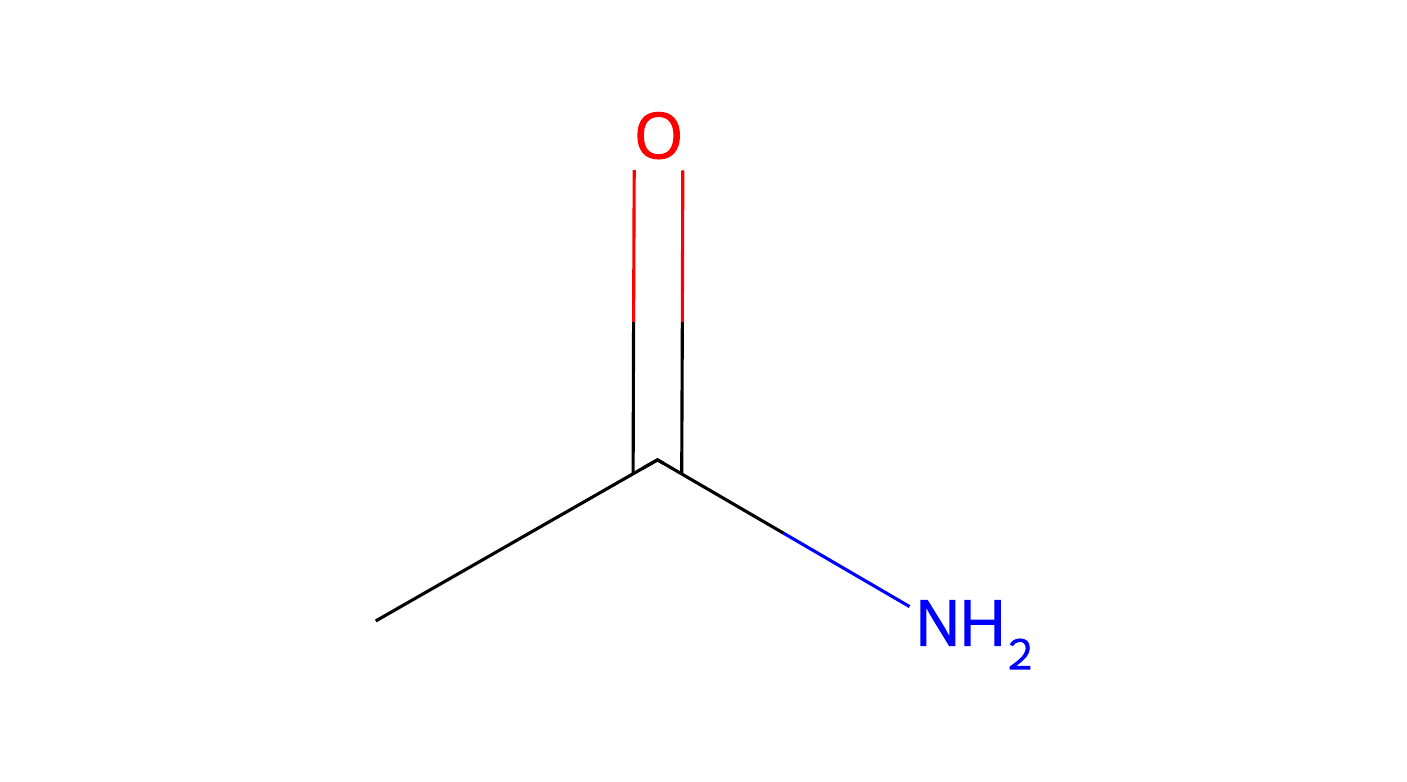}} \\
\parbox[c][1.48cm][c]{0.47\columnwidth}{\raggedright\scriptsize\ttfamily CC(\allowbreak{}=\allowbreak{}O)\allowbreak{}C}
 & \parbox[c][1.48cm][c]{0.27\columnwidth}{\raggedright\scriptsize\ttfamily mobley\_\allowbreak{}3867265}
 & \parbox[c][1.48cm][c]{0.18\columnwidth}{\centering\includegraphics[width=\linewidth,height=1.32cm,keepaspectratio]{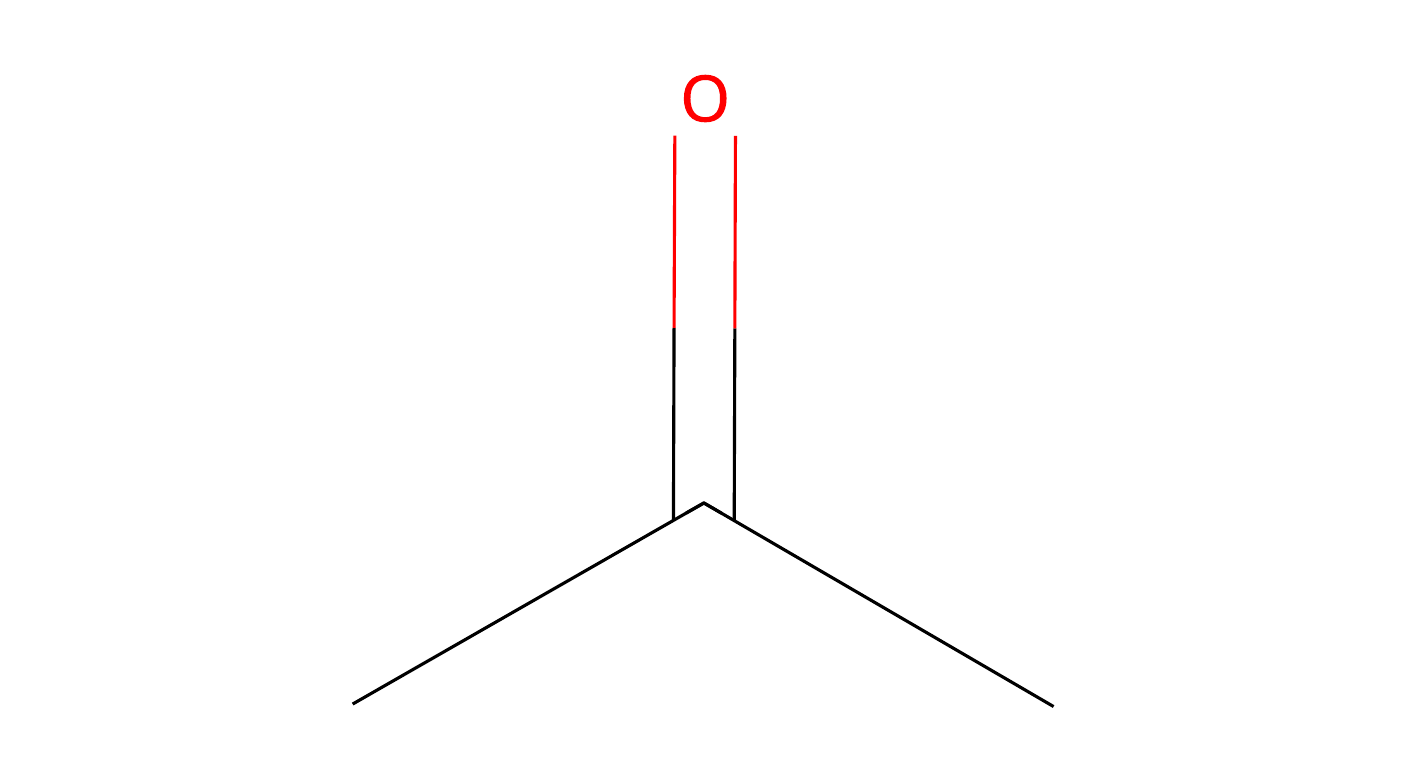}} \\
\parbox[c][1.48cm][c]{0.47\columnwidth}{\raggedright\scriptsize\ttfamily CC\#\allowbreak{}N}
 & \parbox[c][1.48cm][c]{0.27\columnwidth}{\raggedright\scriptsize\ttfamily mobley\_\allowbreak{}7532833}
 & \parbox[c][1.48cm][c]{0.18\columnwidth}{\centering\includegraphics[width=\linewidth,height=1.32cm,keepaspectratio]{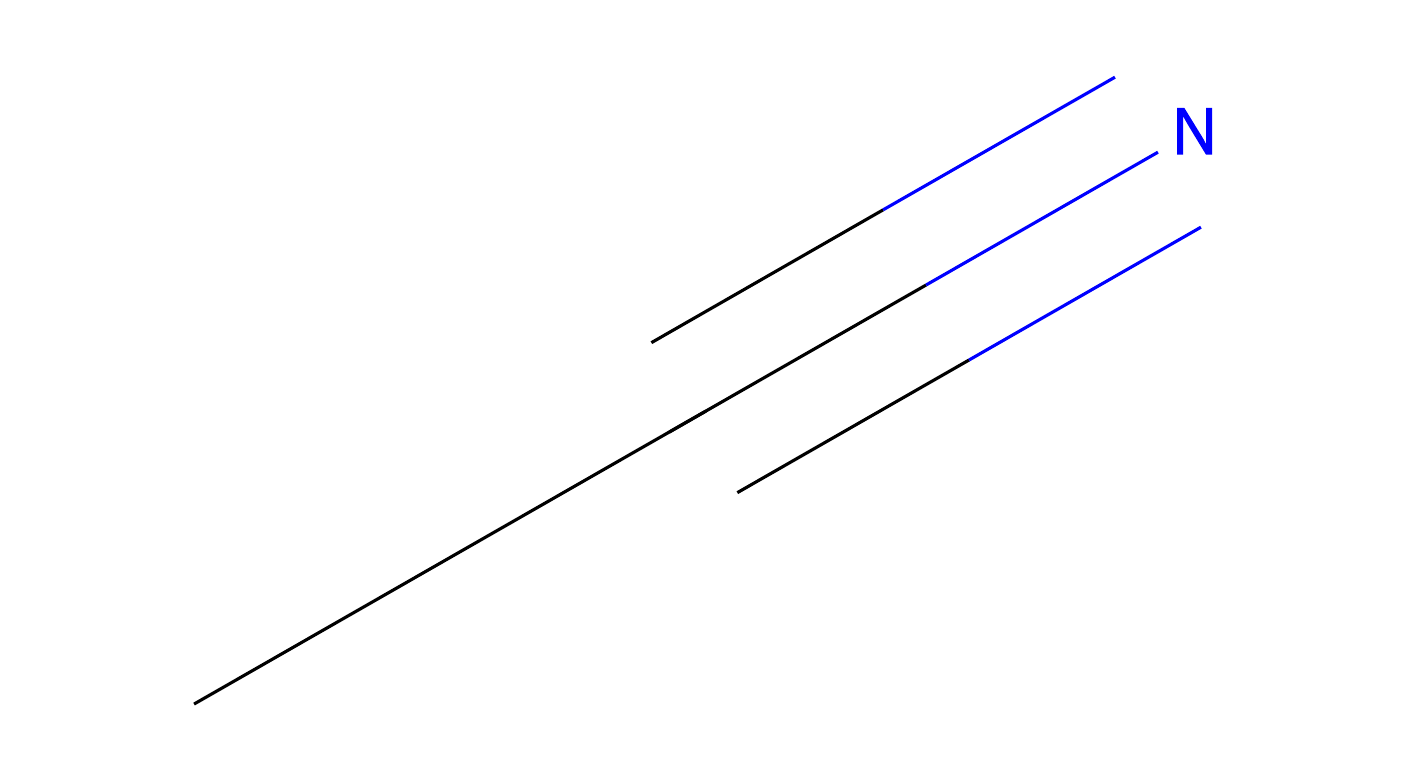}} \\
\parbox[c][1.48cm][c]{0.47\columnwidth}{\raggedright\tiny\ttfamily C\allowbreak{}C\allowbreak{}c\allowbreak{}1\allowbreak{}c\allowbreak{}c\allowbreak{}c\allowbreak{}c\allowbreak{}(\allowbreak{}c\allowbreak{}1\allowbreak{}N\allowbreak{}(\allowbreak{}C\allowbreak{}O\allowbreak{}C\allowbreak{})\allowbreak{}C\allowbreak{}(\allowbreak{}=\allowbreak{}O\allowbreak{})\allowbreak{}C\allowbreak{}C\allowbreak{}l\allowbreak{})\allowbreak{}C\allowbreak{}C}
 & \parbox[c][1.48cm][c]{0.27\columnwidth}{\raggedright\scriptsize\ttfamily mobley\_\allowbreak{}8124669}
 & \parbox[c][1.48cm][c]{0.18\columnwidth}{\centering\includegraphics[width=\linewidth,height=1.32cm,keepaspectratio]{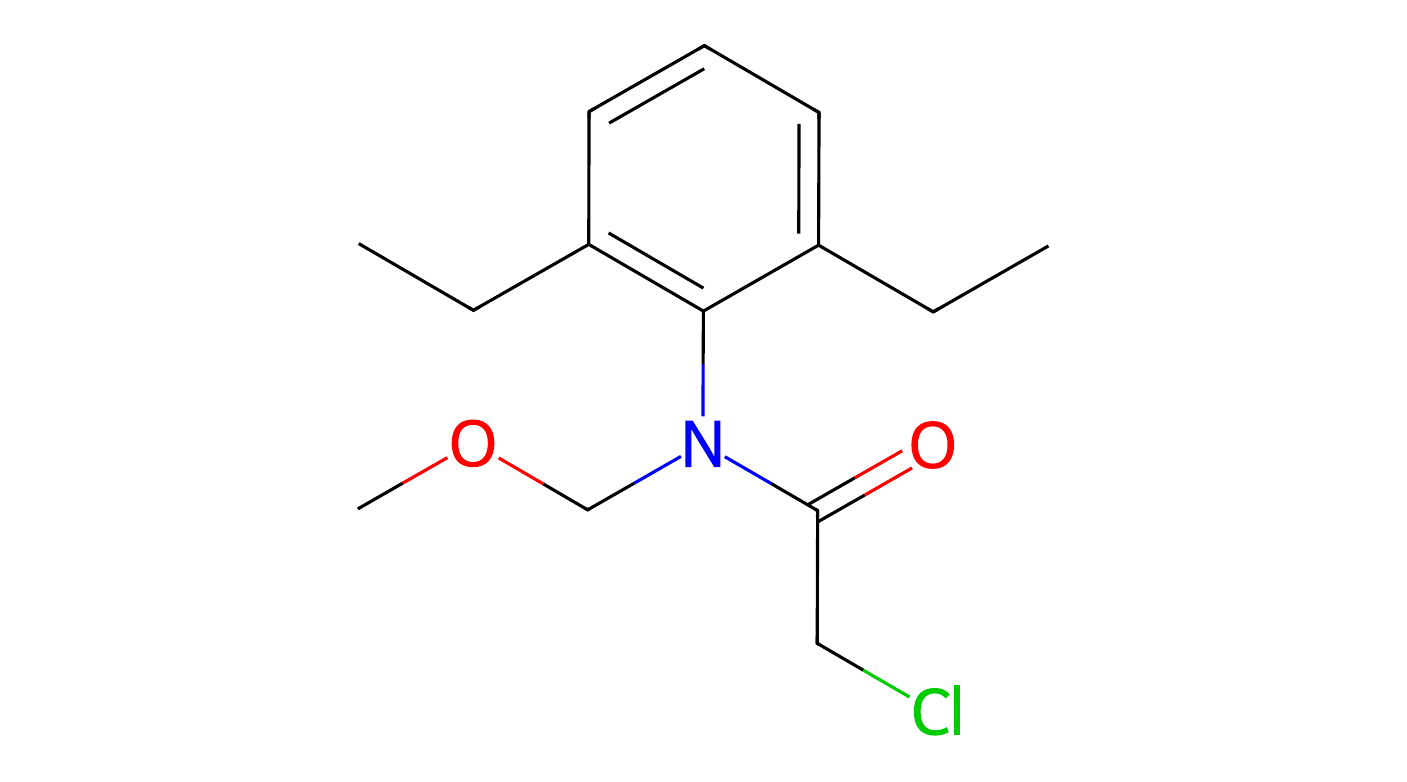}} \\
\parbox[c][1.48cm][c]{0.47\columnwidth}{\raggedright\scriptsize\ttfamily c1ccc(\allowbreak{}cc1)\allowbreak{}N}
 & \parbox[c][1.48cm][c]{0.27\columnwidth}{\raggedright\scriptsize\ttfamily mobley\_\allowbreak{}4883284}
 & \parbox[c][1.48cm][c]{0.18\columnwidth}{\centering\includegraphics[width=\linewidth,height=1.32cm,keepaspectratio]{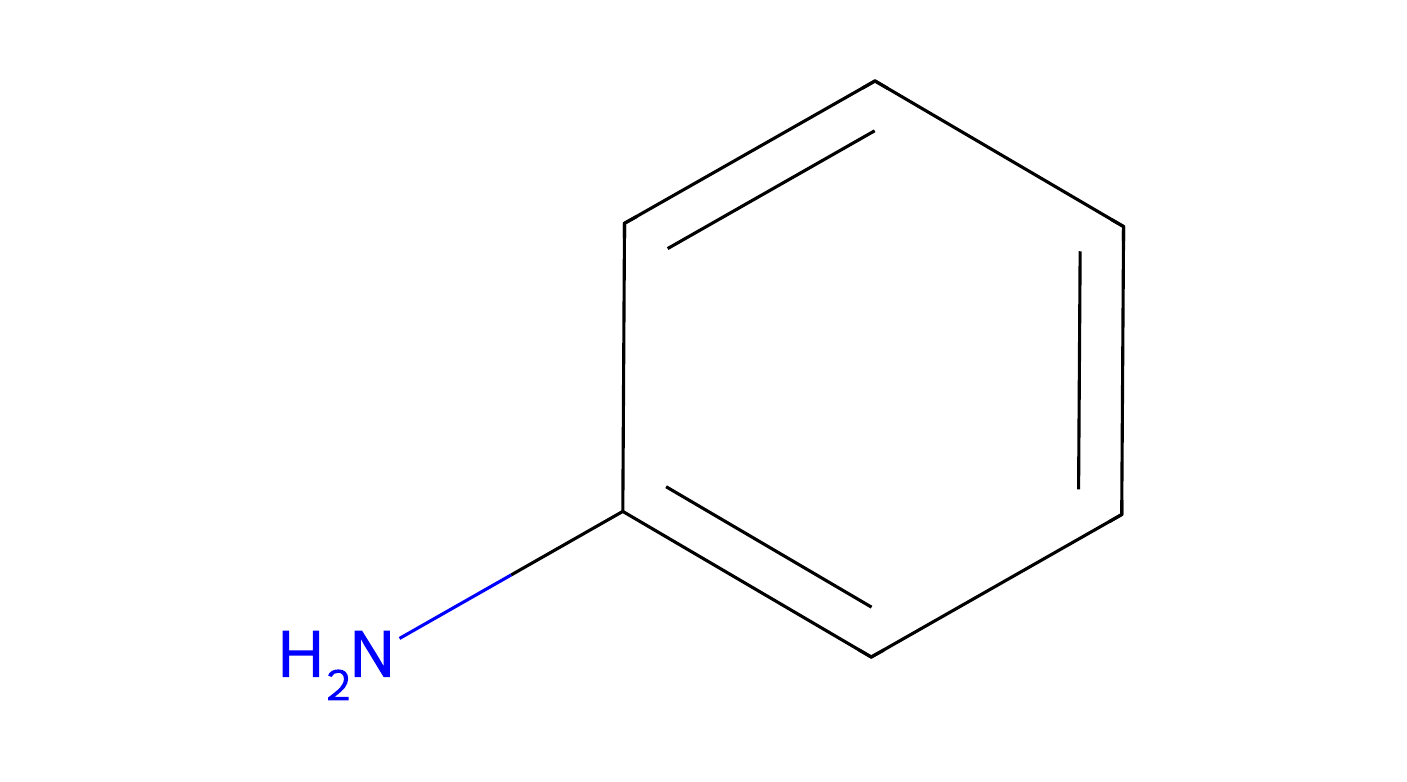}} \\
\parbox[c][1.48cm][c]{0.47\columnwidth}{\raggedright\scriptsize\ttfamily COc1ccccc1}
 & \parbox[c][1.48cm][c]{0.27\columnwidth}{\raggedright\scriptsize\ttfamily mobley\_\allowbreak{}7295828}
 & \parbox[c][1.48cm][c]{0.18\columnwidth}{\centering\includegraphics[width=\linewidth,height=1.32cm,keepaspectratio]{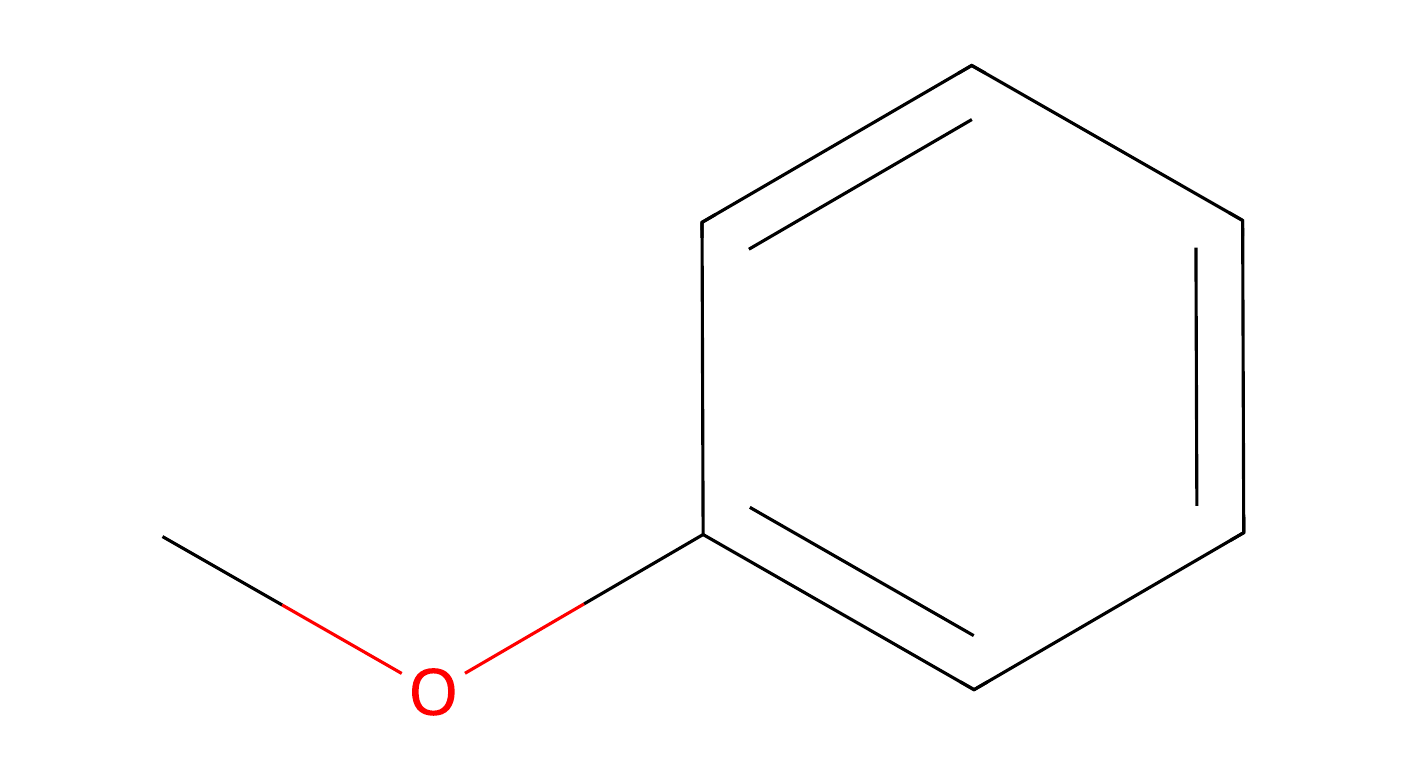}} \\
\parbox[c][1.48cm][c]{0.47\columnwidth}{\raggedright\scriptsize\ttfamily c1ccc(\allowbreak{}cc1)\allowbreak{}Br}
 & \parbox[c][1.48cm][c]{0.27\columnwidth}{\raggedright\scriptsize\ttfamily mobley\_\allowbreak{}7599023}
 & \parbox[c][1.48cm][c]{0.18\columnwidth}{\centering\includegraphics[width=\linewidth,height=1.32cm,keepaspectratio]{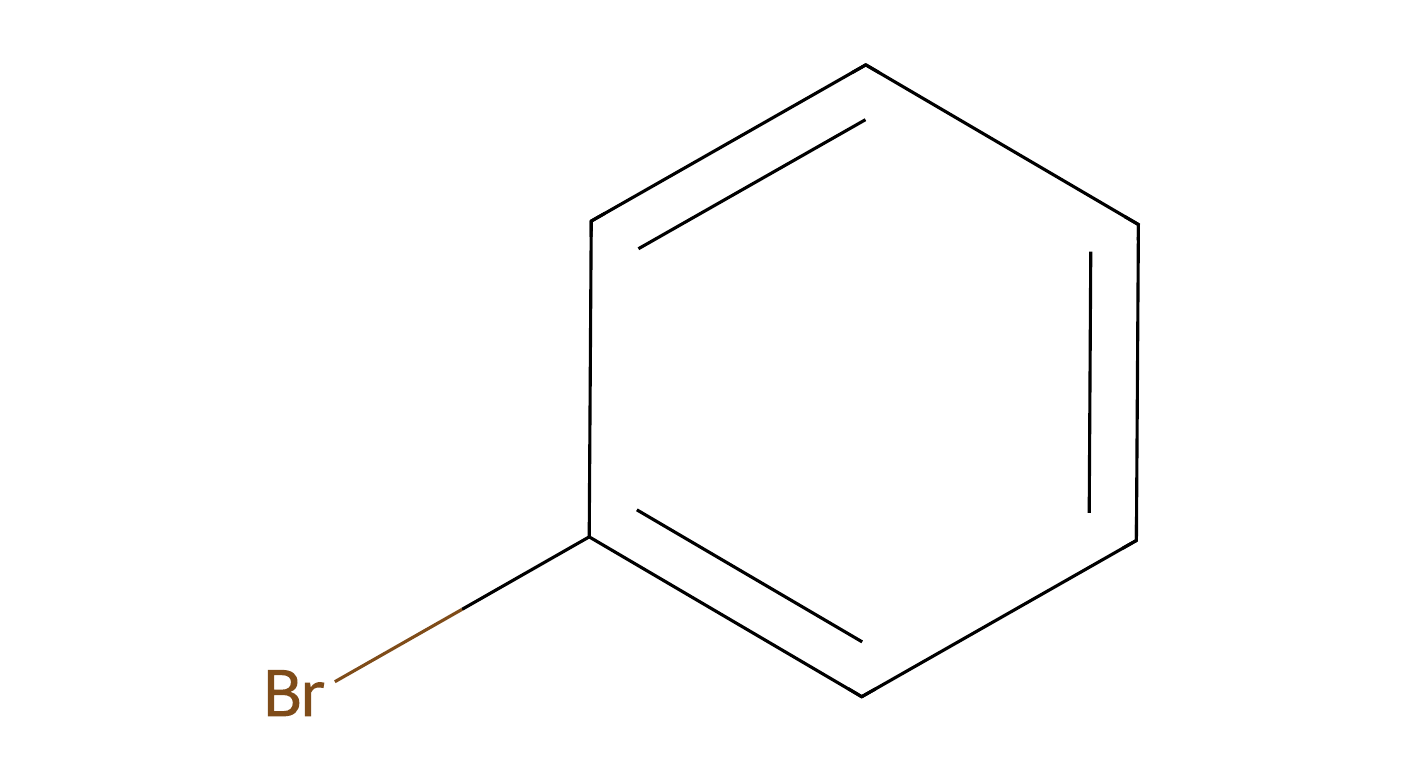}} \\
\parbox[c][1.48cm][c]{0.47\columnwidth}{\raggedright\scriptsize\ttfamily CBr}
 & \parbox[c][1.48cm][c]{0.27\columnwidth}{\raggedright\scriptsize\ttfamily mobley\_\allowbreak{}8983100}
 & \parbox[c][1.48cm][c]{0.18\columnwidth}{\centering\includegraphics[width=\linewidth,height=1.32cm,keepaspectratio]{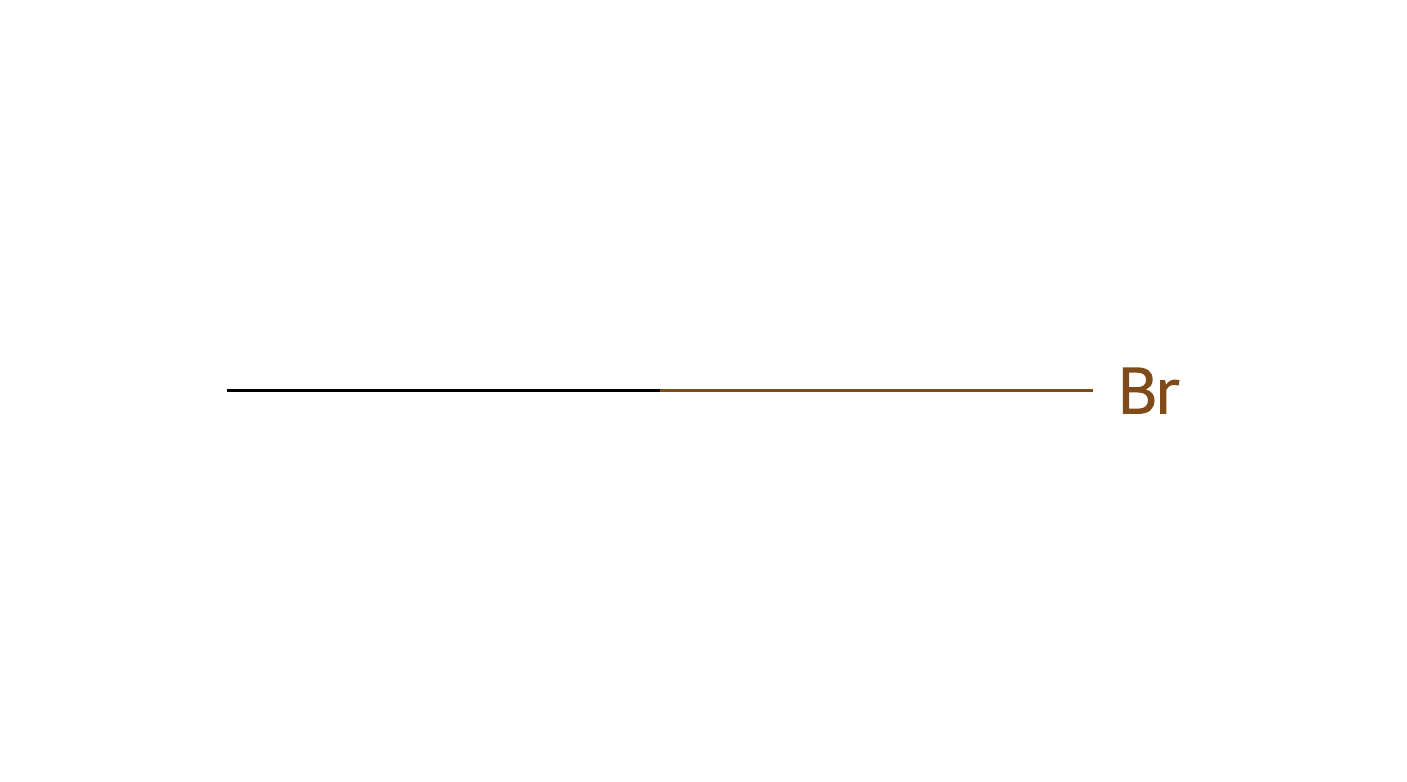}} \\
\bottomrule
\end{tabular}

\newpage
\noindent\textbf{\tablename~\thetable{} -- continued}
\par\vspace{2pt}
\noindent\begin{tabular}{@{}lll@{}}
\toprule
\parbox[t]{0.47\columnwidth}{\raggedright\bfseries SMILES} &
\parbox[t]{0.27\columnwidth}{\raggedright\bfseries freesolv\_id} &
\parbox[t]{0.18\columnwidth}{\centering\bfseries Molecule} \\
\midrule
\parbox[c][1.48cm][c]{0.47\columnwidth}{\raggedright\scriptsize\ttfamily CCl}
 & \parbox[c][1.48cm][c]{0.27\columnwidth}{\raggedright\scriptsize\ttfamily mobley\_\allowbreak{}4434915}
 & \parbox[c][1.48cm][c]{0.18\columnwidth}{\centering\includegraphics[width=\linewidth,height=1.32cm,keepaspectratio]{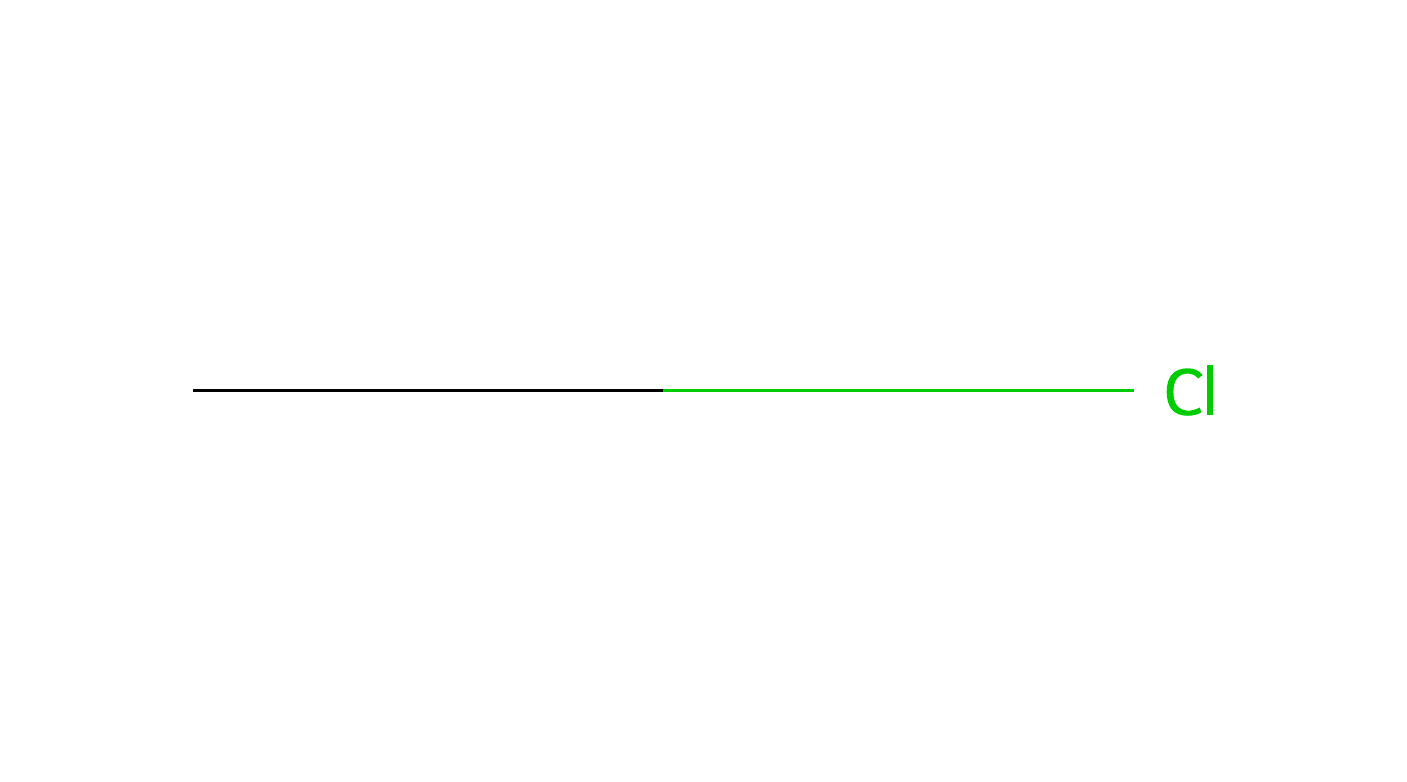}} \\
\parbox[c][1.48cm][c]{0.47\columnwidth}{\raggedright\scriptsize\ttfamily c1ccc(\allowbreak{}cc1)\allowbreak{}Oc2ccccc2}
 & \parbox[c][1.48cm][c]{0.27\columnwidth}{\raggedright\scriptsize\ttfamily mobley\_\allowbreak{}6102880}
 & \parbox[c][1.48cm][c]{0.18\columnwidth}{\centering\includegraphics[width=\linewidth,height=1.32cm,keepaspectratio]{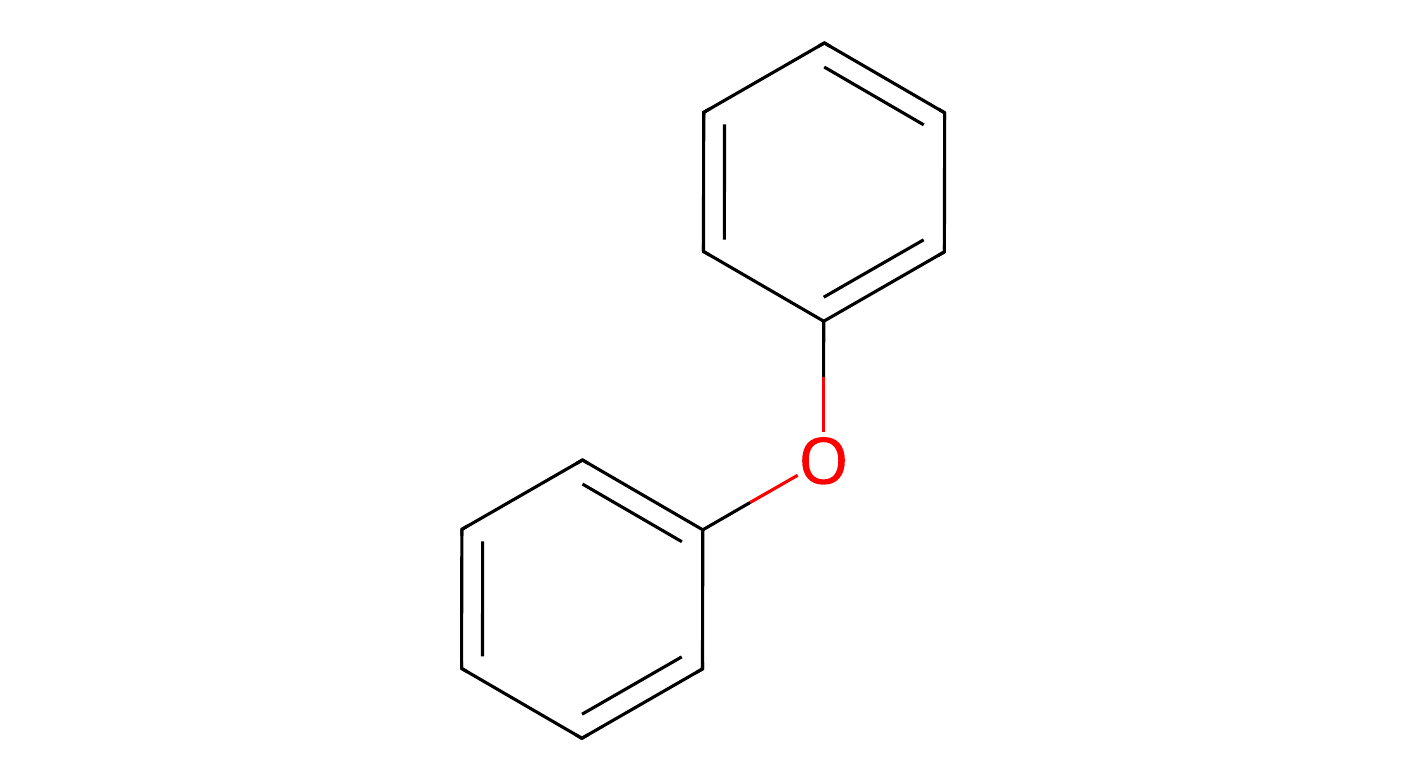}} \\
\parbox[c][1.48cm][c]{0.47\columnwidth}{\raggedright\scriptsize\ttfamily C=\allowbreak{}C}
 & \parbox[c][1.48cm][c]{0.27\columnwidth}{\raggedright\scriptsize\ttfamily mobley\_\allowbreak{}6091882}
 & \parbox[c][1.48cm][c]{0.18\columnwidth}{\centering\includegraphics[width=\linewidth,height=1.32cm,keepaspectratio]{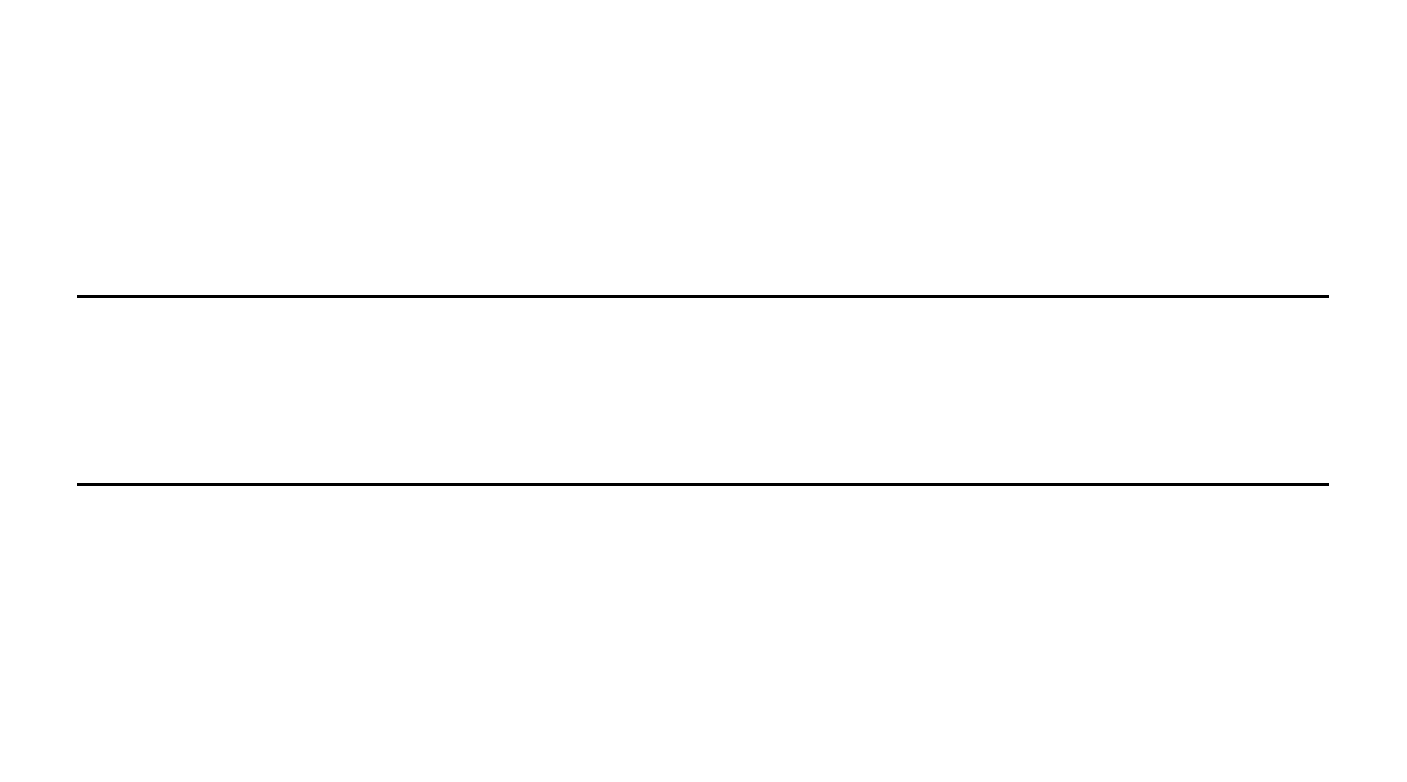}} \\
\parbox[c][1.48cm][c]{0.47\columnwidth}{\raggedright\scriptsize\ttfamily C(\allowbreak{}CO)\allowbreak{}O}
 & \parbox[c][1.48cm][c]{0.27\columnwidth}{\raggedright\scriptsize\ttfamily mobley\_\allowbreak{}4639255}
 & \parbox[c][1.48cm][c]{0.18\columnwidth}{\centering\includegraphics[width=\linewidth,height=1.32cm,keepaspectratio]{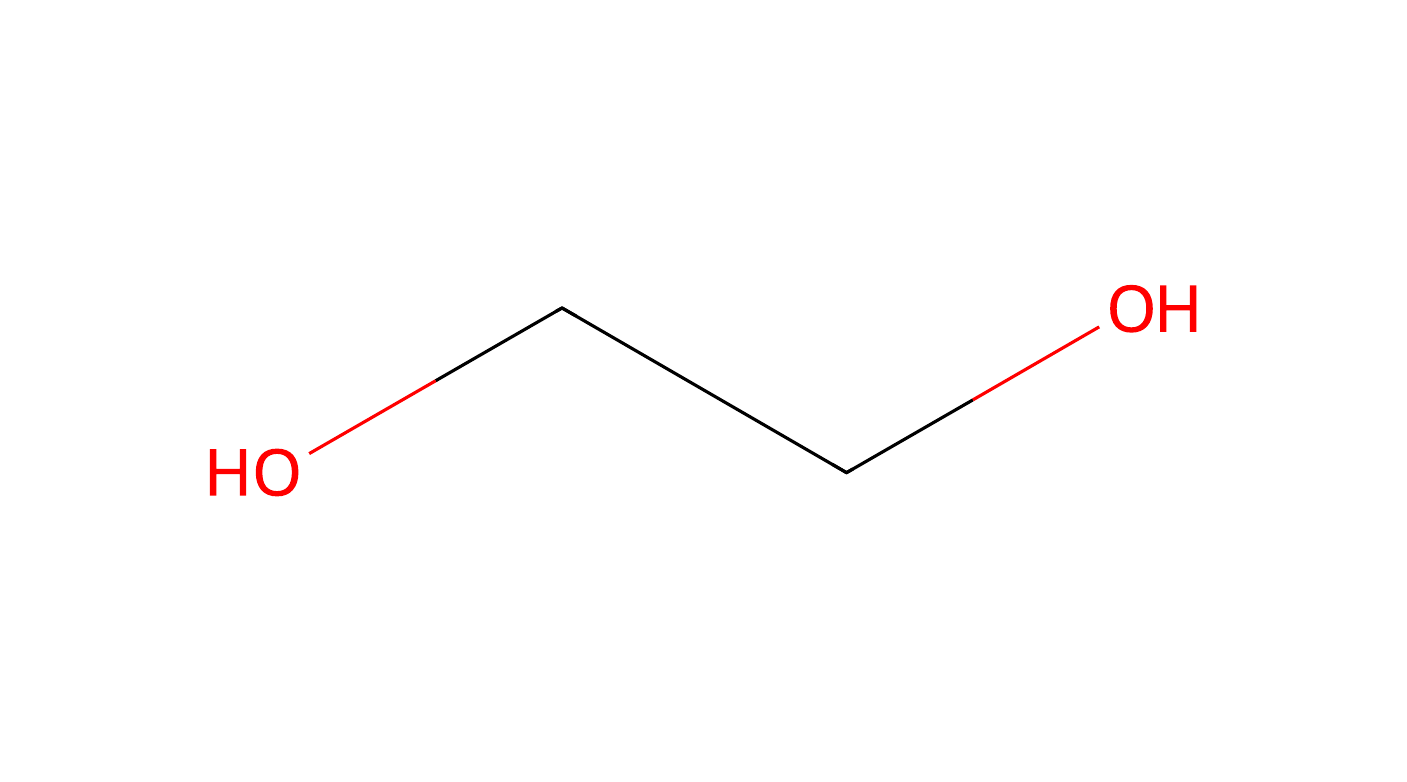}} \\
\parbox[c][1.48cm][c]{0.47\columnwidth}{\raggedright\scriptsize\ttfamily CN(\allowbreak{}C)\allowbreak{}C(\allowbreak{}=\allowbreak{}O)\allowbreak{}Nc1ccccc1}
 & \parbox[c][1.48cm][c]{0.27\columnwidth}{\raggedright\scriptsize\ttfamily mobley\_\allowbreak{}6861308}
 & \parbox[c][1.48cm][c]{0.18\columnwidth}{\centering\includegraphics[width=\linewidth,height=1.32cm,keepaspectratio]{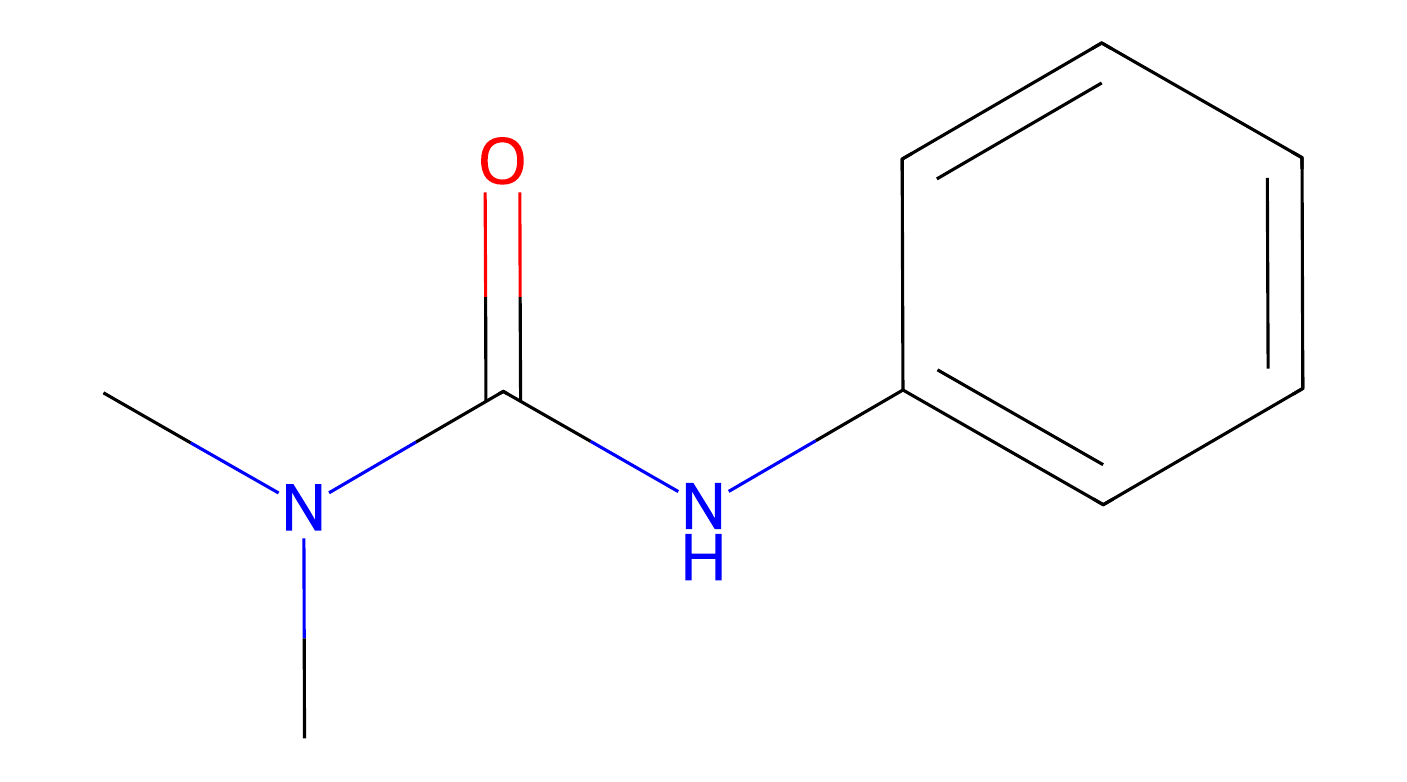}} \\
\parbox[c][1.48cm][c]{0.47\columnwidth}{\raggedright\scriptsize\ttfamily c1ccc(\allowbreak{}cc1)\allowbreak{}F}
 & \parbox[c][1.48cm][c]{0.27\columnwidth}{\raggedright\scriptsize\ttfamily mobley\_\allowbreak{}4483973}
 & \parbox[c][1.48cm][c]{0.18\columnwidth}{\centering\includegraphics[width=\linewidth,height=1.32cm,keepaspectratio]{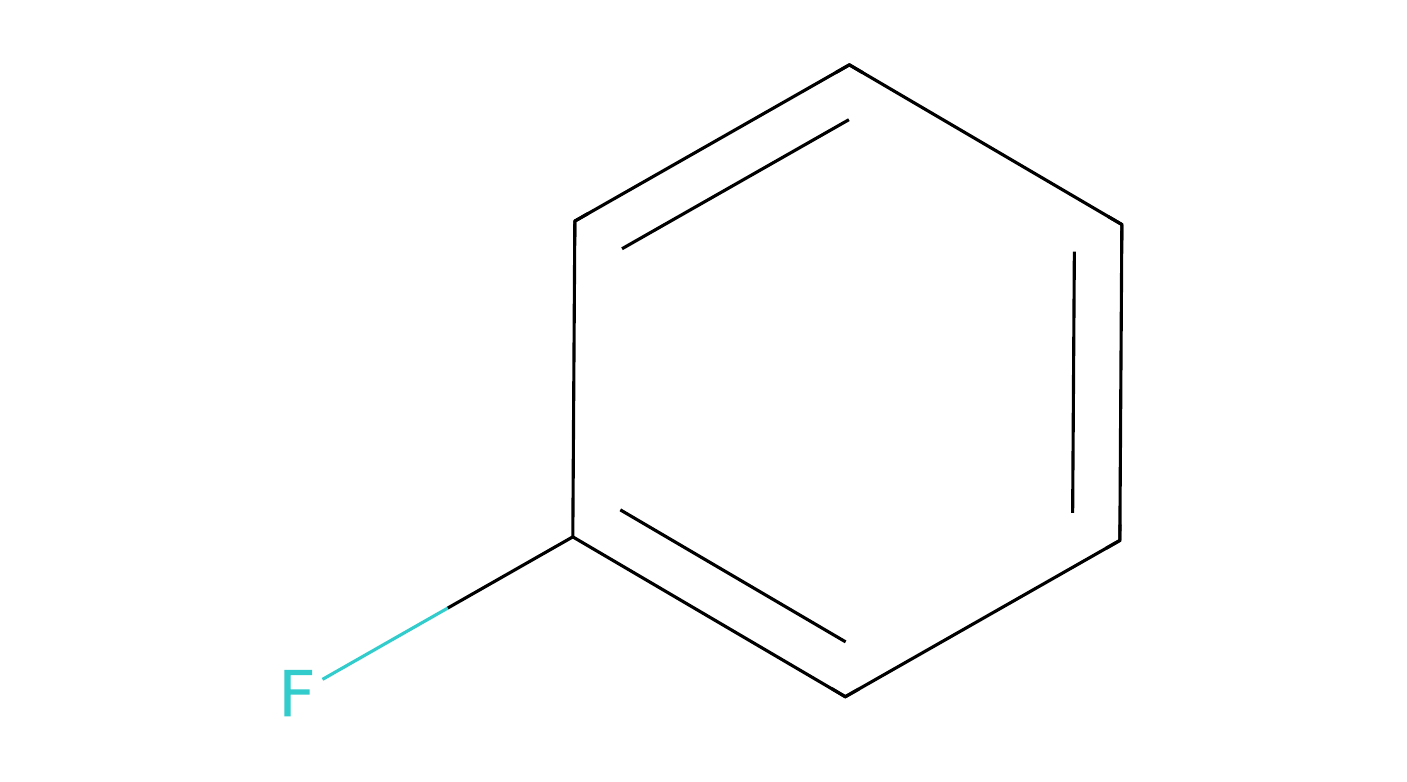}} \\
\parbox[c][1.48cm][c]{0.47\columnwidth}{\raggedright\scriptsize\ttfamily CF}
 & \parbox[c][1.48cm][c]{0.27\columnwidth}{\raggedright\scriptsize\ttfamily mobley\_\allowbreak{}9029594}
 & \parbox[c][1.48cm][c]{0.18\columnwidth}{\centering\includegraphics[width=\linewidth,height=1.32cm,keepaspectratio]{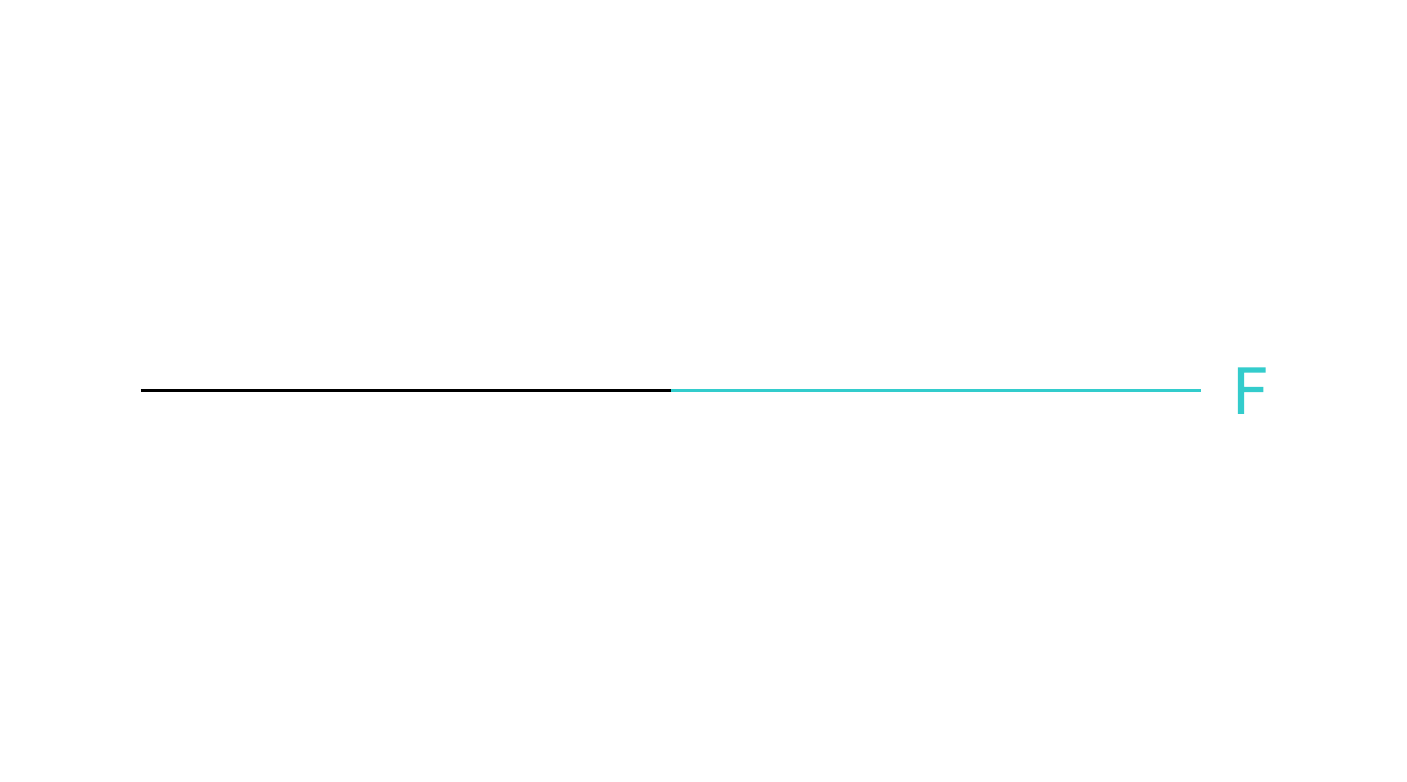}} \\
\parbox[c][1.48cm][c]{0.47\columnwidth}{\raggedright\scriptsize\ttfamily C=\allowbreak{}O}
 & \parbox[c][1.48cm][c]{0.27\columnwidth}{\raggedright\scriptsize\ttfamily mobley\_\allowbreak{}2146331}
 & \parbox[c][1.48cm][c]{0.18\columnwidth}{\centering\includegraphics[width=\linewidth,height=1.32cm,keepaspectratio]{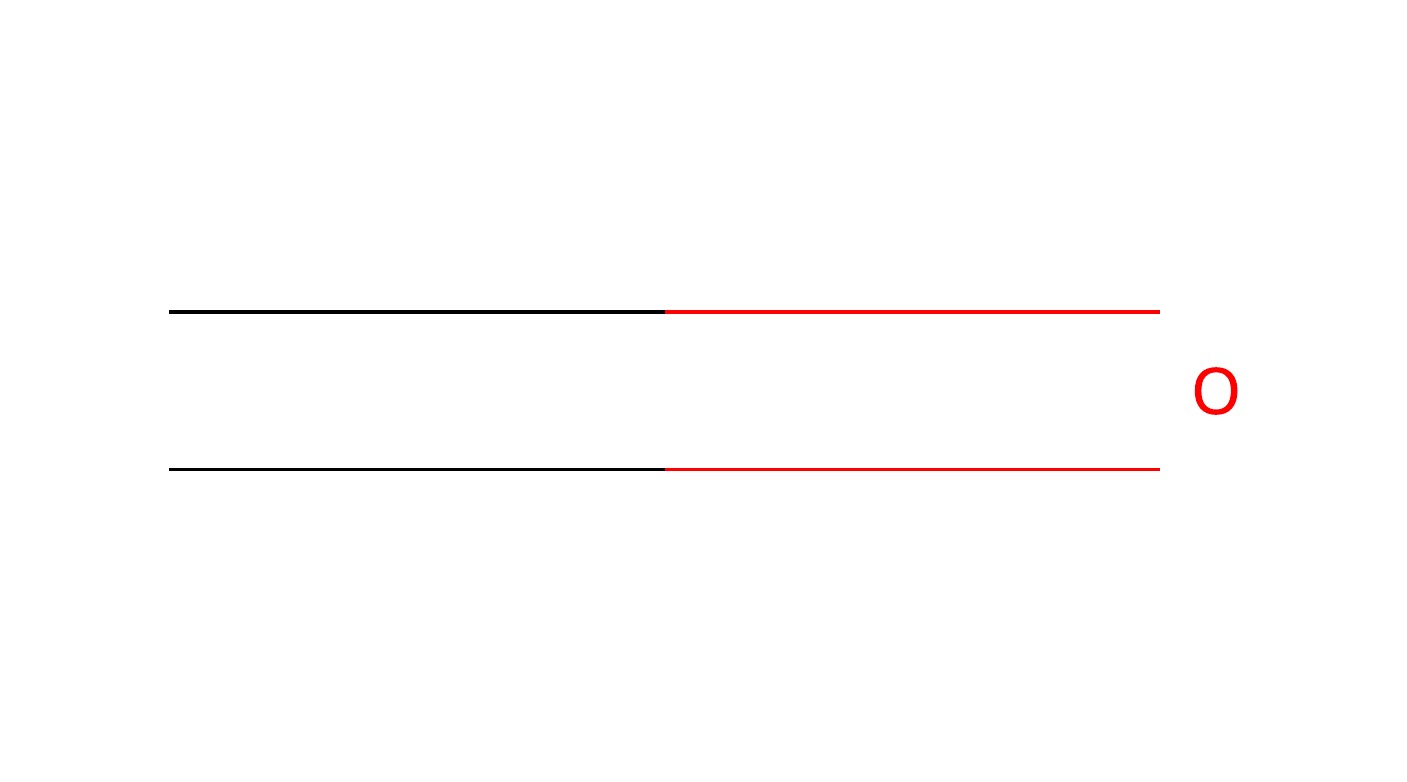}} \\
\parbox[c][1.48cm][c]{0.47\columnwidth}{\raggedright\scriptsize\ttfamily CS}
 & \parbox[c][1.48cm][c]{0.27\columnwidth}{\raggedright\scriptsize\ttfamily mobley\_\allowbreak{}525934}
 & \parbox[c][1.48cm][c]{0.18\columnwidth}{\centering\includegraphics[width=\linewidth,height=1.32cm,keepaspectratio]{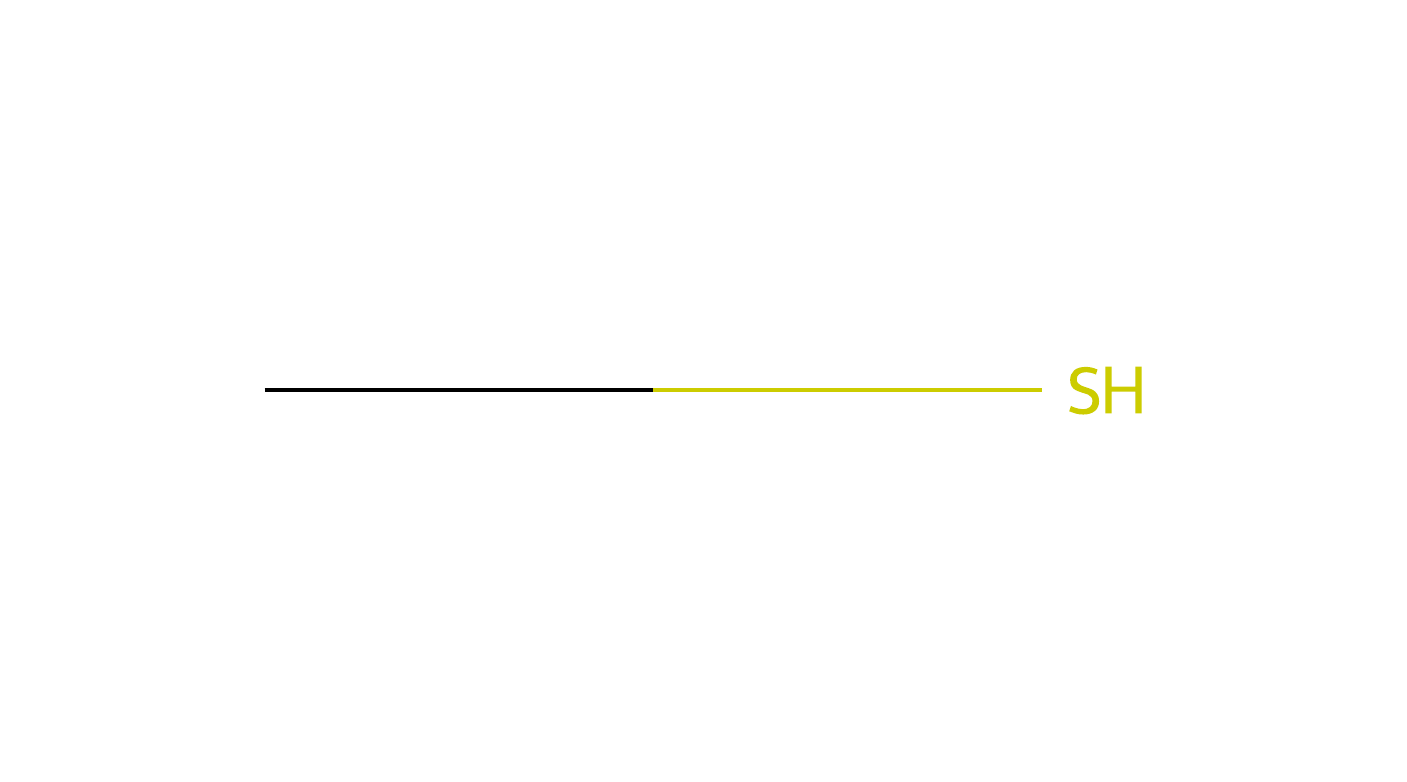}} \\
\parbox[c][1.48cm][c]{0.47\columnwidth}{\raggedright\scriptsize\ttfamily CO}
 & \parbox[c][1.48cm][c]{0.27\columnwidth}{\raggedright\scriptsize\ttfamily mobley\_\allowbreak{}1636752}
 & \parbox[c][1.48cm][c]{0.18\columnwidth}{\centering\includegraphics[width=\linewidth,height=1.32cm,keepaspectratio]{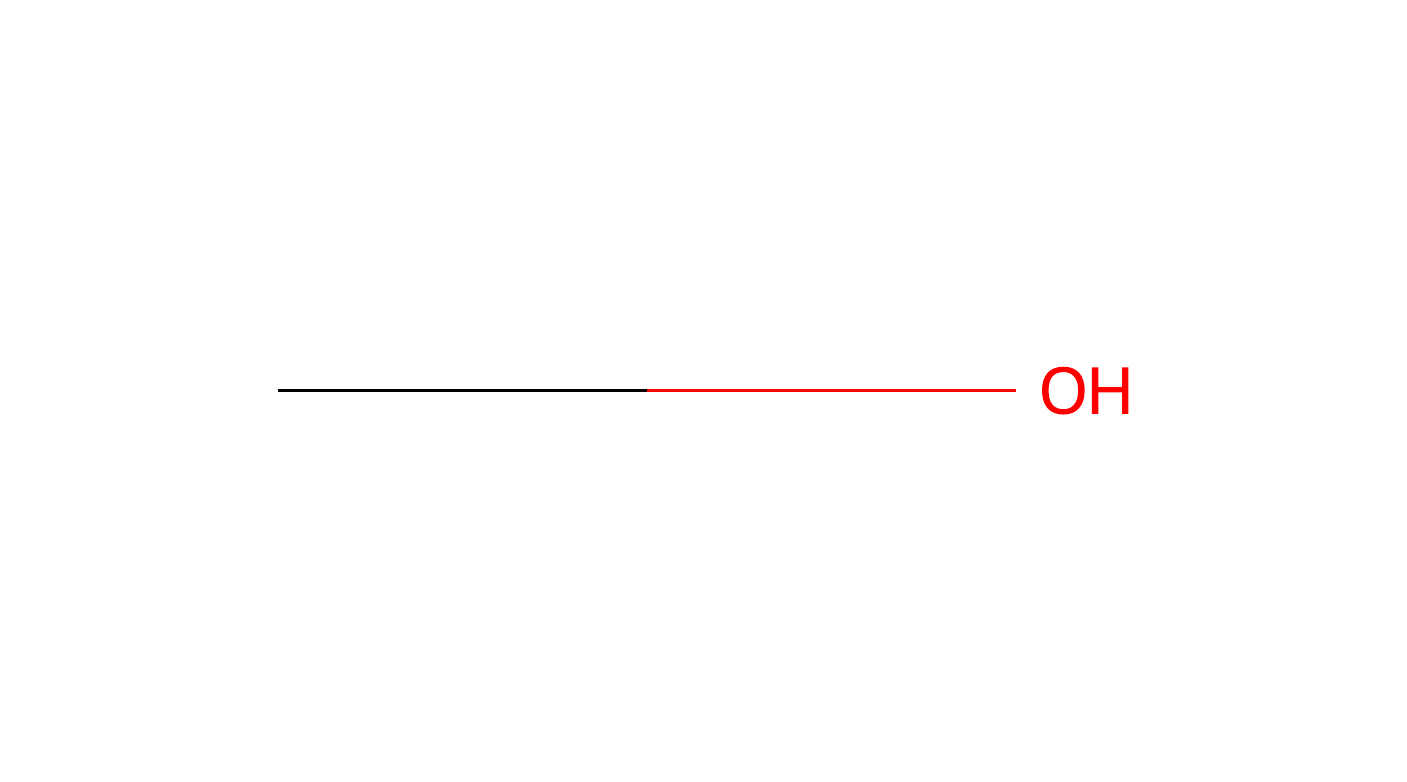}} \\
\parbox[c][1.48cm][c]{0.47\columnwidth}{\raggedright\scriptsize\ttfamily COC}
 & \parbox[c][1.48cm][c]{0.27\columnwidth}{\raggedright\scriptsize\ttfamily mobley\_\allowbreak{}7015518}
 & \parbox[c][1.48cm][c]{0.18\columnwidth}{\centering\includegraphics[width=\linewidth,height=1.32cm,keepaspectratio]{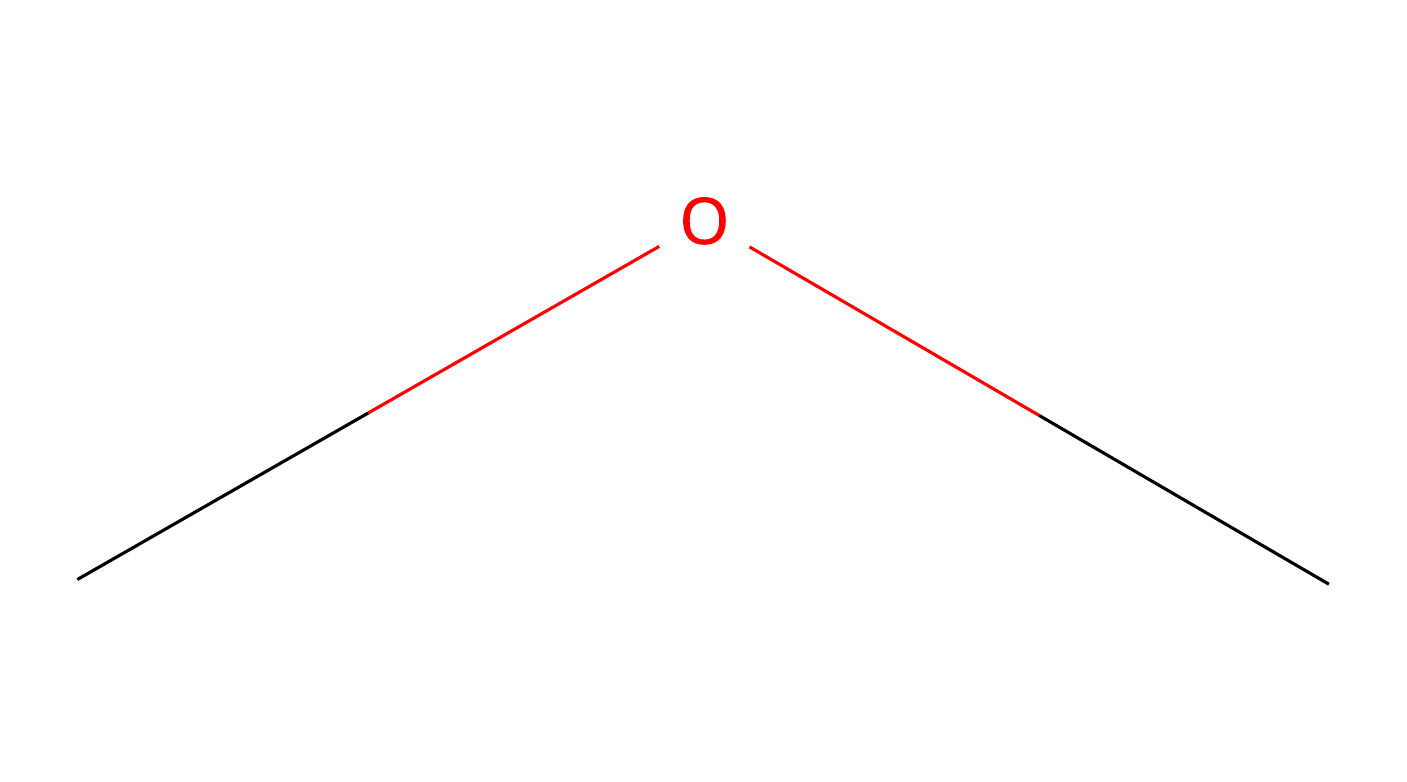}} \\
\parbox[c][1.48cm][c]{0.47\columnwidth}{\raggedright\scriptsize\ttfamily COC=\allowbreak{}O}
 & \parbox[c][1.48cm][c]{0.27\columnwidth}{\raggedright\scriptsize\ttfamily mobley\_\allowbreak{}1717215}
 & \parbox[c][1.48cm][c]{0.18\columnwidth}{\centering\includegraphics[width=\linewidth,height=1.32cm,keepaspectratio]{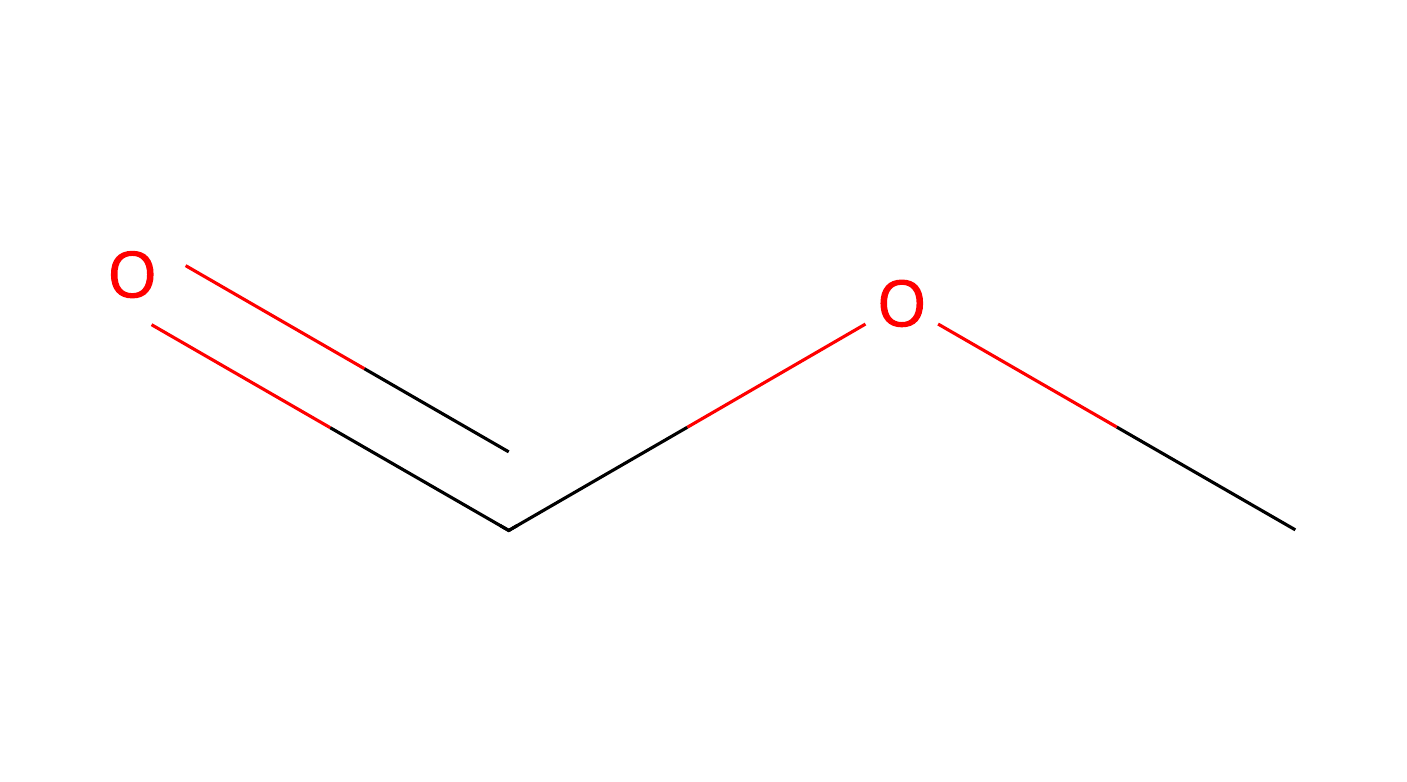}} \\
\bottomrule
\end{tabular}

\newpage
\noindent\textbf{\tablename~\thetable{} -- continued}
\par\vspace{2pt}
\noindent\begin{tabular}{@{}lll@{}}
\toprule
\parbox[t]{0.47\columnwidth}{\raggedright\bfseries SMILES} &
\parbox[t]{0.27\columnwidth}{\raggedright\bfseries freesolv\_id} &
\parbox[t]{0.18\columnwidth}{\centering\bfseries Molecule} \\
\midrule
\parbox[c][1.48cm][c]{0.47\columnwidth}{\raggedright\scriptsize\ttfamily CSC}
 & \parbox[c][1.48cm][c]{0.27\columnwidth}{\raggedright\scriptsize\ttfamily mobley\_\allowbreak{}9073553}
 & \parbox[c][1.48cm][c]{0.18\columnwidth}{\centering\includegraphics[width=\linewidth,height=1.32cm,keepaspectratio]{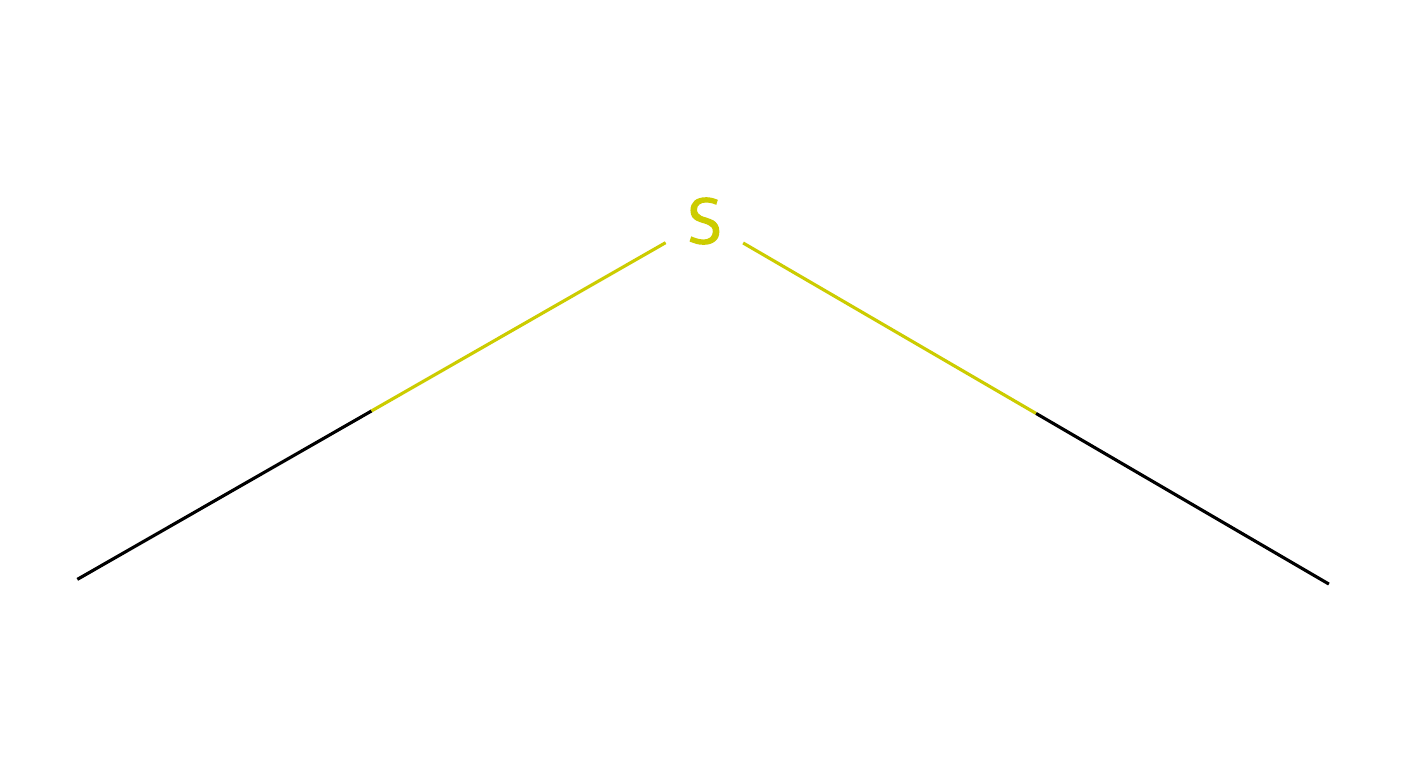}} \\
\parbox[c][1.48cm][c]{0.47\columnwidth}{\raggedright\scriptsize\ttfamily CS(\allowbreak{}=\allowbreak{}O)\allowbreak{}C}
 & \parbox[c][1.48cm][c]{0.27\columnwidth}{\raggedright\scriptsize\ttfamily mobley\_\allowbreak{}8578590}
 & \parbox[c][1.48cm][c]{0.18\columnwidth}{\centering\includegraphics[width=\linewidth,height=1.32cm,keepaspectratio]{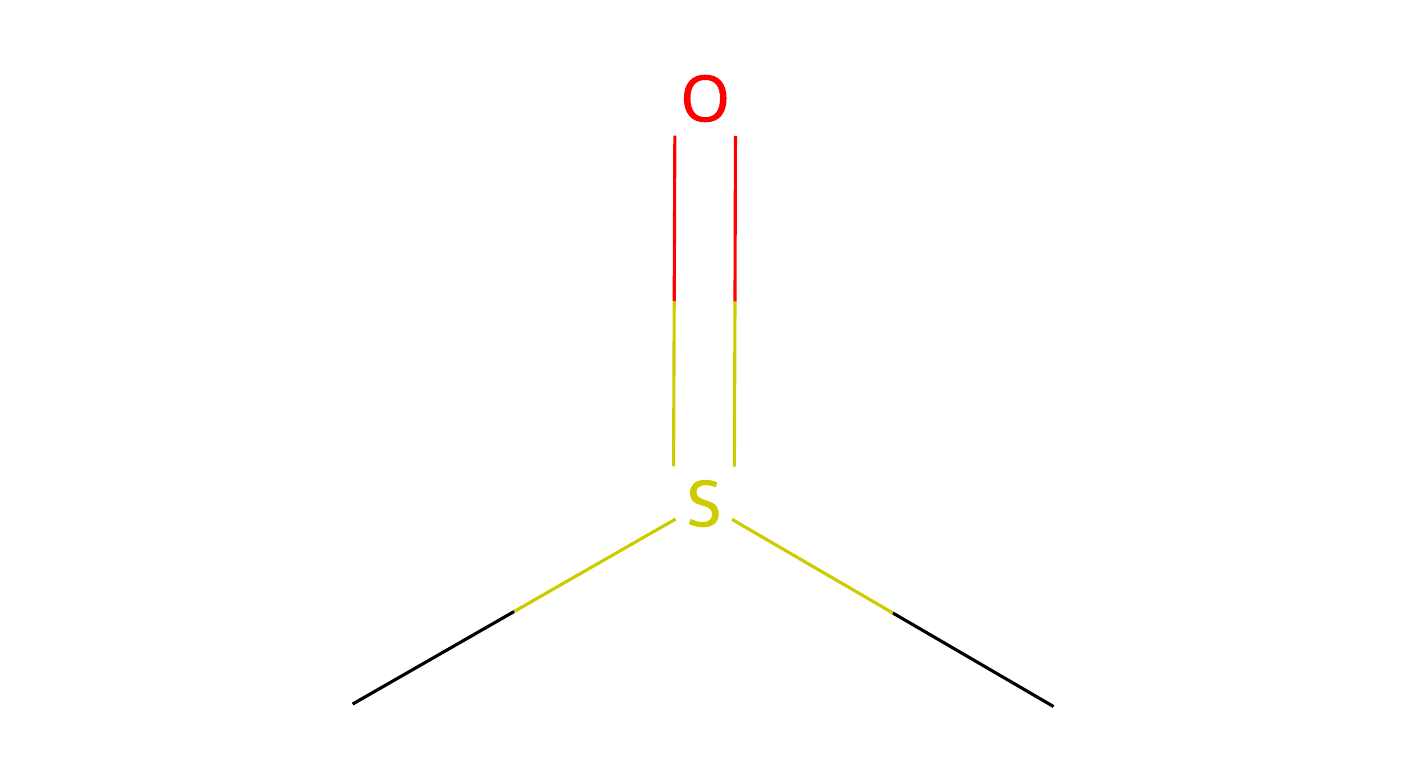}} \\
\parbox[c][1.48cm][c]{0.47\columnwidth}{\raggedright\scriptsize\ttfamily CS(\allowbreak{}=\allowbreak{}O)\allowbreak{}(\allowbreak{}=\allowbreak{}O)\allowbreak{}C}
 & \parbox[c][1.48cm][c]{0.27\columnwidth}{\raggedright\scriptsize\ttfamily mobley\_\allowbreak{}1708457}
 & \parbox[c][1.48cm][c]{0.18\columnwidth}{\centering\includegraphics[width=\linewidth,height=1.32cm,keepaspectratio]{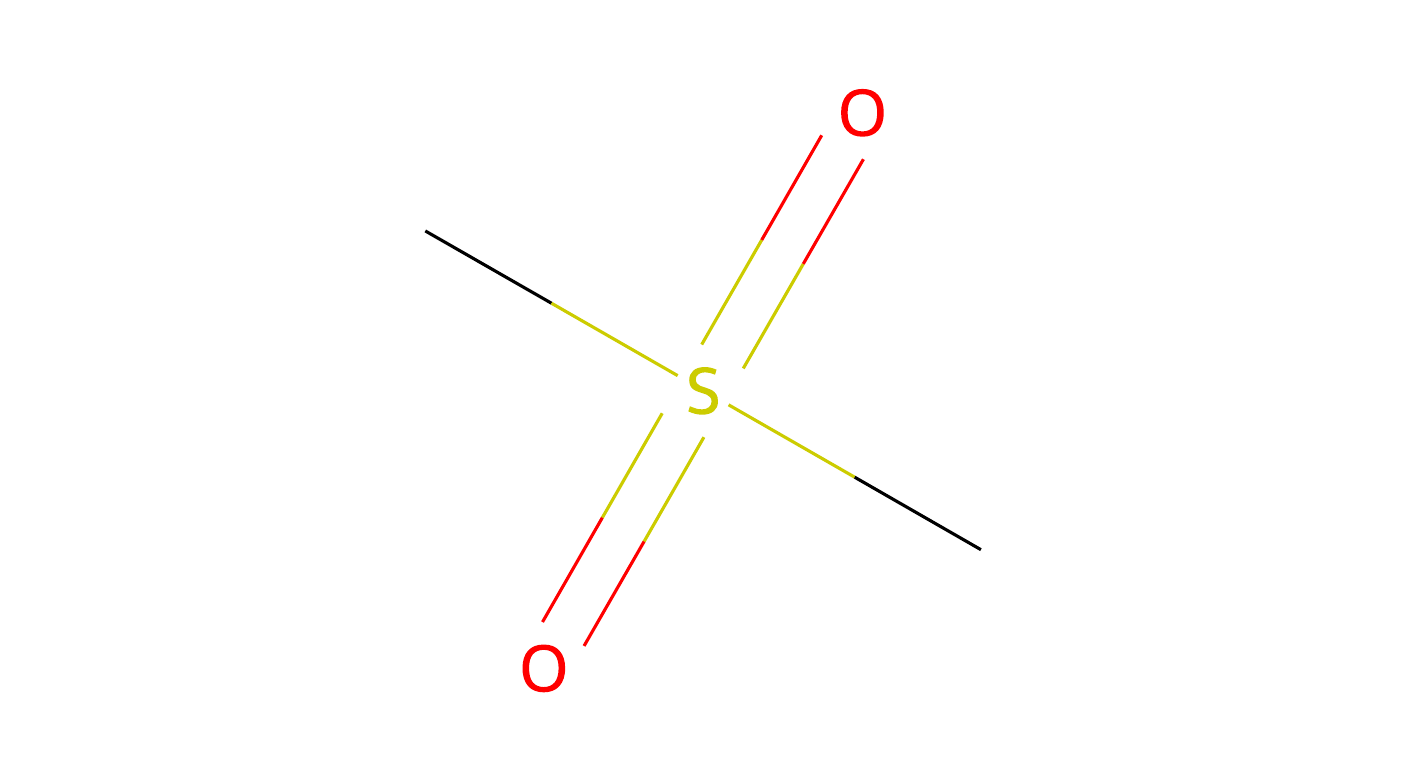}} \\
\parbox[c][1.48cm][c]{0.47\columnwidth}{\raggedright\scriptsize\ttfamily CC(\allowbreak{}=\allowbreak{}O)\allowbreak{}NC}
 & \parbox[c][1.48cm][c]{0.27\columnwidth}{\raggedright\scriptsize\ttfamily mobley\_\allowbreak{}1963873}
 & \parbox[c][1.48cm][c]{0.18\columnwidth}{\centering\includegraphics[width=\linewidth,height=1.32cm,keepaspectratio]{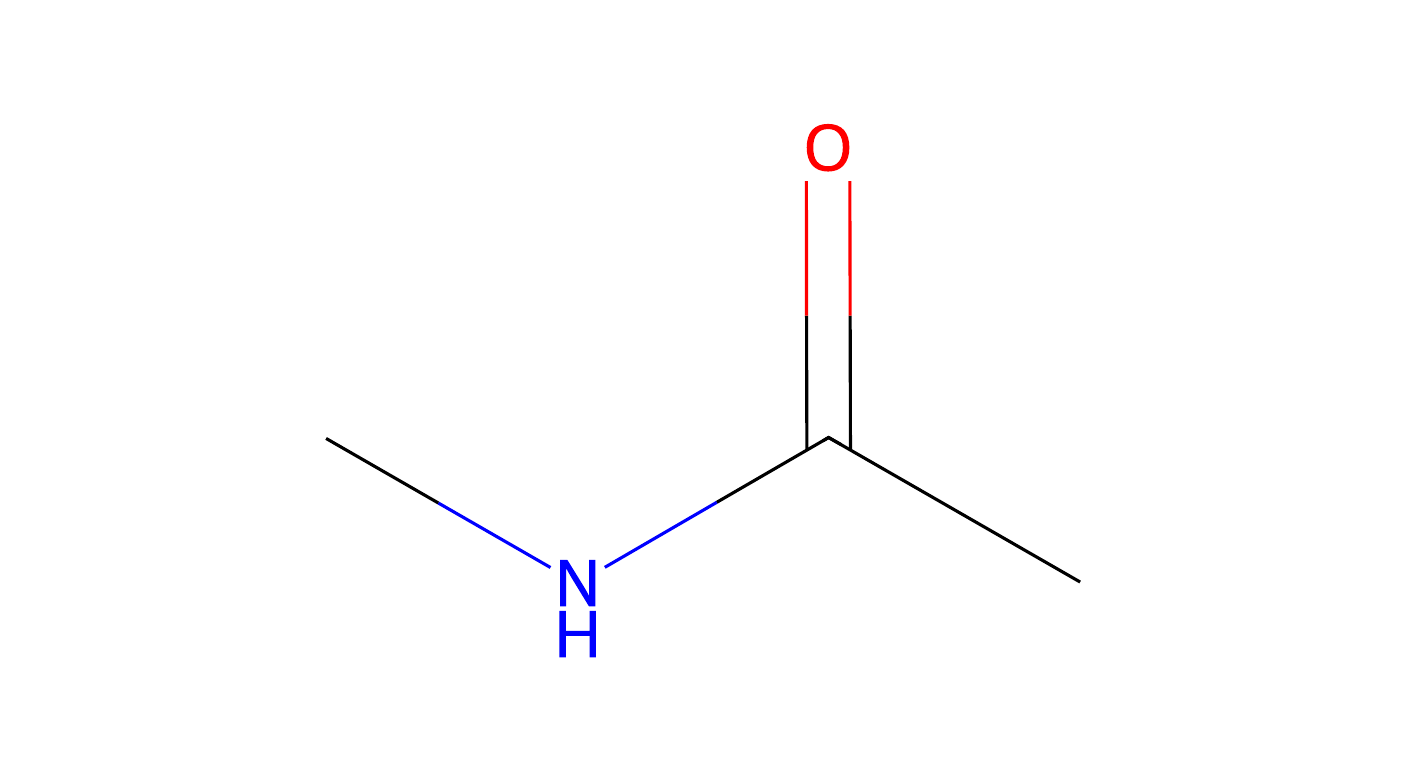}} \\
\parbox[c][1.48cm][c]{0.47\columnwidth}{\raggedright\scriptsize\ttfamily CNc1ccccc1}
 & \parbox[c][1.48cm][c]{0.27\columnwidth}{\raggedright\scriptsize\ttfamily mobley\_\allowbreak{}7774695}
 & \parbox[c][1.48cm][c]{0.18\columnwidth}{\centering\includegraphics[width=\linewidth,height=1.32cm,keepaspectratio]{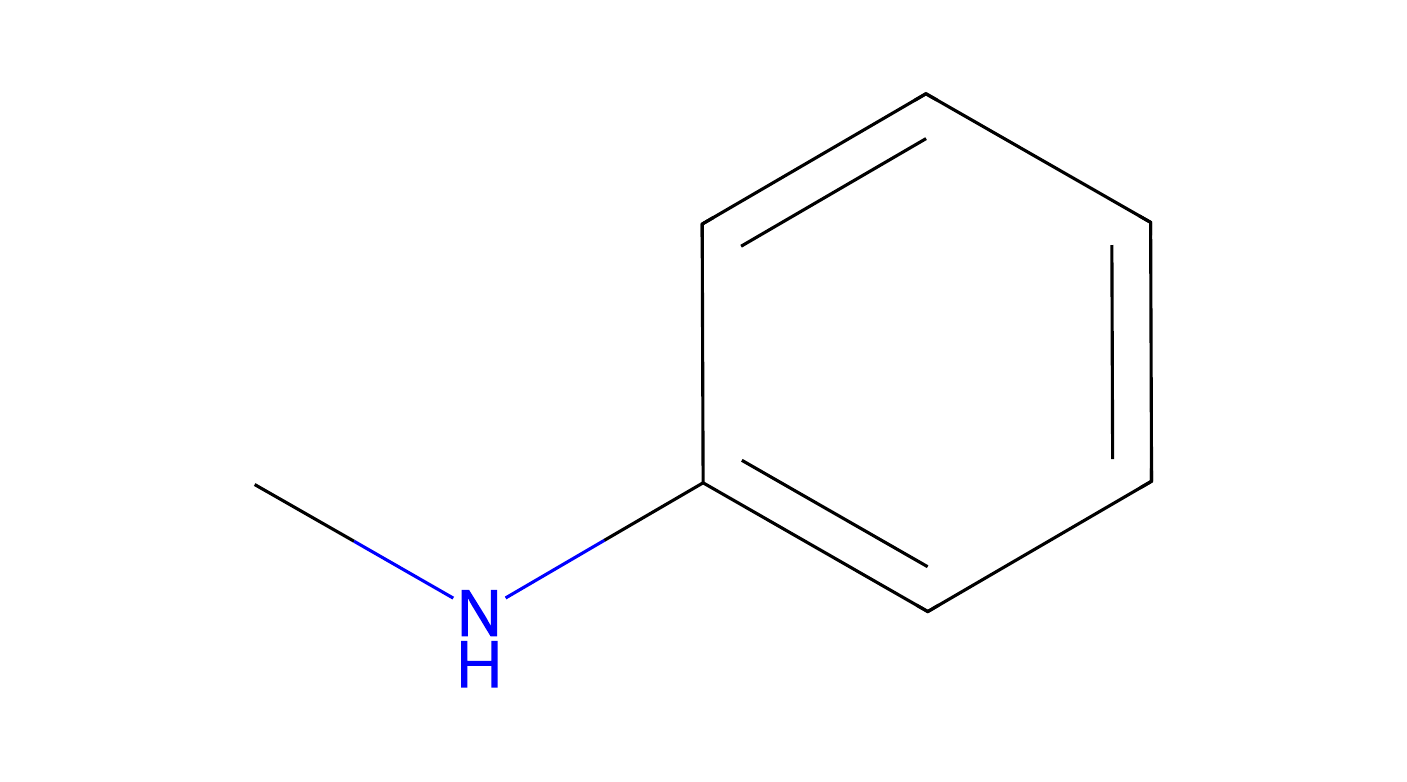}} \\
\parbox[c][1.48cm][c]{0.47\columnwidth}{\raggedright\scriptsize\ttfamily CN(\allowbreak{}C)\allowbreak{}c1ccccc1}
 & \parbox[c][1.48cm][c]{0.27\columnwidth}{\raggedright\scriptsize\ttfamily mobley\_\allowbreak{}5326154}
 & \parbox[c][1.48cm][c]{0.18\columnwidth}{\centering\includegraphics[width=\linewidth,height=1.32cm,keepaspectratio]{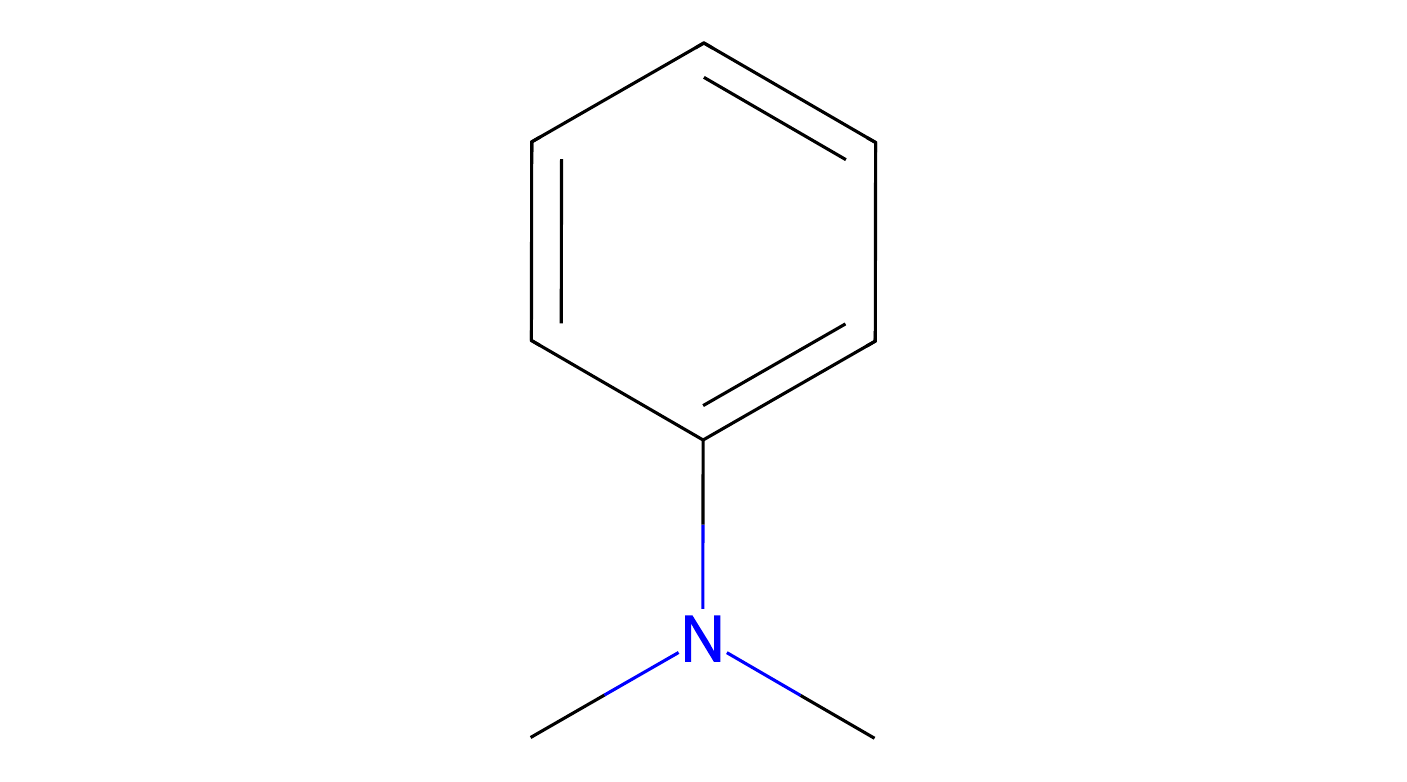}} \\
\parbox[c][1.48cm][c]{0.47\columnwidth}{\raggedright\scriptsize\ttfamily CN(\allowbreak{}C)\allowbreak{}C=\allowbreak{}O}
 & \parbox[c][1.48cm][c]{0.27\columnwidth}{\raggedright\scriptsize\ttfamily mobley\_\allowbreak{}8011706}
 & \parbox[c][1.48cm][c]{0.18\columnwidth}{\centering\includegraphics[width=\linewidth,height=1.32cm,keepaspectratio]{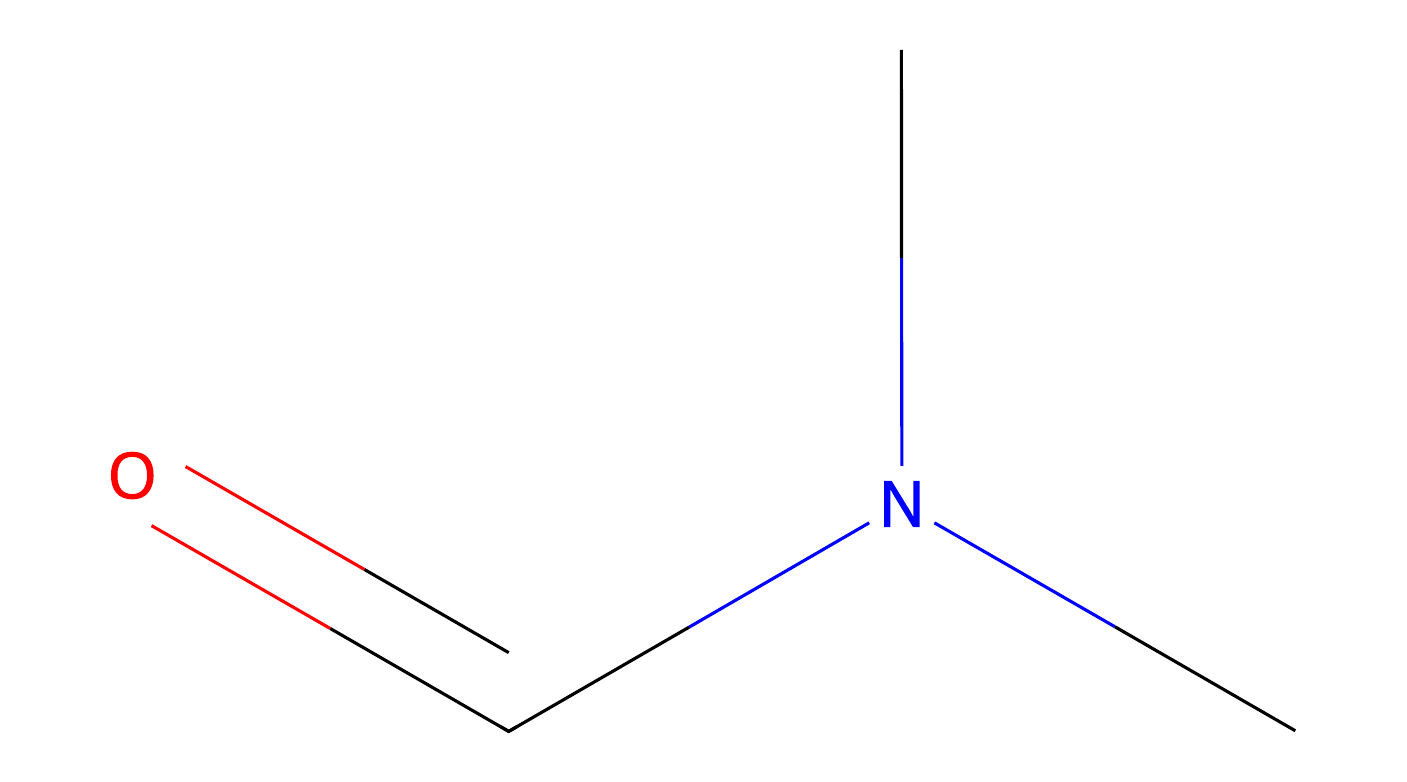}} \\
\parbox[c][1.48cm][c]{0.47\columnwidth}{\raggedright\scriptsize\ttfamily C\allowbreak{}N\allowbreak{}C\allowbreak{}(\allowbreak{}=\allowbreak{}O\allowbreak{})\allowbreak{}O\allowbreak{}c\allowbreak{}1\allowbreak{}c\allowbreak{}c\allowbreak{}c\allowbreak{}c\allowbreak{}2\allowbreak{}c\allowbreak{}1\allowbreak{}c\allowbreak{}c\allowbreak{}c\allowbreak{}c\allowbreak{}2}
 & \parbox[c][1.48cm][c]{0.27\columnwidth}{\raggedright\scriptsize\ttfamily mobley\_\allowbreak{}9007496}
 & \parbox[c][1.48cm][c]{0.18\columnwidth}{\centering\includegraphics[width=\linewidth,height=1.32cm,keepaspectratio]{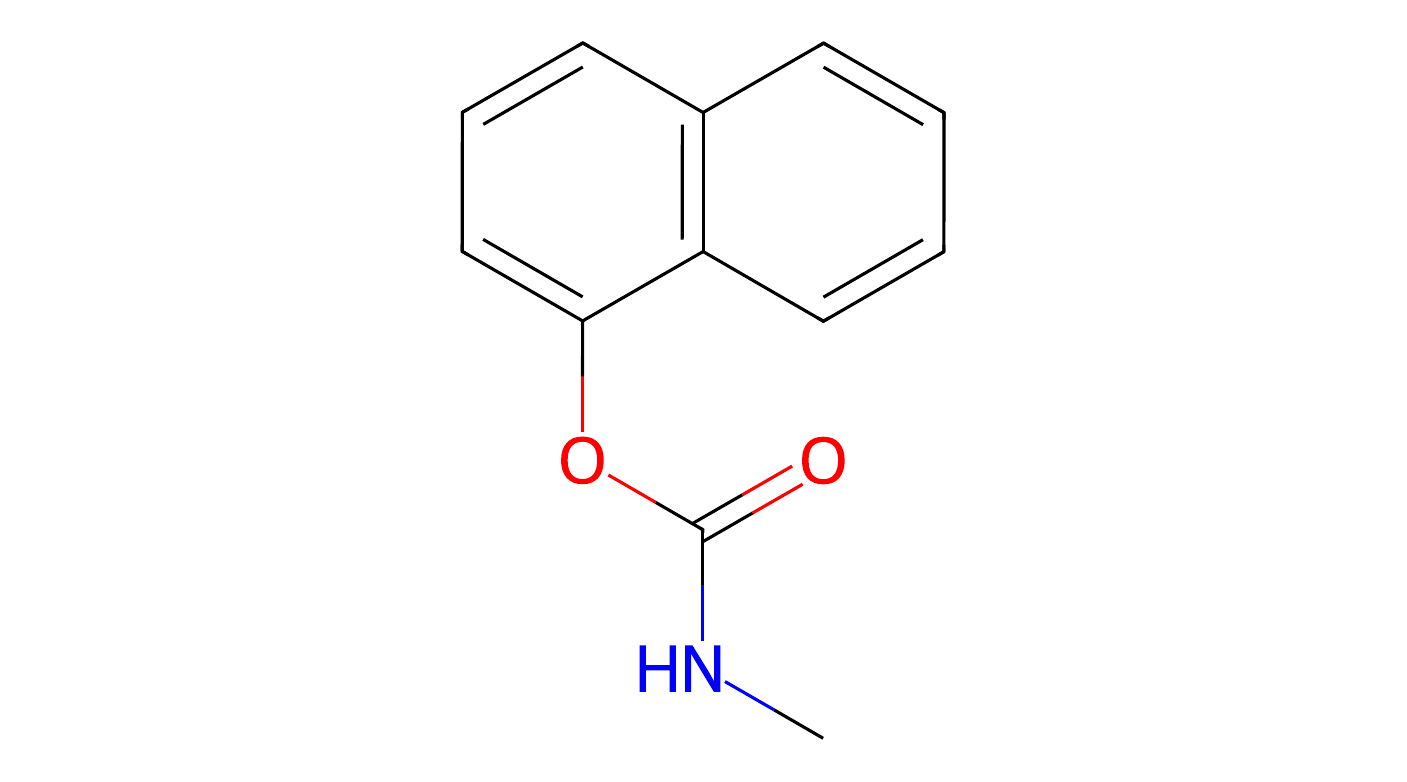}} \\
\parbox[c][1.48cm][c]{0.47\columnwidth}{\raggedright\scriptsize\ttfamily CCCCN(\allowbreak{}CC)\allowbreak{}C(\allowbreak{}=\allowbreak{}O)\allowbreak{}SCCC}
 & \parbox[c][1.48cm][c]{0.27\columnwidth}{\raggedright\scriptsize\ttfamily mobley\_\allowbreak{}1352110}
 & \parbox[c][1.48cm][c]{0.18\columnwidth}{\centering\includegraphics[width=\linewidth,height=1.32cm,keepaspectratio]{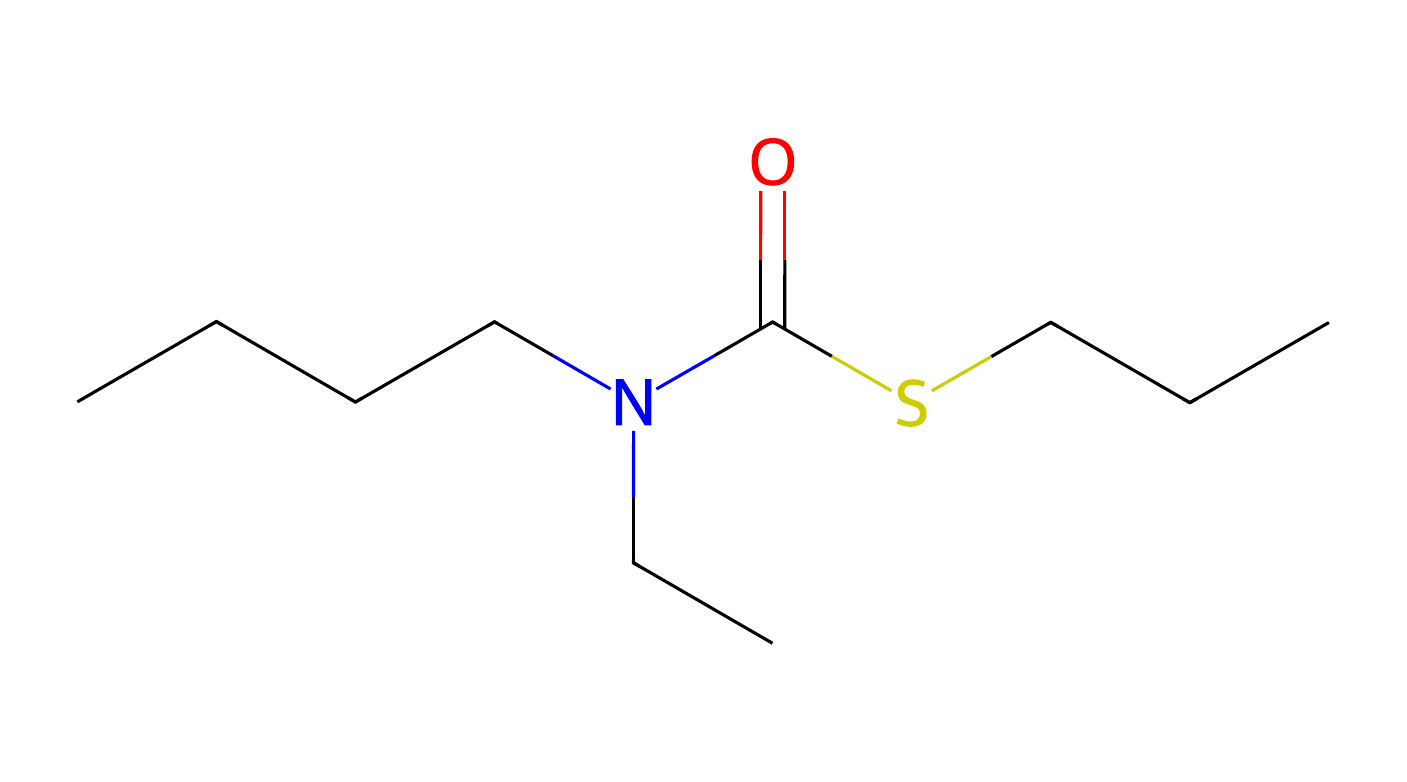}} \\
\parbox[c][1.48cm][c]{0.47\columnwidth}{\raggedright\scriptsize\ttfamily c1ccc(\allowbreak{}cc1)\allowbreak{}O}
 & \parbox[c][1.48cm][c]{0.27\columnwidth}{\raggedright\scriptsize\ttfamily mobley\_\allowbreak{}20524}
 & \parbox[c][1.48cm][c]{0.18\columnwidth}{\centering\includegraphics[width=\linewidth,height=1.32cm,keepaspectratio]{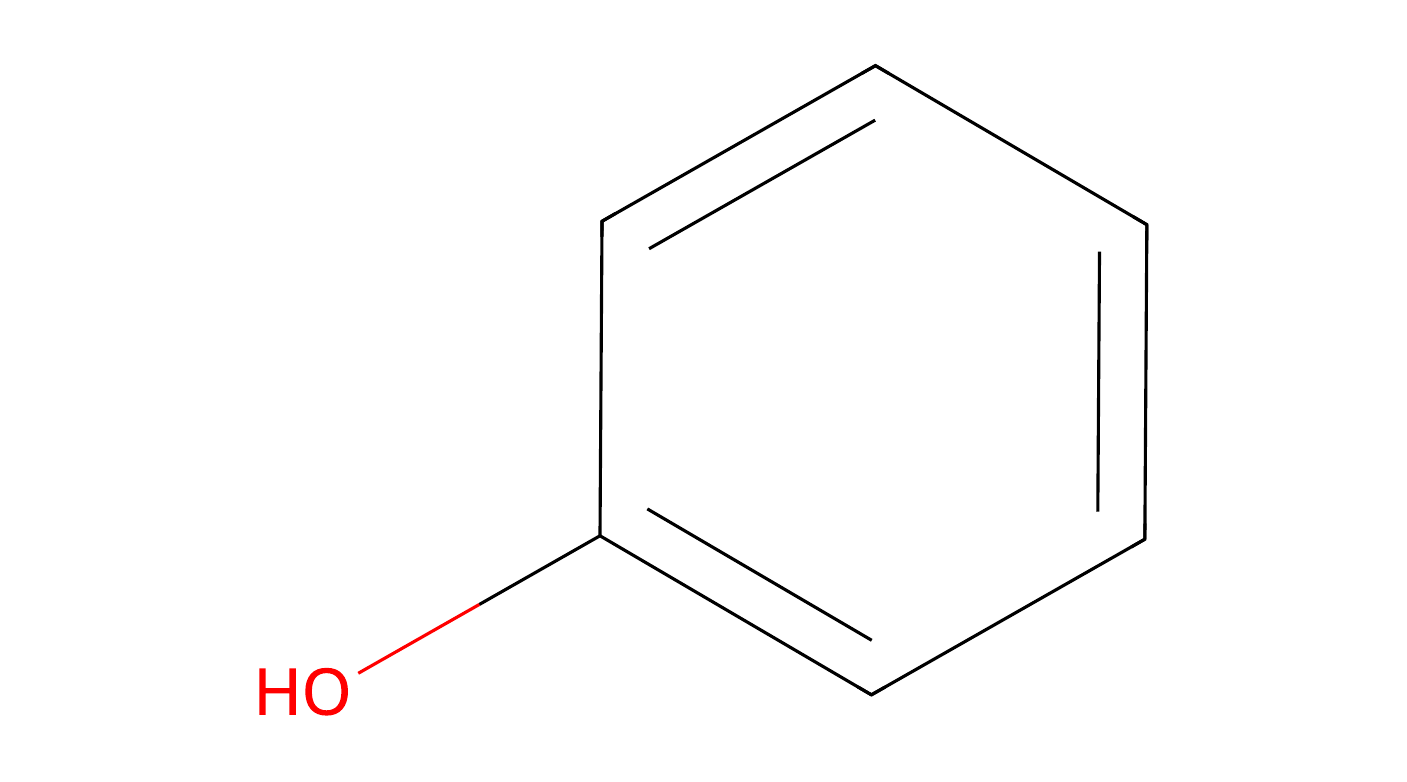}} \\
\parbox[c][1.48cm][c]{0.47\columnwidth}{\raggedright\scriptsize\ttfamily c\allowbreak{}1\allowbreak{}c\allowbreak{}c\allowbreak{}c\allowbreak{}2\allowbreak{}c\allowbreak{}(\allowbreak{}c\allowbreak{}1\allowbreak{})\allowbreak{}C\allowbreak{}(\allowbreak{}=\allowbreak{}O\allowbreak{})\allowbreak{}N\allowbreak{}C\allowbreak{}2\allowbreak{}=\allowbreak{}O}
 & \parbox[c][1.48cm][c]{0.27\columnwidth}{\raggedright\scriptsize\ttfamily mobley\_\allowbreak{}2782339}
 & \parbox[c][1.48cm][c]{0.18\columnwidth}{\centering\includegraphics[width=\linewidth,height=1.32cm,keepaspectratio]{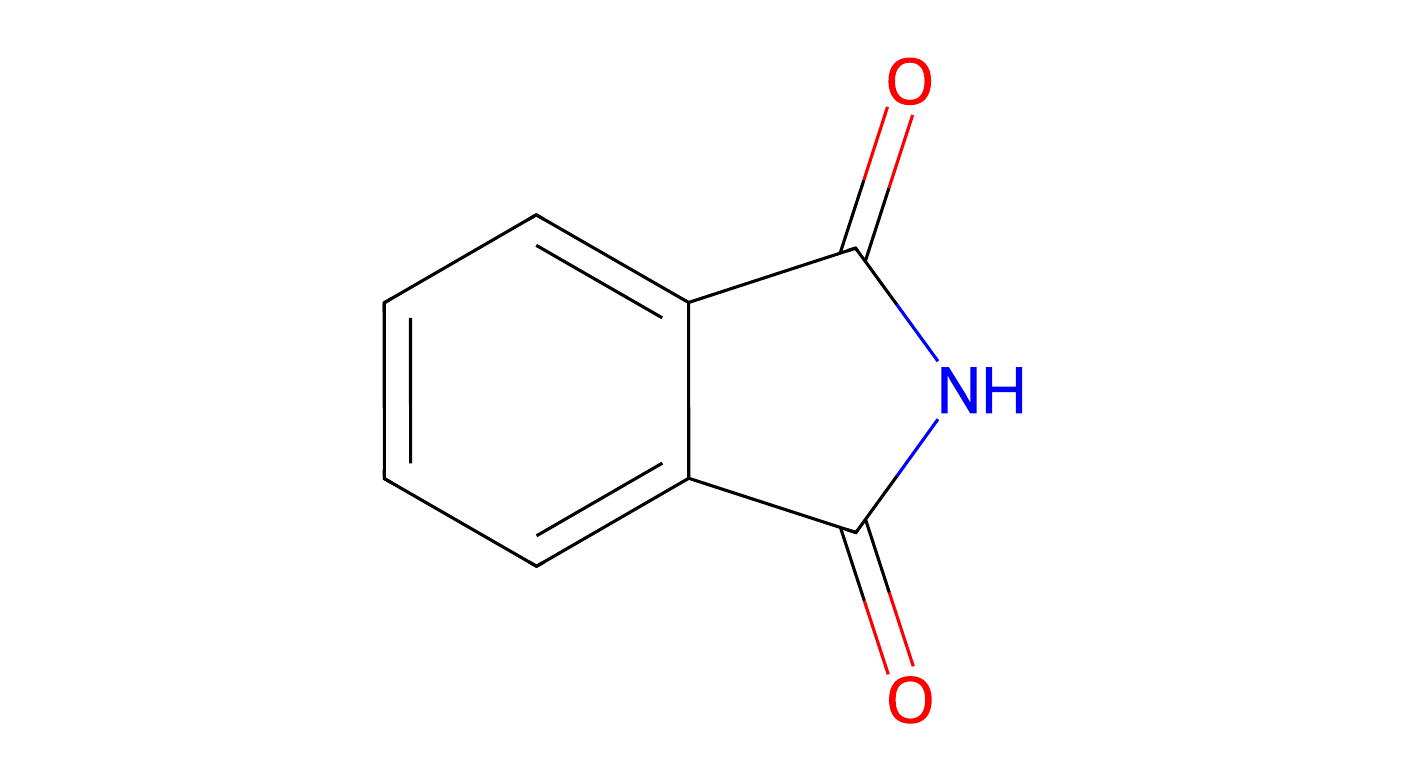}} \\
\parbox[c][1.48cm][c]{0.47\columnwidth}{\raggedright\scriptsize\ttfamily CC\#\allowbreak{}C}
 & \parbox[c][1.48cm][c]{0.27\columnwidth}{\raggedright\scriptsize\ttfamily mobley\_\allowbreak{}8260524}
 & \parbox[c][1.48cm][c]{0.18\columnwidth}{\centering\includegraphics[width=\linewidth,height=1.32cm,keepaspectratio]{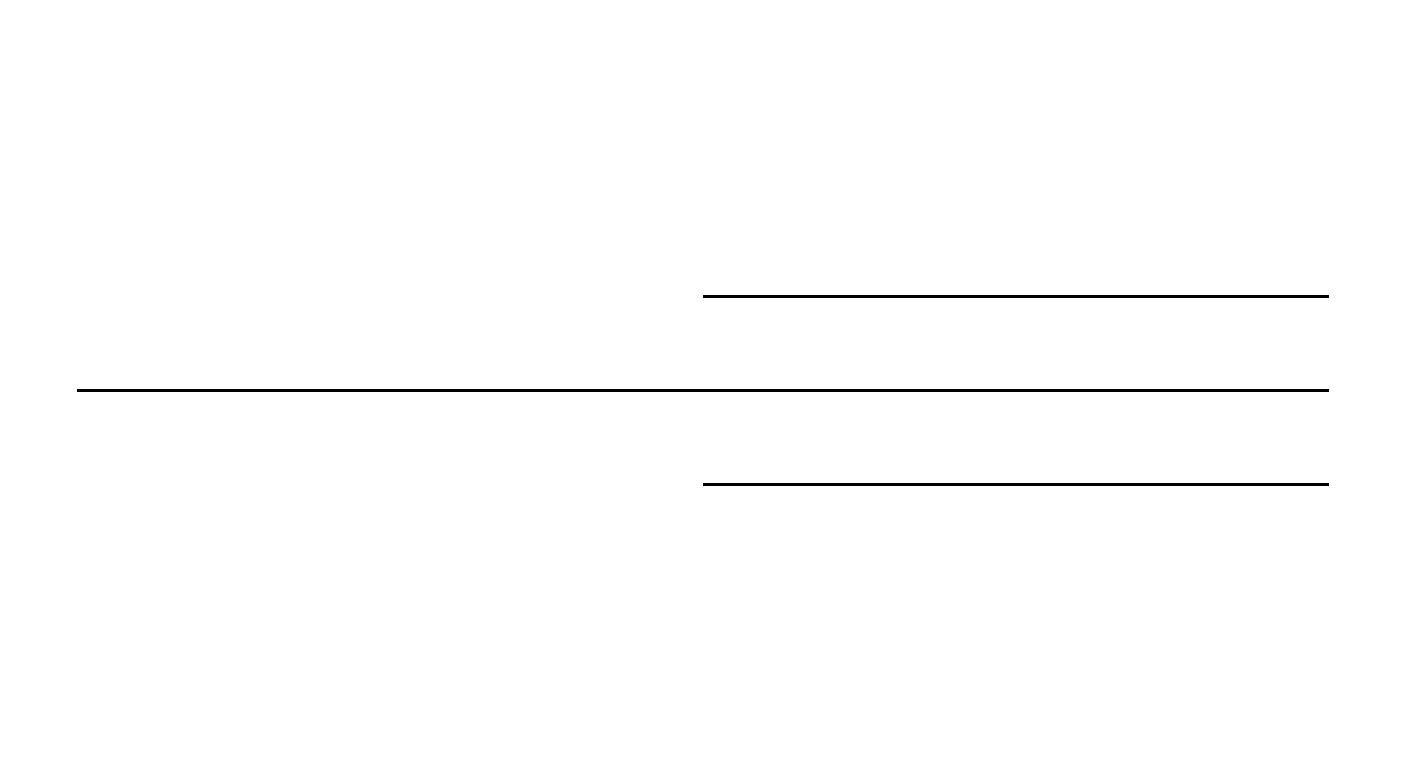}} \\
\bottomrule
\end{tabular}

\newpage
\noindent\textbf{\tablename~\thetable{} -- continued}
\par\vspace{2pt}
\noindent\begin{tabular}{@{}lll@{}}
\toprule
\parbox[t]{0.47\columnwidth}{\raggedright\bfseries SMILES} &
\parbox[t]{0.27\columnwidth}{\raggedright\bfseries freesolv\_id} &
\parbox[t]{0.18\columnwidth}{\centering\bfseries Molecule} \\
\midrule
\parbox[c][1.48cm][c]{0.47\columnwidth}{\raggedright\scriptsize\ttfamily CC(\allowbreak{}C)\allowbreak{}O}
 & \parbox[c][1.48cm][c]{0.27\columnwidth}{\raggedright\scriptsize\ttfamily mobley\_\allowbreak{}7326982}
 & \parbox[c][1.48cm][c]{0.18\columnwidth}{\centering\includegraphics[width=\linewidth,height=1.32cm,keepaspectratio]{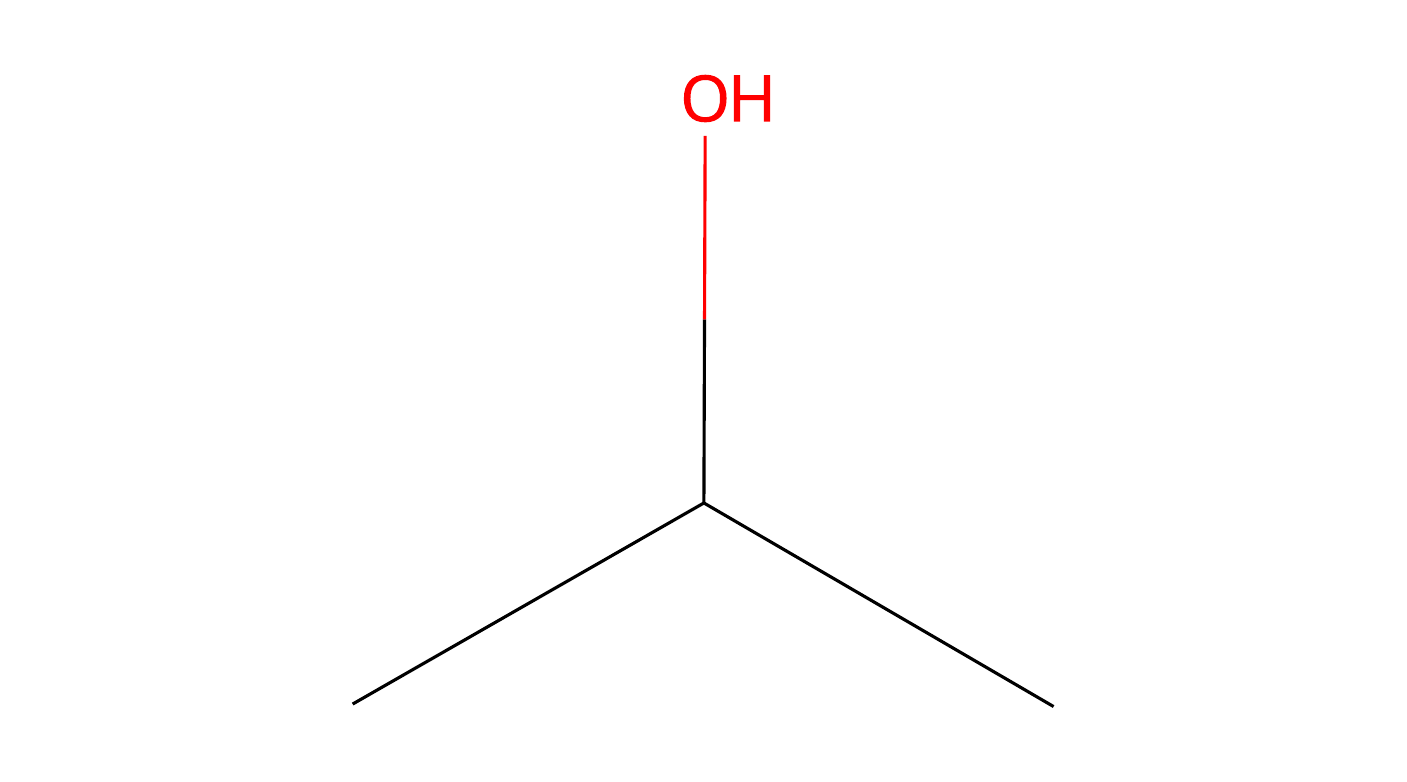}} \\
\parbox[c][1.48cm][c]{0.47\columnwidth}{\raggedright\scriptsize\ttfamily c1ccsc1}
 & \parbox[c][1.48cm][c]{0.27\columnwidth}{\raggedright\scriptsize\ttfamily mobley\_\allowbreak{}2972906}
 & \parbox[c][1.48cm][c]{0.18\columnwidth}{\centering\includegraphics[width=\linewidth,height=1.32cm,keepaspectratio]{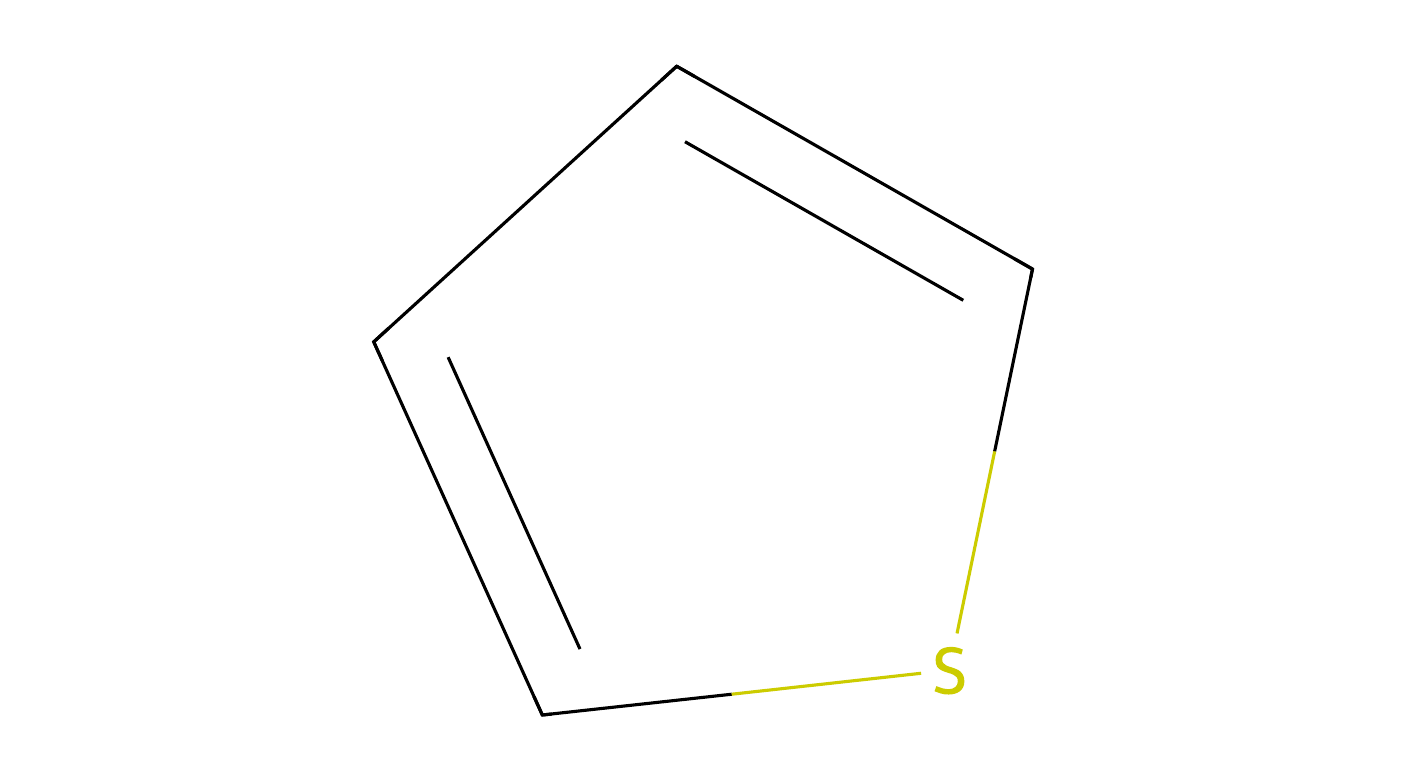}} \\
\parbox[c][1.48cm][c]{0.47\columnwidth}{\raggedright\scriptsize\ttfamily CC(\allowbreak{}=\allowbreak{}O)\allowbreak{}c1cccnc1}
 & \parbox[c][1.48cm][c]{0.27\columnwidth}{\raggedright\scriptsize\ttfamily mobley\_\allowbreak{}6353617}
 & \parbox[c][1.48cm][c]{0.18\columnwidth}{\centering\includegraphics[width=\linewidth,height=1.32cm,keepaspectratio]{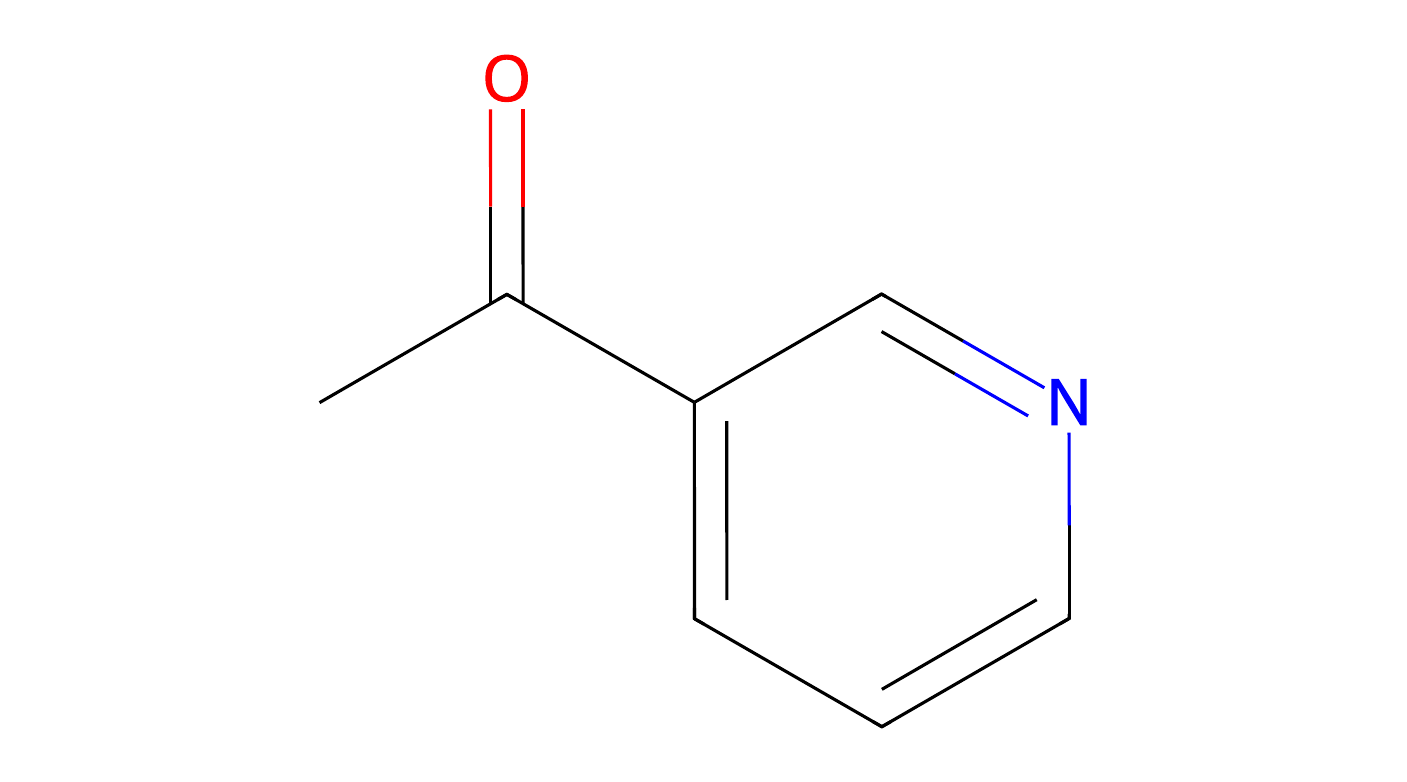}} \\
\parbox[c][1.48cm][c]{0.47\columnwidth}{\raggedright\scriptsize\ttfamily CC(\allowbreak{}=\allowbreak{}O)\allowbreak{}c1ccncc1}
 & \parbox[c][1.48cm][c]{0.27\columnwidth}{\raggedright\scriptsize\ttfamily mobley\_\allowbreak{}6794076}
 & \parbox[c][1.48cm][c]{0.18\columnwidth}{\centering\includegraphics[width=\linewidth,height=1.32cm,keepaspectratio]{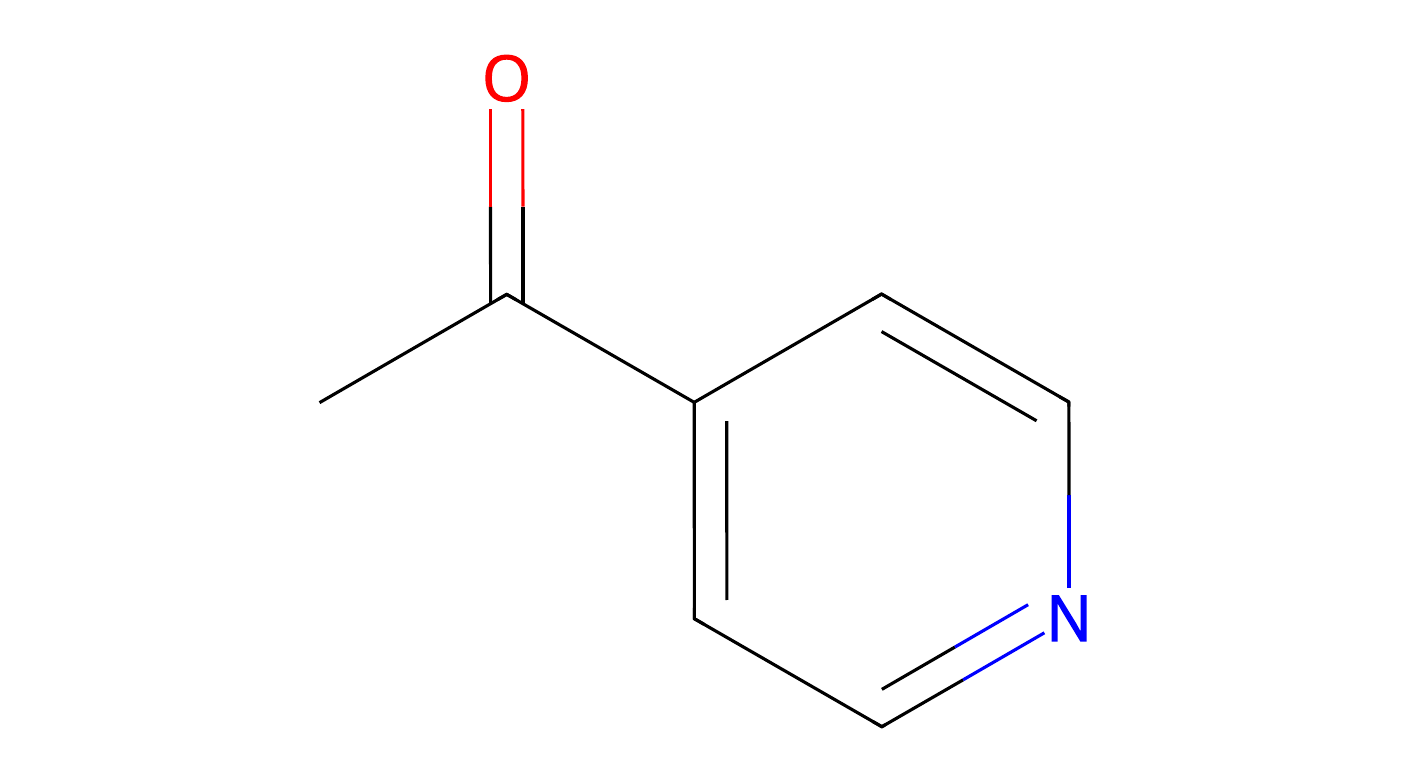}} \\
\parbox[c][1.48cm][c]{0.47\columnwidth}{\raggedright\scriptsize\ttfamily CC(\allowbreak{}=\allowbreak{}O)\allowbreak{}C1CCCCC1}
 & \parbox[c][1.48cm][c]{0.27\columnwidth}{\raggedright\scriptsize\ttfamily mobley\_\allowbreak{}2043882}
 & \parbox[c][1.48cm][c]{0.18\columnwidth}{\centering\includegraphics[width=\linewidth,height=1.32cm,keepaspectratio]{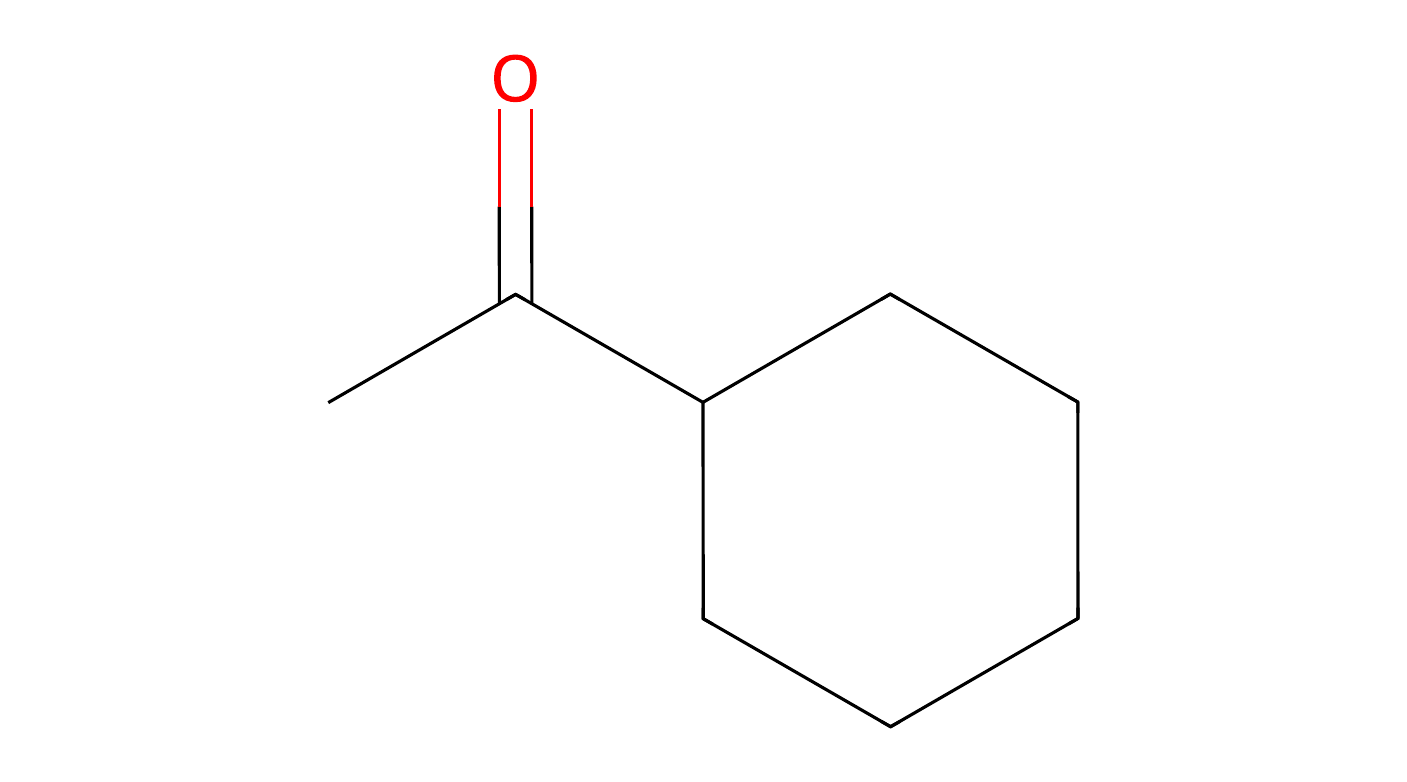}} \\
\parbox[c][1.48cm][c]{0.47\columnwidth}{\raggedright\scriptsize\ttfamily Cn1cccc1}
 & \parbox[c][1.48cm][c]{0.27\columnwidth}{\raggedright\scriptsize\ttfamily mobley\_\allowbreak{}3775790}
 & \parbox[c][1.48cm][c]{0.18\columnwidth}{\centering\includegraphics[width=\linewidth,height=1.32cm,keepaspectratio]{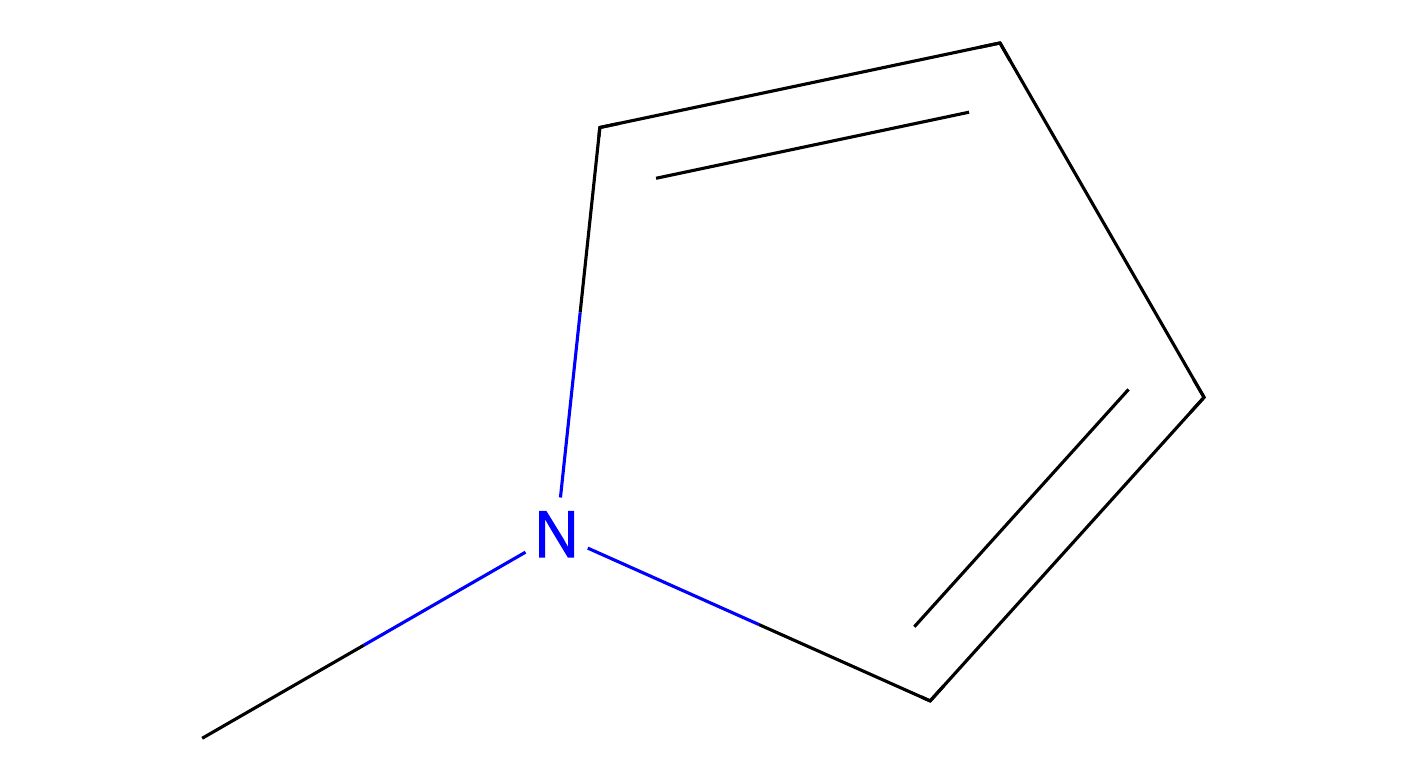}} \\
\parbox[c][1.48cm][c]{0.47\columnwidth}{\raggedright\scriptsize\ttfamily CC(\allowbreak{}=\allowbreak{}O)\allowbreak{}c1ccccc1}
 & \parbox[c][1.48cm][c]{0.27\columnwidth}{\raggedright\scriptsize\ttfamily mobley\_\allowbreak{}7497999}
 & \parbox[c][1.48cm][c]{0.18\columnwidth}{\centering\includegraphics[width=\linewidth,height=1.32cm,keepaspectratio]{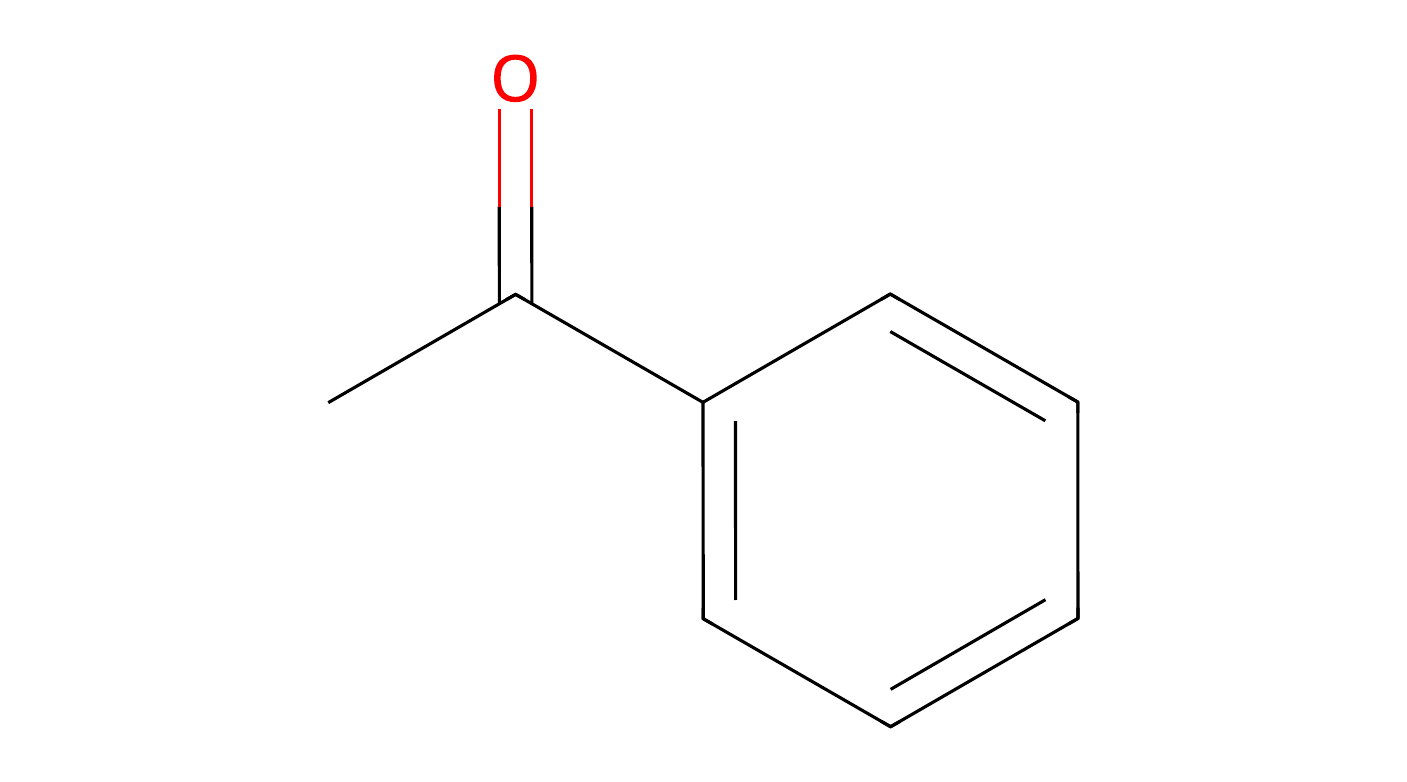}} \\
\parbox[c][1.48cm][c]{0.47\columnwidth}{\raggedright\scriptsize\ttfamily C1COCCO1}
 & \parbox[c][1.48cm][c]{0.27\columnwidth}{\raggedright\scriptsize\ttfamily mobley\_\allowbreak{}6739648}
 & \parbox[c][1.48cm][c]{0.18\columnwidth}{\centering\includegraphics[width=\linewidth,height=1.32cm,keepaspectratio]{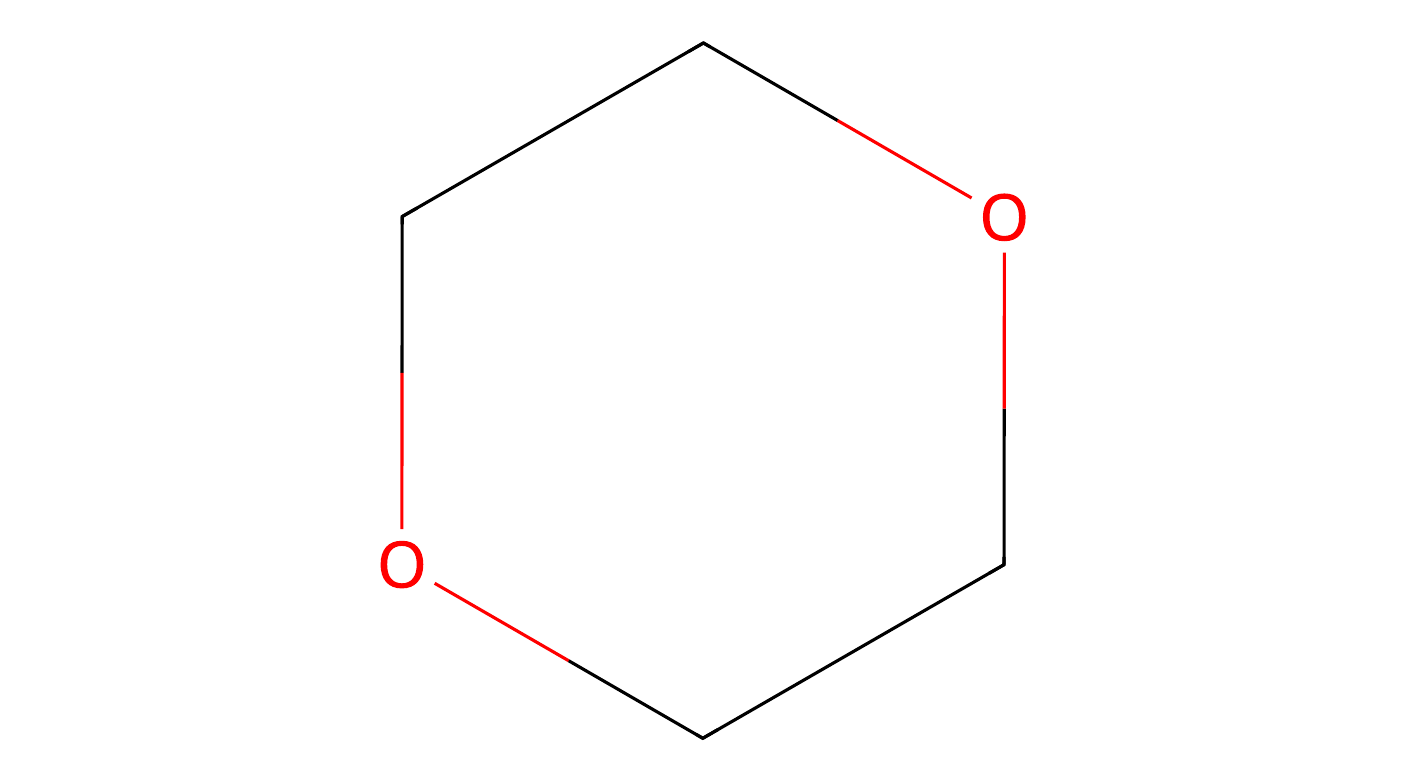}} \\
\parbox[c][1.48cm][c]{0.47\columnwidth}{\raggedright\scriptsize\ttfamily c1ccnc(\allowbreak{}c1)\allowbreak{}Cl}
 & \parbox[c][1.48cm][c]{0.27\columnwidth}{\raggedright\scriptsize\ttfamily mobley\_\allowbreak{}4291494}
 & \parbox[c][1.48cm][c]{0.18\columnwidth}{\centering\includegraphics[width=\linewidth,height=1.32cm,keepaspectratio]{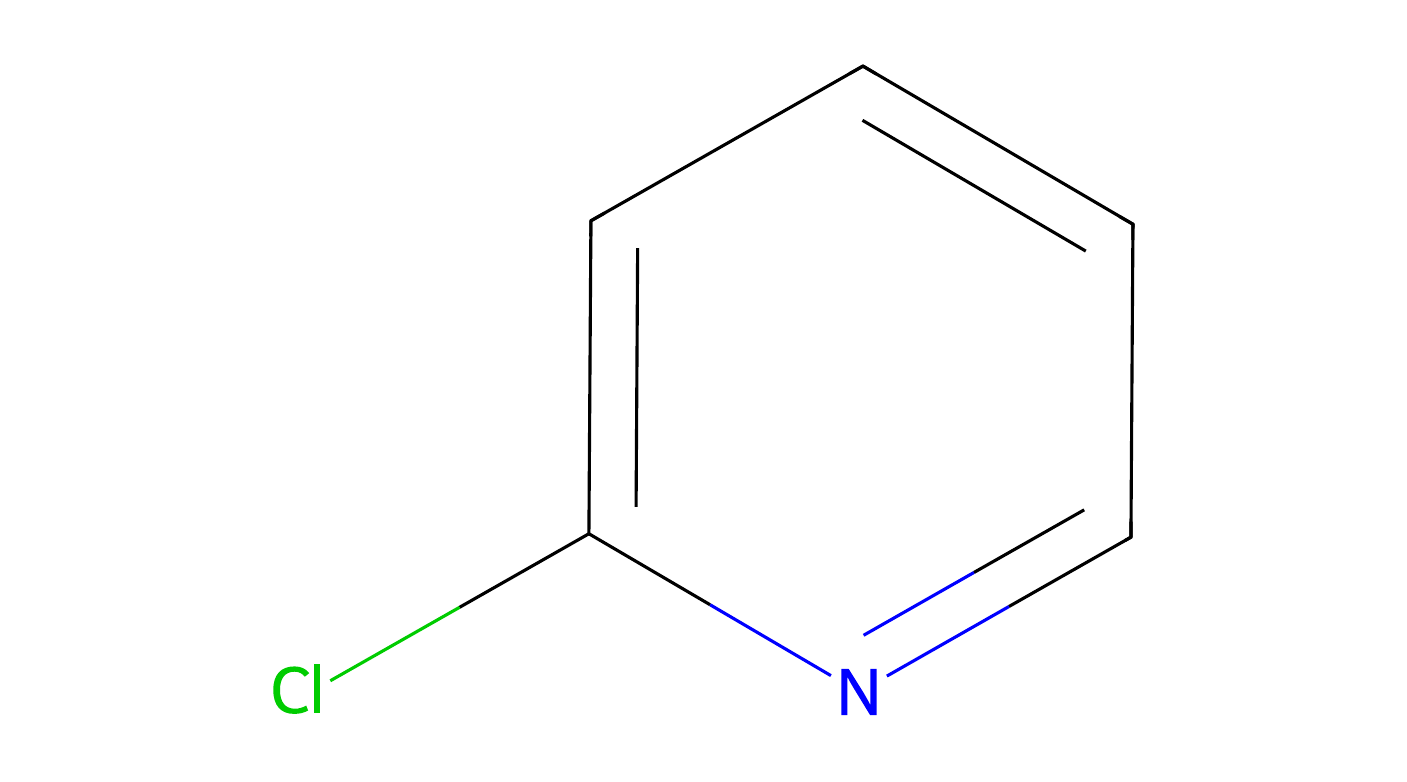}} \\
\parbox[c][1.48cm][c]{0.47\columnwidth}{\raggedright\scriptsize\ttfamily CCc1cnccn1}
 & \parbox[c][1.48cm][c]{0.27\columnwidth}{\raggedright\scriptsize\ttfamily mobley\_\allowbreak{}5220185}
 & \parbox[c][1.48cm][c]{0.18\columnwidth}{\centering\includegraphics[width=\linewidth,height=1.32cm,keepaspectratio]{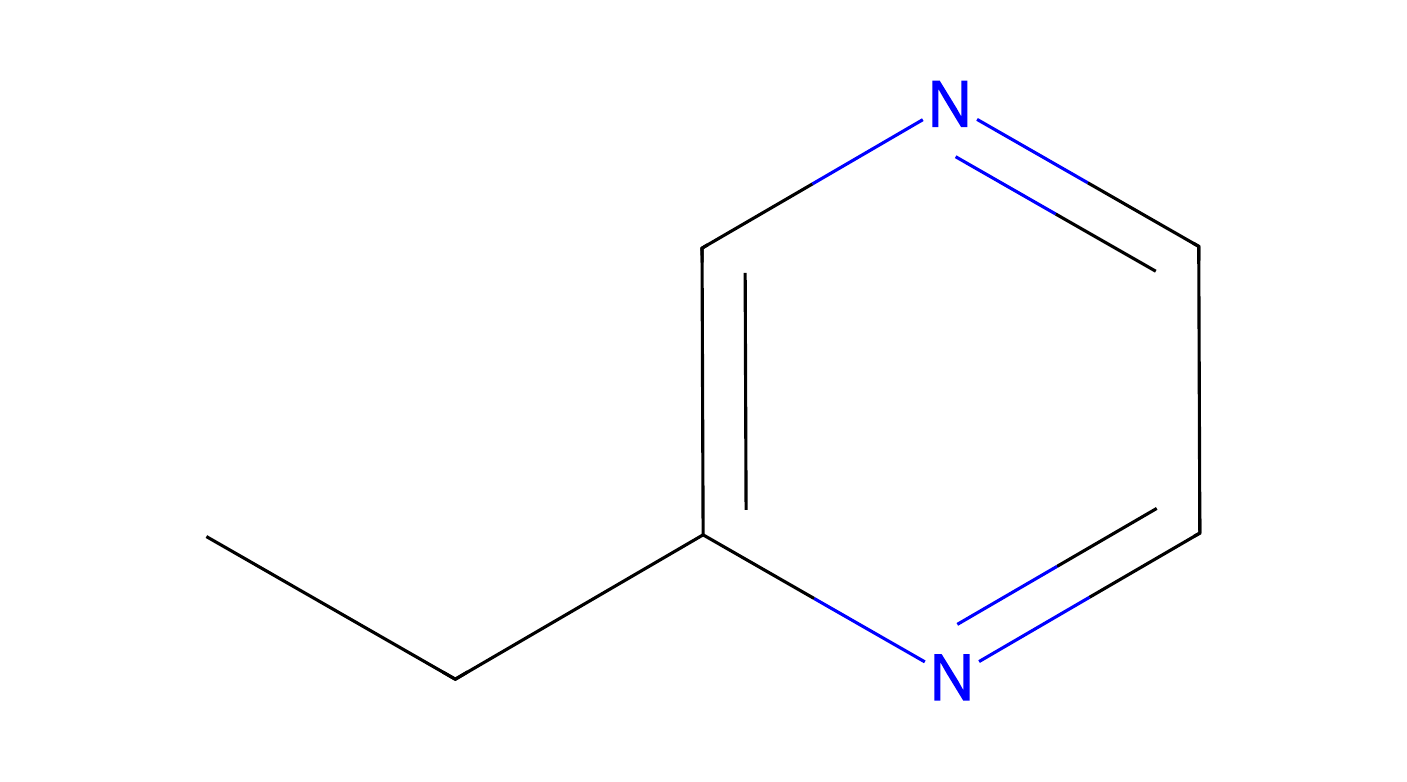}} \\
\parbox[c][1.48cm][c]{0.47\columnwidth}{\raggedright\scriptsize\ttfamily CCc1ccccn1}
 & \parbox[c][1.48cm][c]{0.27\columnwidth}{\raggedright\scriptsize\ttfamily mobley\_\allowbreak{}6988468}
 & \parbox[c][1.48cm][c]{0.18\columnwidth}{\centering\includegraphics[width=\linewidth,height=1.32cm,keepaspectratio]{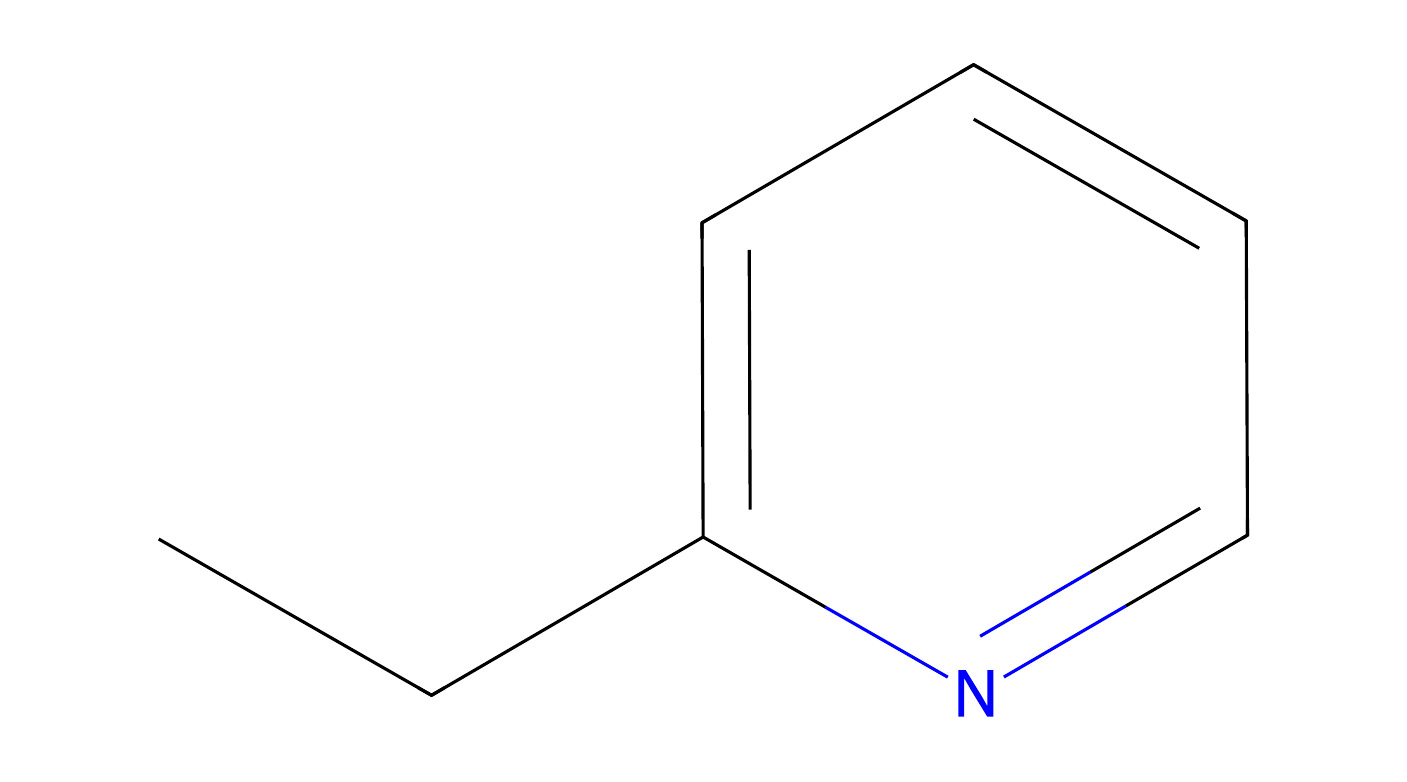}} \\
\parbox[c][1.48cm][c]{0.47\columnwidth}{\raggedright\scriptsize\ttfamily Cc1cnccn1}
 & \parbox[c][1.48cm][c]{0.27\columnwidth}{\raggedright\scriptsize\ttfamily mobley\_\allowbreak{}8966374}
 & \parbox[c][1.48cm][c]{0.18\columnwidth}{\centering\includegraphics[width=\linewidth,height=1.32cm,keepaspectratio]{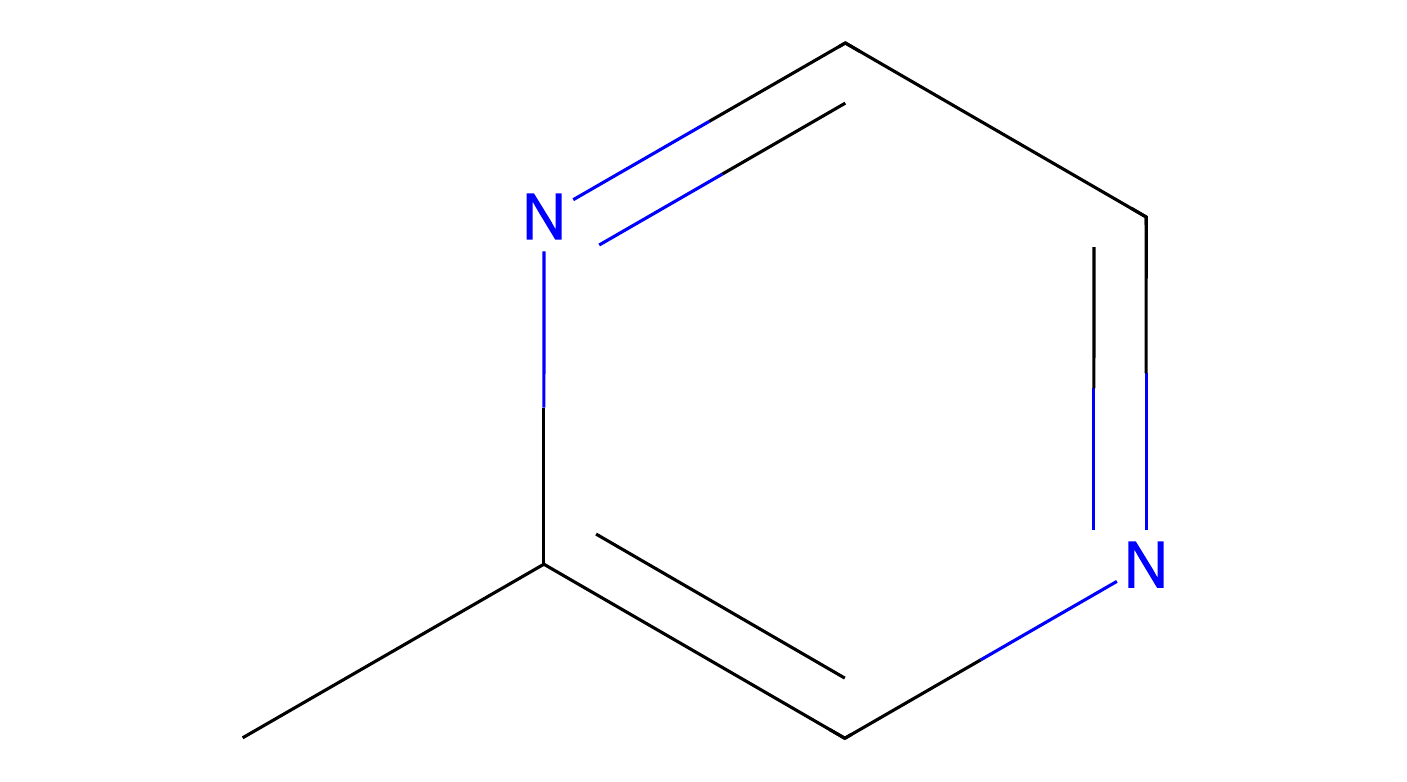}} \\
\bottomrule
\end{tabular}

\newpage
\noindent\textbf{\tablename~\thetable{} -- continued}
\par\vspace{2pt}
\noindent\begin{tabular}{@{}lll@{}}
\toprule
\parbox[t]{0.47\columnwidth}{\raggedright\bfseries SMILES} &
\parbox[t]{0.27\columnwidth}{\raggedright\bfseries freesolv\_id} &
\parbox[t]{0.18\columnwidth}{\centering\bfseries Molecule} \\
\midrule
\parbox[c][1.48cm][c]{0.47\columnwidth}{\raggedright\scriptsize\ttfamily Cc1ccccn1}
 & \parbox[c][1.48cm][c]{0.27\columnwidth}{\raggedright\scriptsize\ttfamily mobley\_\allowbreak{}4287564}
 & \parbox[c][1.48cm][c]{0.18\columnwidth}{\centering\includegraphics[width=\linewidth,height=1.32cm,keepaspectratio]{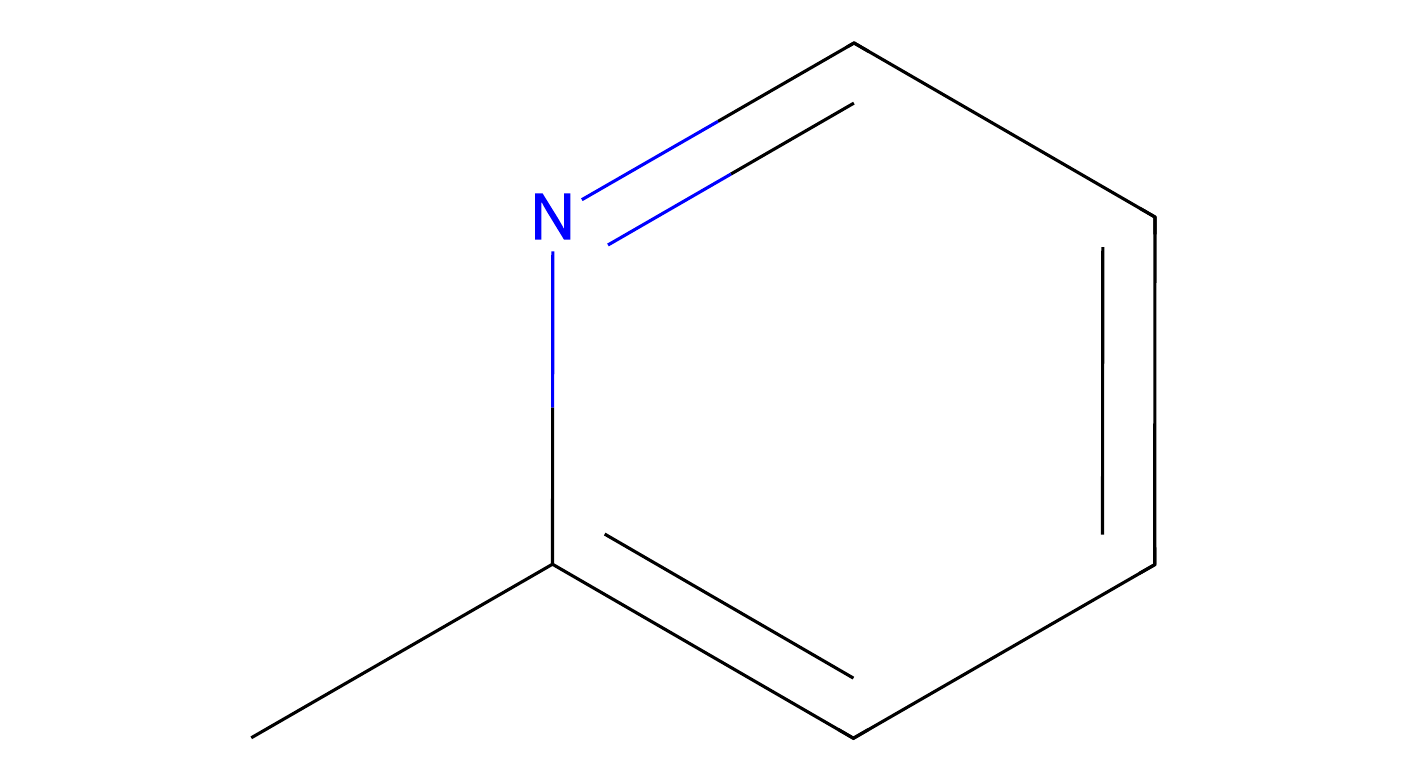}} \\
\parbox[c][1.48cm][c]{0.47\columnwidth}{\raggedright\scriptsize\ttfamily c1cc(\allowbreak{}cc(\allowbreak{}c1)\allowbreak{}Cl)\allowbreak{}N}
 & \parbox[c][1.48cm][c]{0.27\columnwidth}{\raggedright\scriptsize\ttfamily mobley\_\allowbreak{}6257907}
 & \parbox[c][1.48cm][c]{0.18\columnwidth}{\centering\includegraphics[width=\linewidth,height=1.32cm,keepaspectratio]{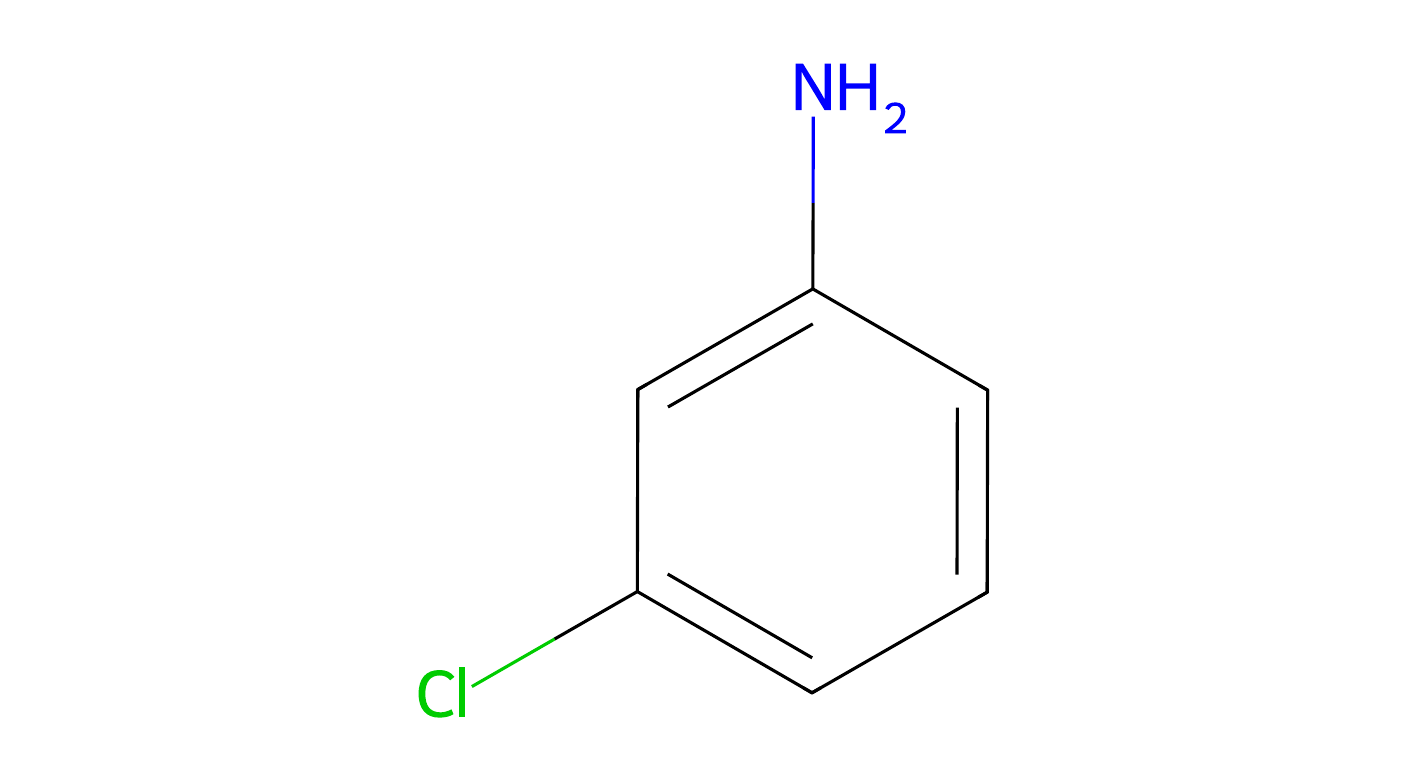}} \\
\parbox[c][1.48cm][c]{0.47\columnwidth}{\raggedright\scriptsize\ttfamily c1cc(\allowbreak{}cc(\allowbreak{}c1)\allowbreak{}Cl)\allowbreak{}O}
 & \parbox[c][1.48cm][c]{0.27\columnwidth}{\raggedright\scriptsize\ttfamily mobley\_\allowbreak{}8057732}
 & \parbox[c][1.48cm][c]{0.18\columnwidth}{\centering\includegraphics[width=\linewidth,height=1.32cm,keepaspectratio]{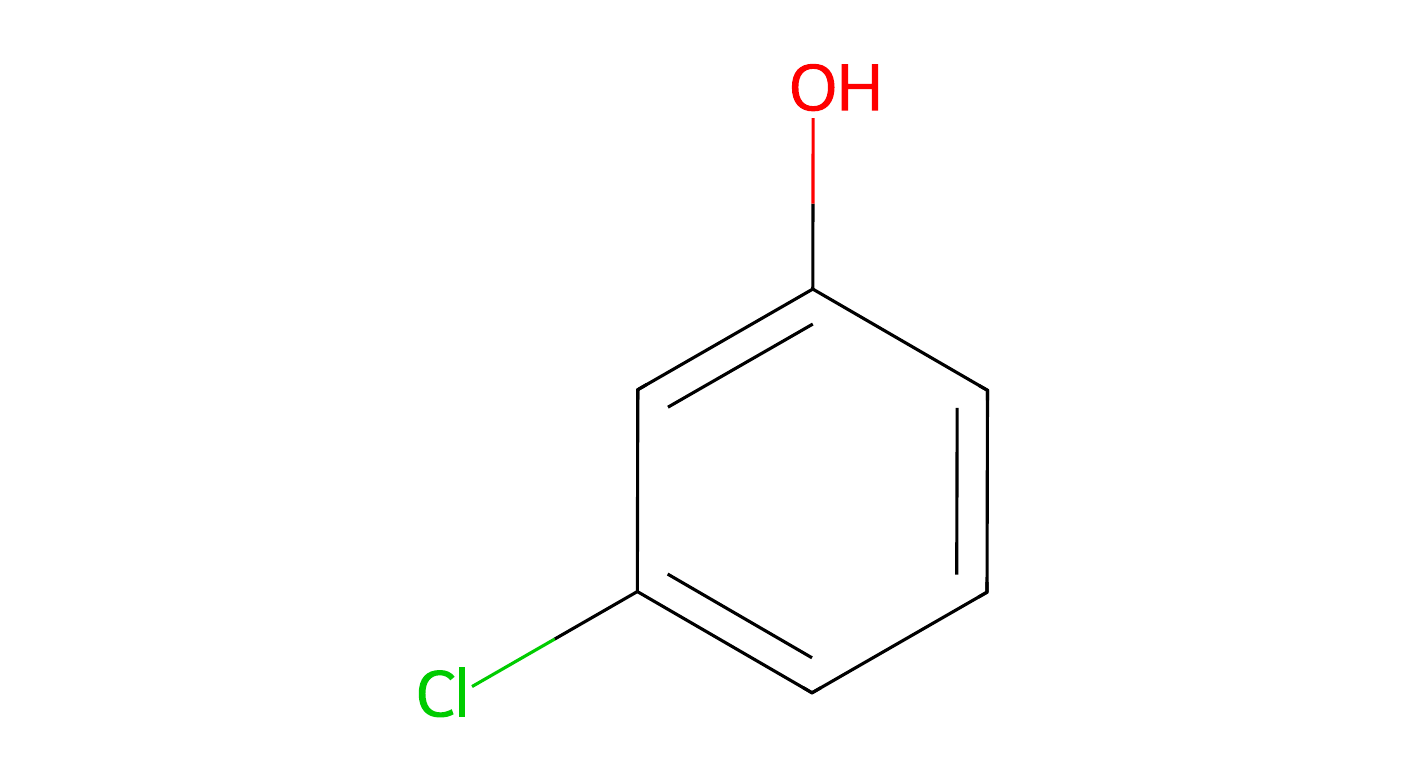}} \\
\parbox[c][1.48cm][c]{0.47\columnwidth}{\raggedright\scriptsize\ttfamily CCc1cccnc1}
 & \parbox[c][1.48cm][c]{0.27\columnwidth}{\raggedright\scriptsize\ttfamily mobley\_\allowbreak{}4884177}
 & \parbox[c][1.48cm][c]{0.18\columnwidth}{\centering\includegraphics[width=\linewidth,height=1.32cm,keepaspectratio]{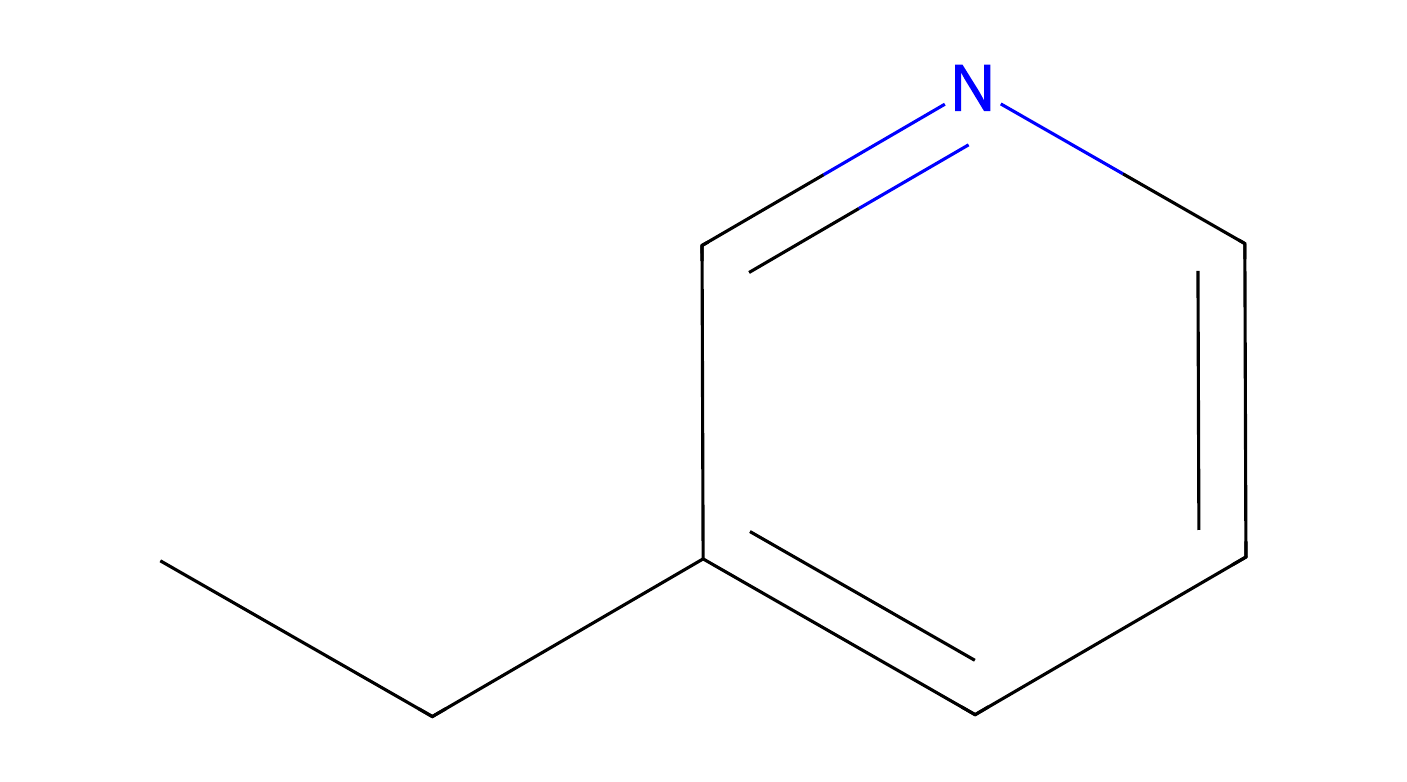}} \\
\parbox[c][1.48cm][c]{0.47\columnwidth}{\raggedright\scriptsize\ttfamily c1cc(\allowbreak{}ccc1N)\allowbreak{}Cl}
 & \parbox[c][1.48cm][c]{0.27\columnwidth}{\raggedright\scriptsize\ttfamily mobley\_\allowbreak{}8723116}
 & \parbox[c][1.48cm][c]{0.18\columnwidth}{\centering\includegraphics[width=\linewidth,height=1.32cm,keepaspectratio]{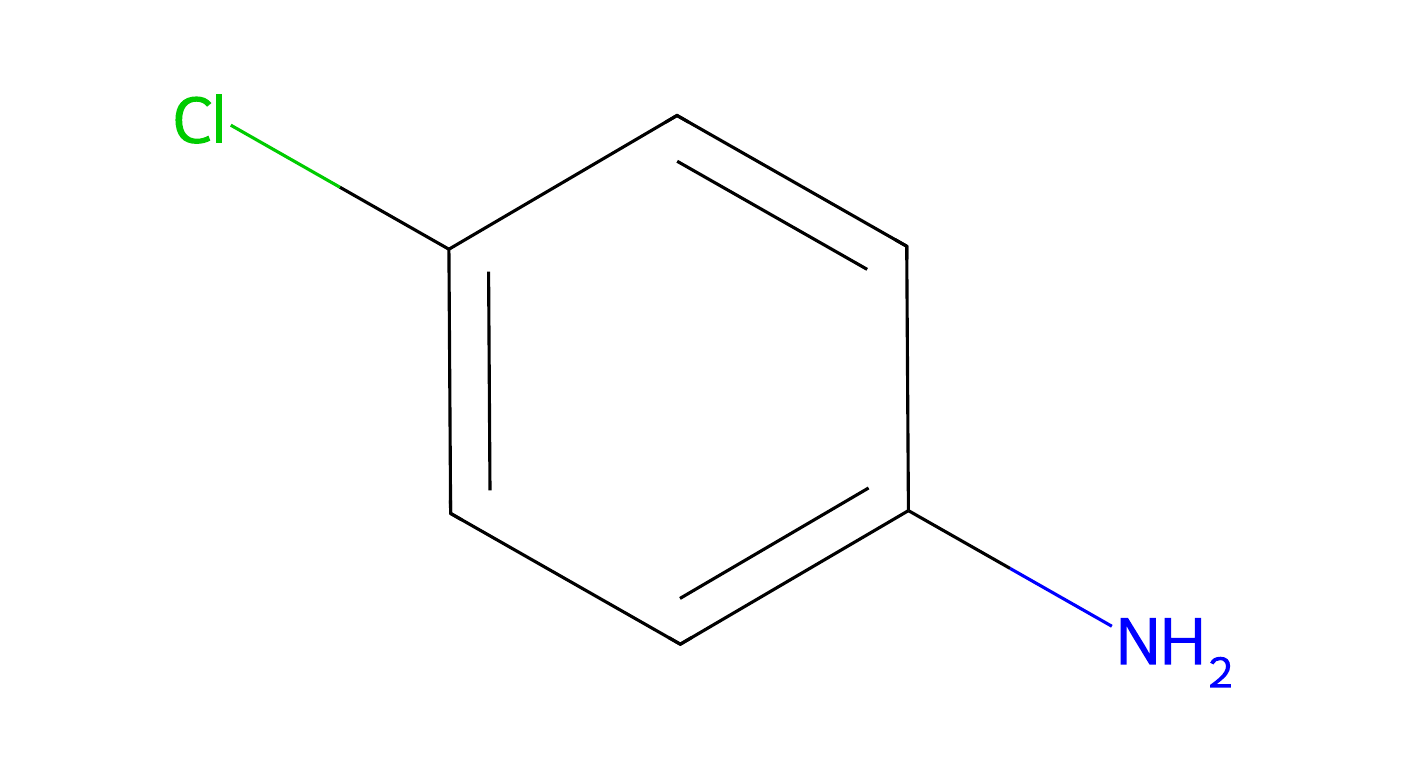}} \\
\parbox[c][1.48cm][c]{0.47\columnwidth}{\raggedright\scriptsize\ttfamily c1cc(\allowbreak{}ccc1O)\allowbreak{}Cl}
 & \parbox[c][1.48cm][c]{0.27\columnwidth}{\raggedright\scriptsize\ttfamily mobley\_\allowbreak{}7066554}
 & \parbox[c][1.48cm][c]{0.18\columnwidth}{\centering\includegraphics[width=\linewidth,height=1.32cm,keepaspectratio]{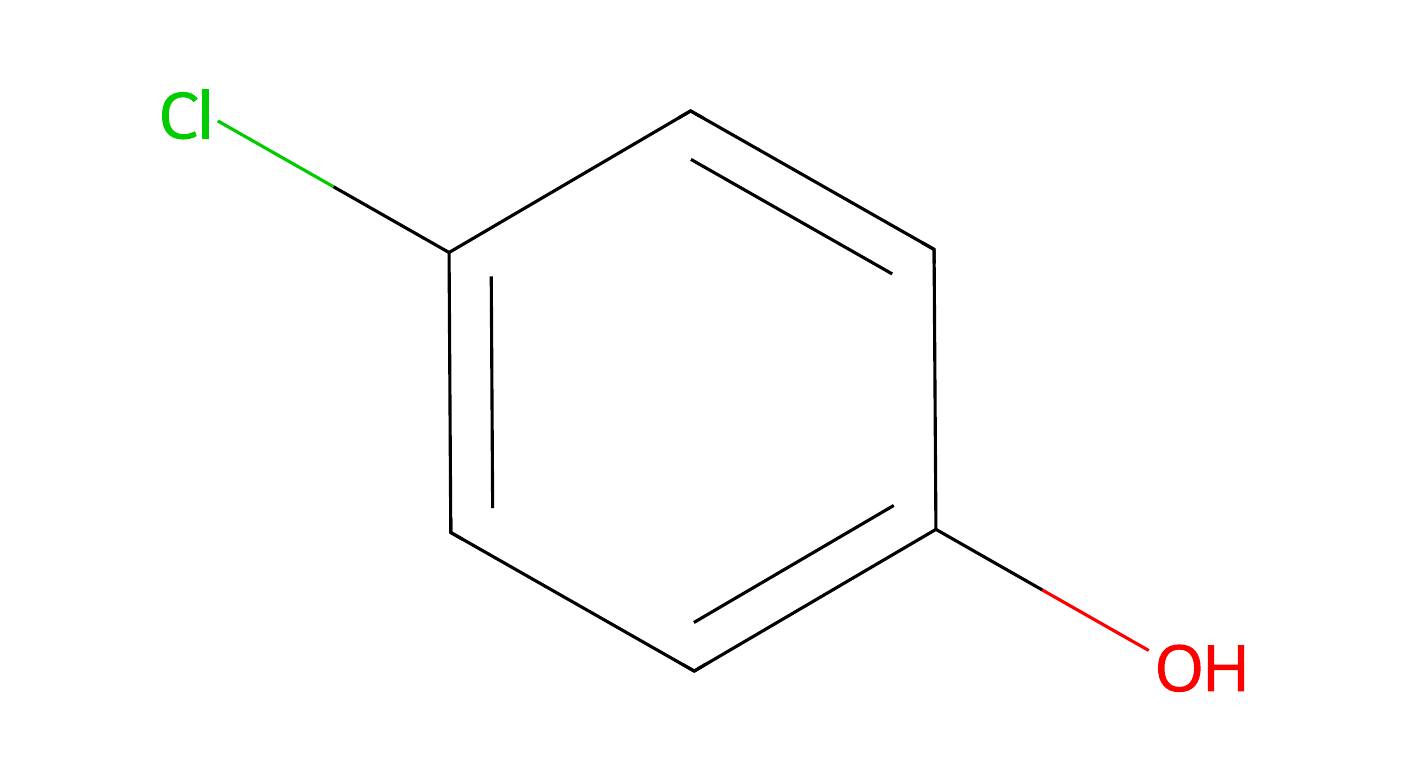}} \\
\parbox[c][1.48cm][c]{0.47\columnwidth}{\raggedright\scriptsize\ttfamily CCc1ccncc1}
 & \parbox[c][1.48cm][c]{0.27\columnwidth}{\raggedright\scriptsize\ttfamily mobley\_\allowbreak{}6303022}
 & \parbox[c][1.48cm][c]{0.18\columnwidth}{\centering\includegraphics[width=\linewidth,height=1.32cm,keepaspectratio]{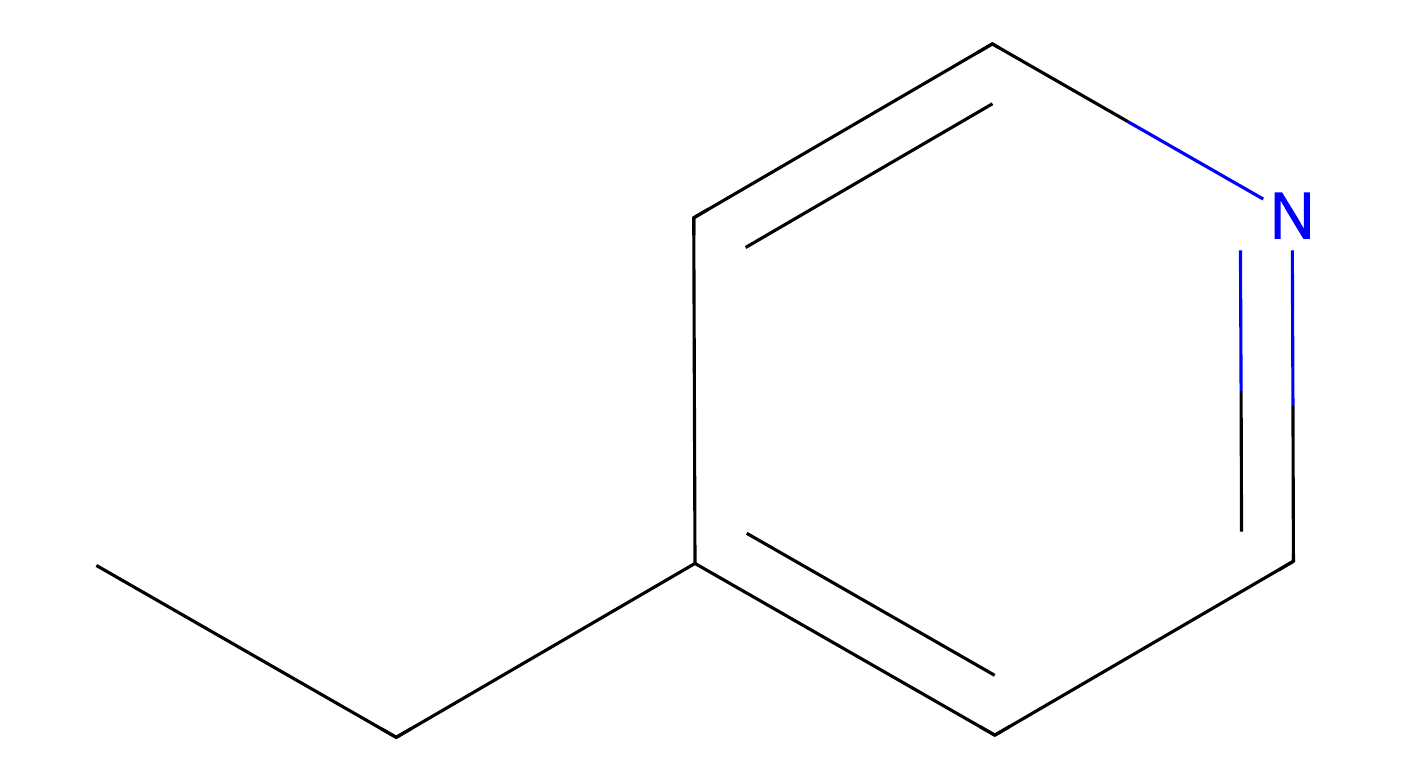}} \\
\parbox[c][1.48cm][c]{0.47\columnwidth}{\raggedright\scriptsize\ttfamily COc1ccc(\allowbreak{}cc1)\allowbreak{}N}
 & \parbox[c][1.48cm][c]{0.27\columnwidth}{\raggedright\scriptsize\ttfamily mobley\_\allowbreak{}4218209}
 & \parbox[c][1.48cm][c]{0.18\columnwidth}{\centering\includegraphics[width=\linewidth,height=1.32cm,keepaspectratio]{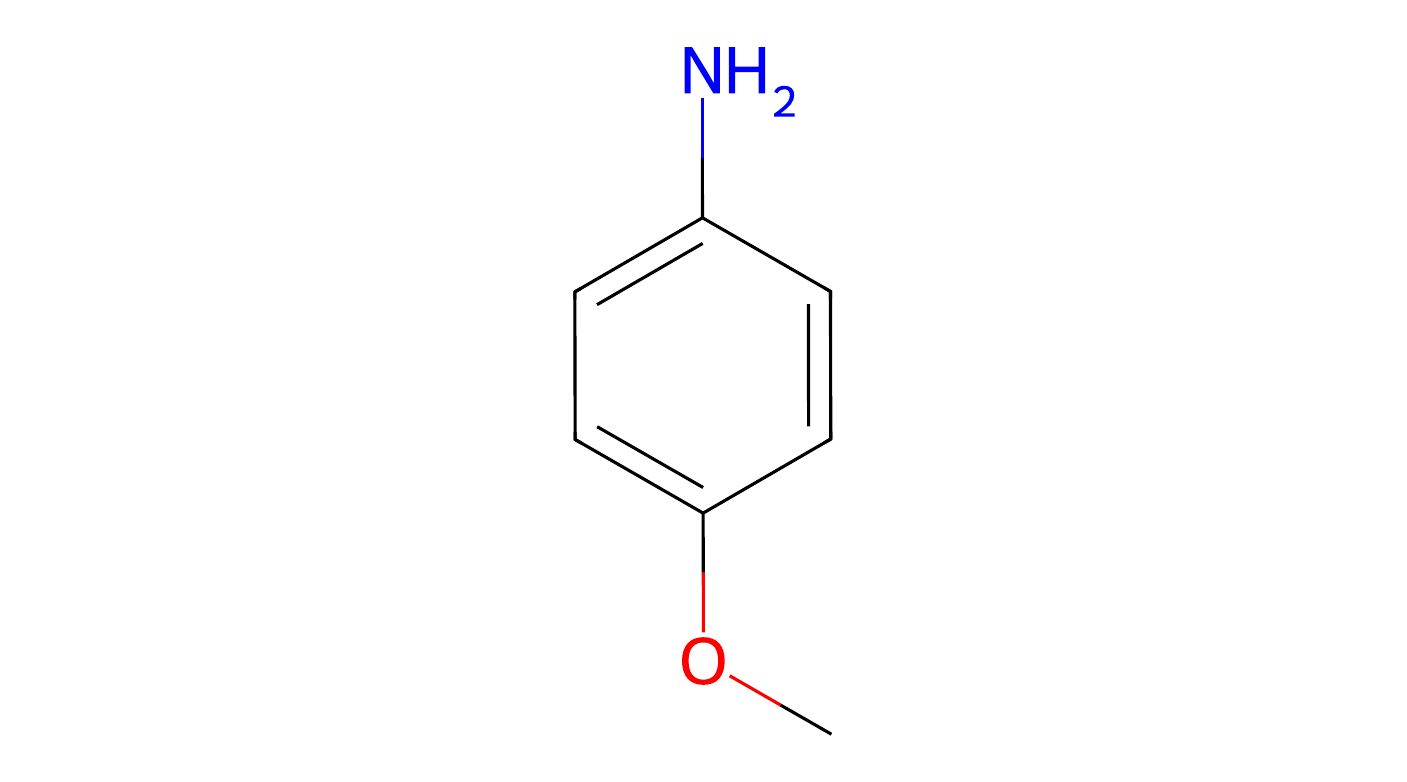}} \\
\parbox[c][1.48cm][c]{0.47\columnwidth}{\raggedright\scriptsize\ttfamily Cc1ccc(\allowbreak{}cc1)\allowbreak{}N}
 & \parbox[c][1.48cm][c]{0.27\columnwidth}{\raggedright\scriptsize\ttfamily mobley\_\allowbreak{}5518547}
 & \parbox[c][1.48cm][c]{0.18\columnwidth}{\centering\includegraphics[width=\linewidth,height=1.32cm,keepaspectratio]{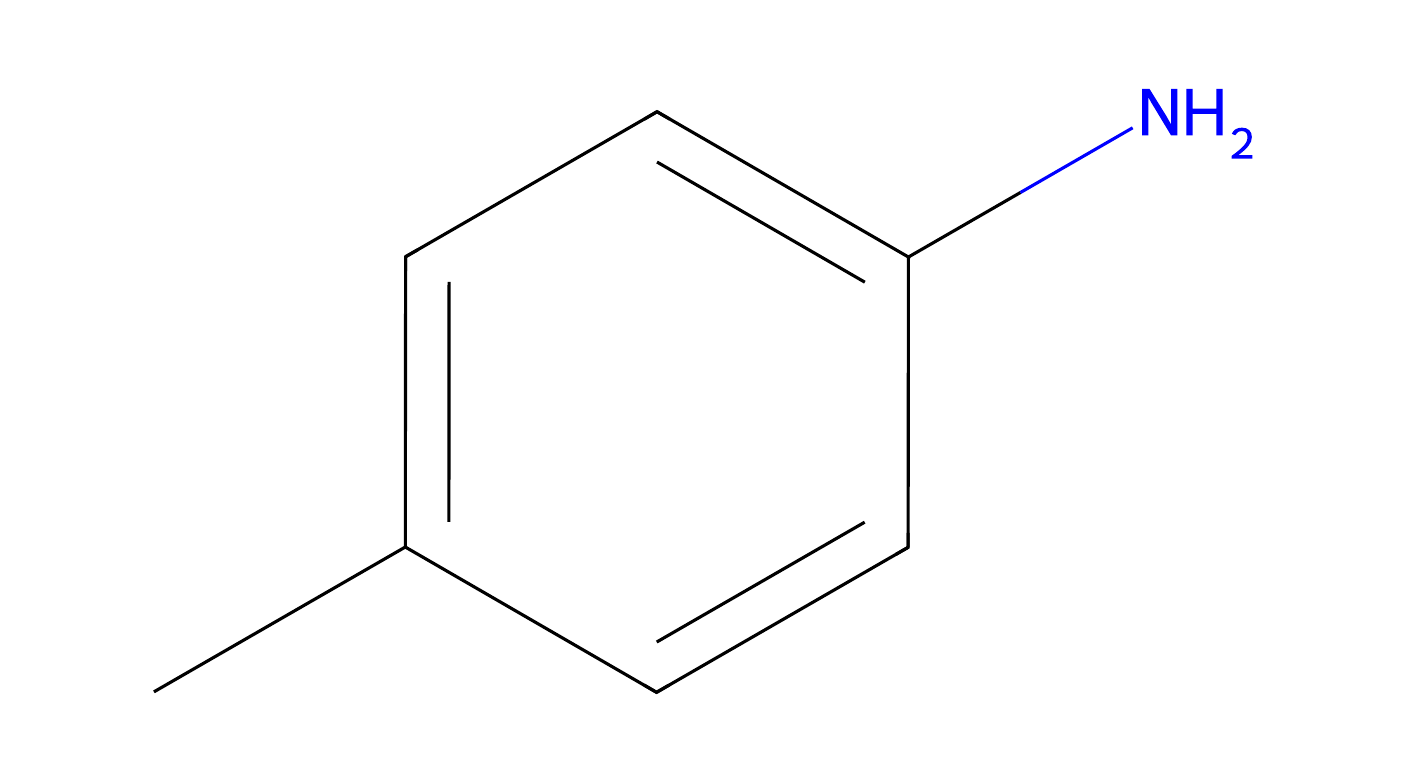}} \\
\parbox[c][1.48cm][c]{0.47\columnwidth}{\raggedright\scriptsize\ttfamily CC[\allowbreak{}C@\allowbreak{}H]\allowbreak{}(\allowbreak{}C)\allowbreak{}O}
 & \parbox[c][1.48cm][c]{0.27\columnwidth}{\raggedright\scriptsize\ttfamily mobley\_\allowbreak{}1903702}
 & \parbox[c][1.48cm][c]{0.18\columnwidth}{\centering\includegraphics[width=\linewidth,height=1.32cm,keepaspectratio]{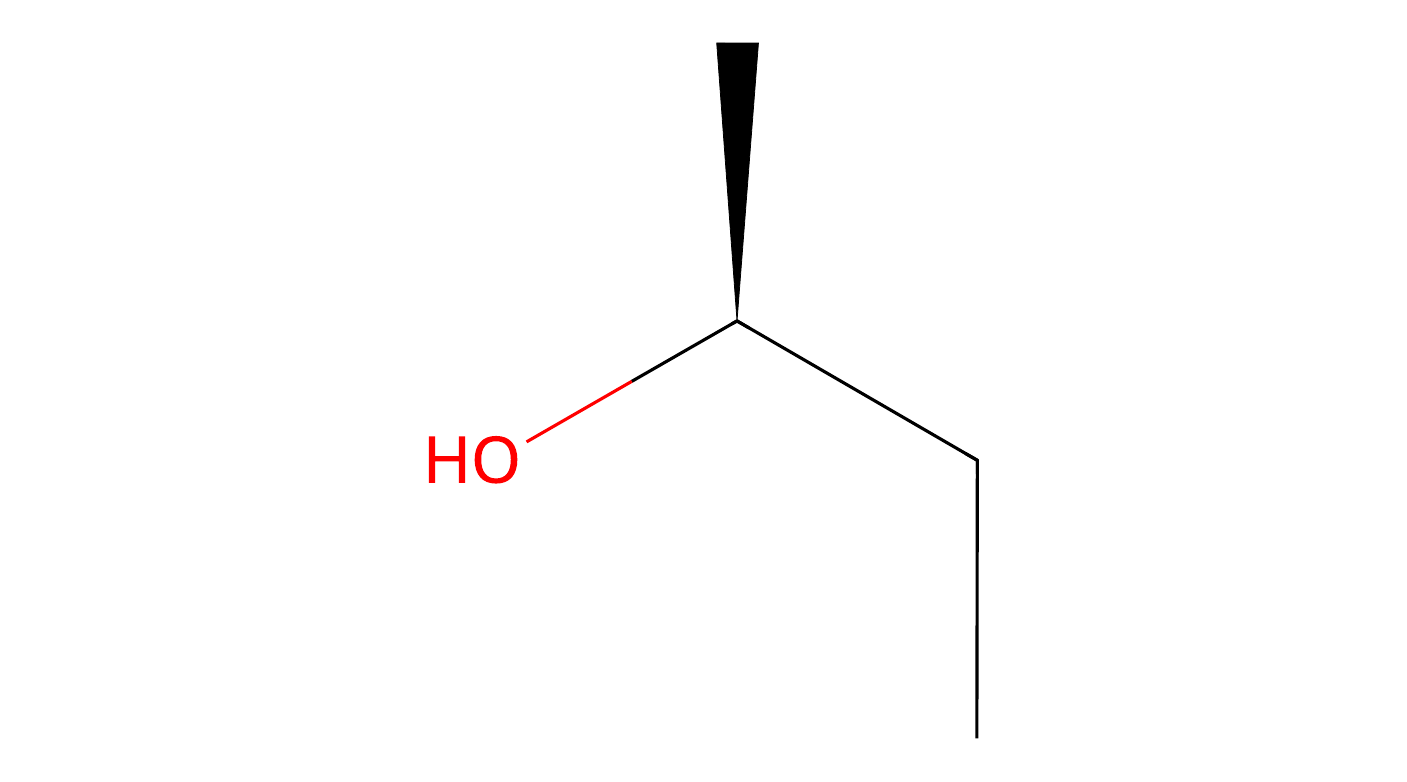}} \\
\parbox[c][1.48cm][c]{0.47\columnwidth}{\raggedright\scriptsize\ttfamily c1ccc(\allowbreak{}cc1)\allowbreak{}Cl}
 & \parbox[c][1.48cm][c]{0.27\columnwidth}{\raggedright\scriptsize\ttfamily mobley\_\allowbreak{}7608462}
 & \parbox[c][1.48cm][c]{0.18\columnwidth}{\centering\includegraphics[width=\linewidth,height=1.32cm,keepaspectratio]{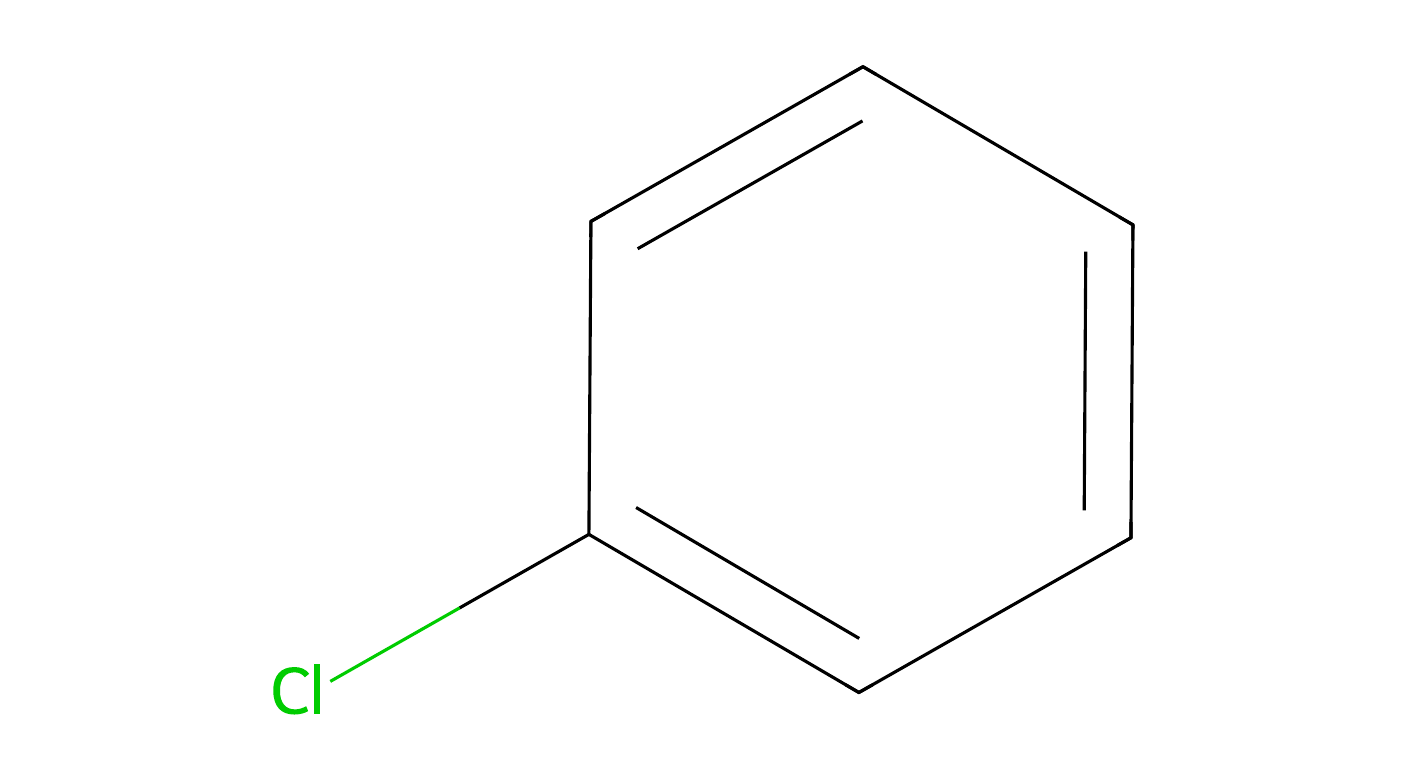}} \\
\bottomrule
\end{tabular}

\newpage
\noindent\textbf{\tablename~\thetable{} -- continued}
\par\vspace{2pt}
\noindent\begin{tabular}{@{}lll@{}}
\toprule
\parbox[t]{0.47\columnwidth}{\raggedright\bfseries SMILES} &
\parbox[t]{0.27\columnwidth}{\raggedright\bfseries freesolv\_id} &
\parbox[t]{0.18\columnwidth}{\centering\bfseries Molecule} \\
\midrule
\parbox[c][1.48cm][c]{0.47\columnwidth}{\raggedright\scriptsize\ttfamily C1CCCC(\allowbreak{}CC1)\allowbreak{}O}
 & \parbox[c][1.48cm][c]{0.27\columnwidth}{\raggedright\scriptsize\ttfamily mobley\_\allowbreak{}1244778}
 & \parbox[c][1.48cm][c]{0.18\columnwidth}{\centering\includegraphics[width=\linewidth,height=1.32cm,keepaspectratio]{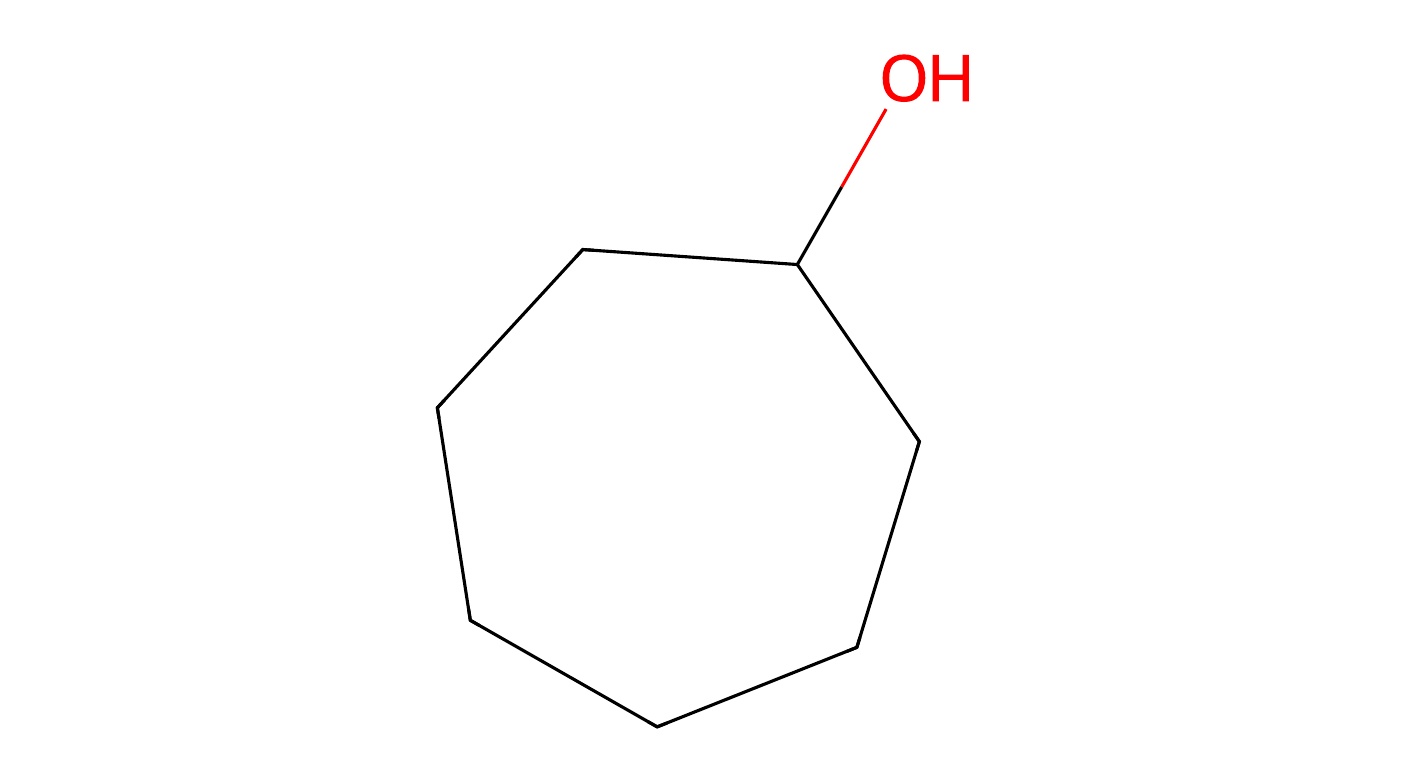}} \\
\parbox[c][1.48cm][c]{0.47\columnwidth}{\raggedright\scriptsize\ttfamily C1CCC(\allowbreak{}CC1)\allowbreak{}O}
 & \parbox[c][1.48cm][c]{0.27\columnwidth}{\raggedright\scriptsize\ttfamily mobley\_\allowbreak{}7010316}
 & \parbox[c][1.48cm][c]{0.18\columnwidth}{\centering\includegraphics[width=\linewidth,height=1.32cm,keepaspectratio]{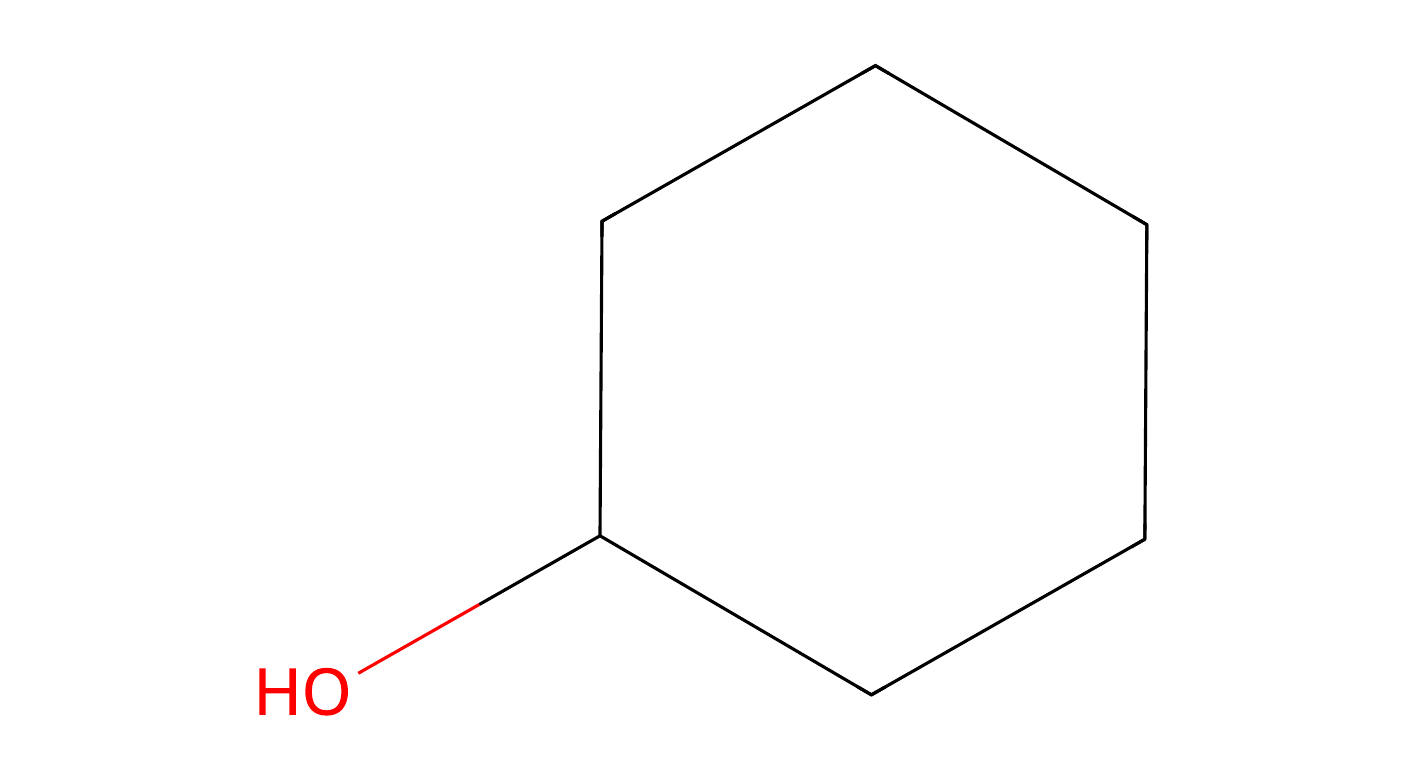}} \\
\parbox[c][1.48cm][c]{0.47\columnwidth}{\raggedright\scriptsize\ttfamily C1CCC(\allowbreak{}=\allowbreak{}O)\allowbreak{}CC1}
 & \parbox[c][1.48cm][c]{0.27\columnwidth}{\raggedright\scriptsize\ttfamily mobley\_\allowbreak{}3682850}
 & \parbox[c][1.48cm][c]{0.18\columnwidth}{\centering\includegraphics[width=\linewidth,height=1.32cm,keepaspectratio]{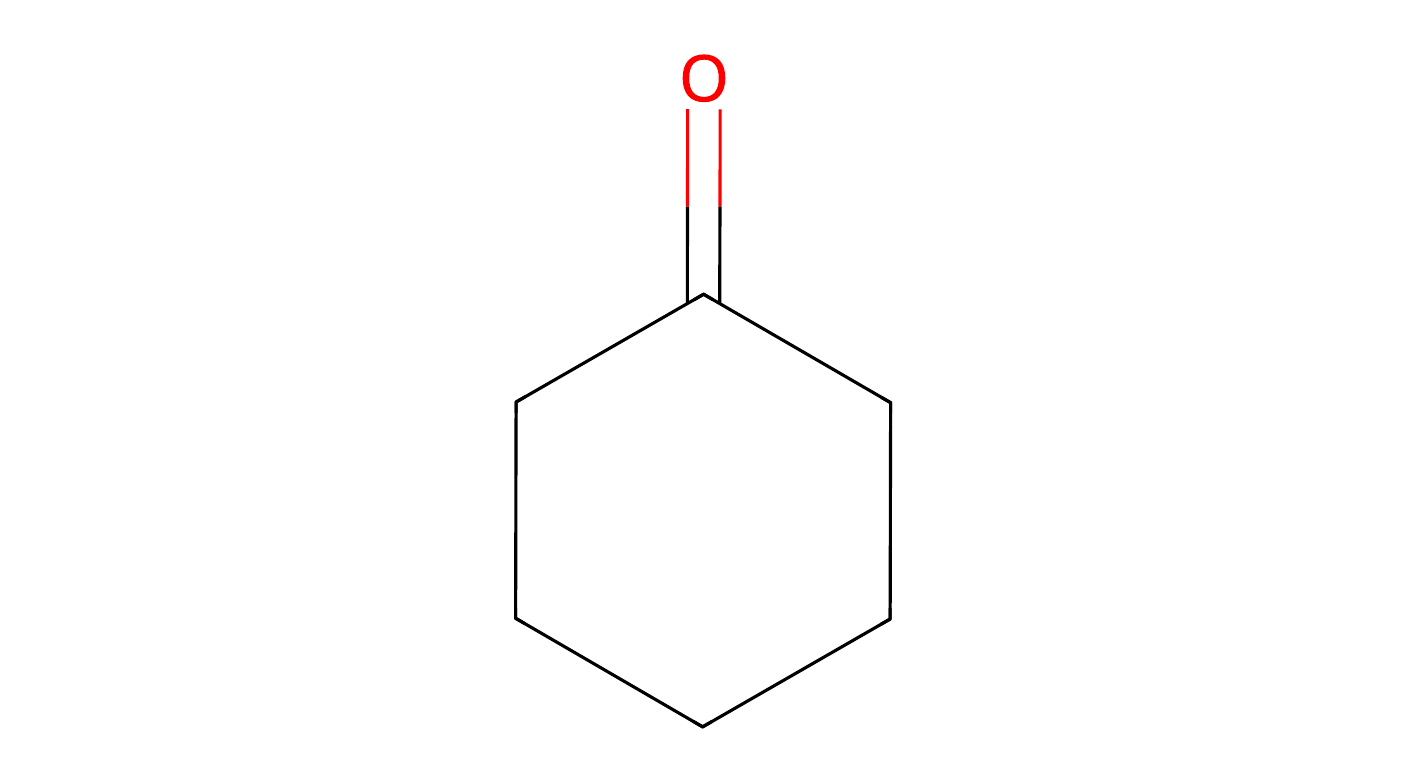}} \\
\parbox[c][1.48cm][c]{0.47\columnwidth}{\raggedright\scriptsize\ttfamily C1CCC(\allowbreak{}C1)\allowbreak{}O}
 & \parbox[c][1.48cm][c]{0.27\columnwidth}{\raggedright\scriptsize\ttfamily mobley\_\allowbreak{}8899867}
 & \parbox[c][1.48cm][c]{0.18\columnwidth}{\centering\includegraphics[width=\linewidth,height=1.32cm,keepaspectratio]{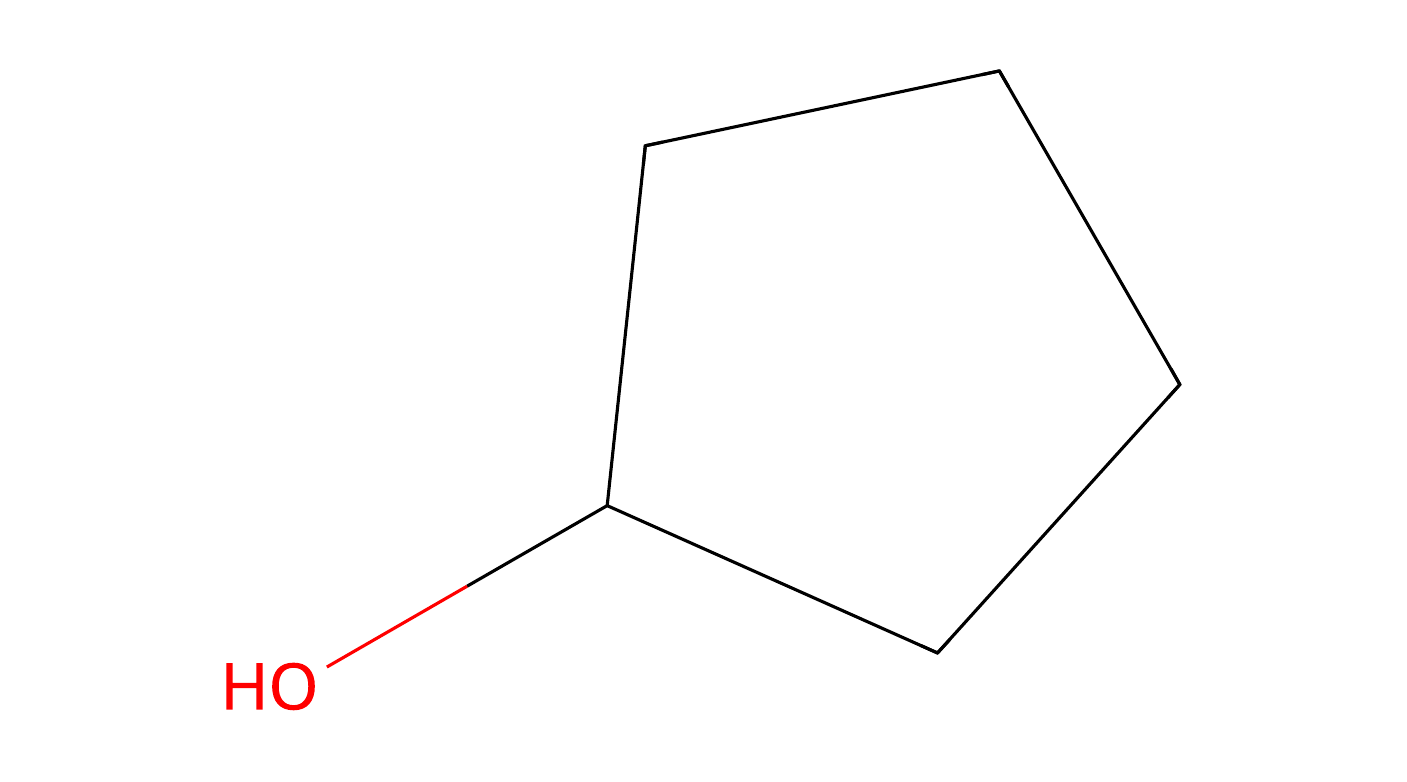}} \\
\parbox[c][1.48cm][c]{0.47\columnwidth}{\raggedright\scriptsize\ttfamily CCSC}
 & \parbox[c][1.48cm][c]{0.27\columnwidth}{\raggedright\scriptsize\ttfamily mobley\_\allowbreak{}2049967}
 & \parbox[c][1.48cm][c]{0.18\columnwidth}{\centering\includegraphics[width=\linewidth,height=1.32cm,keepaspectratio]{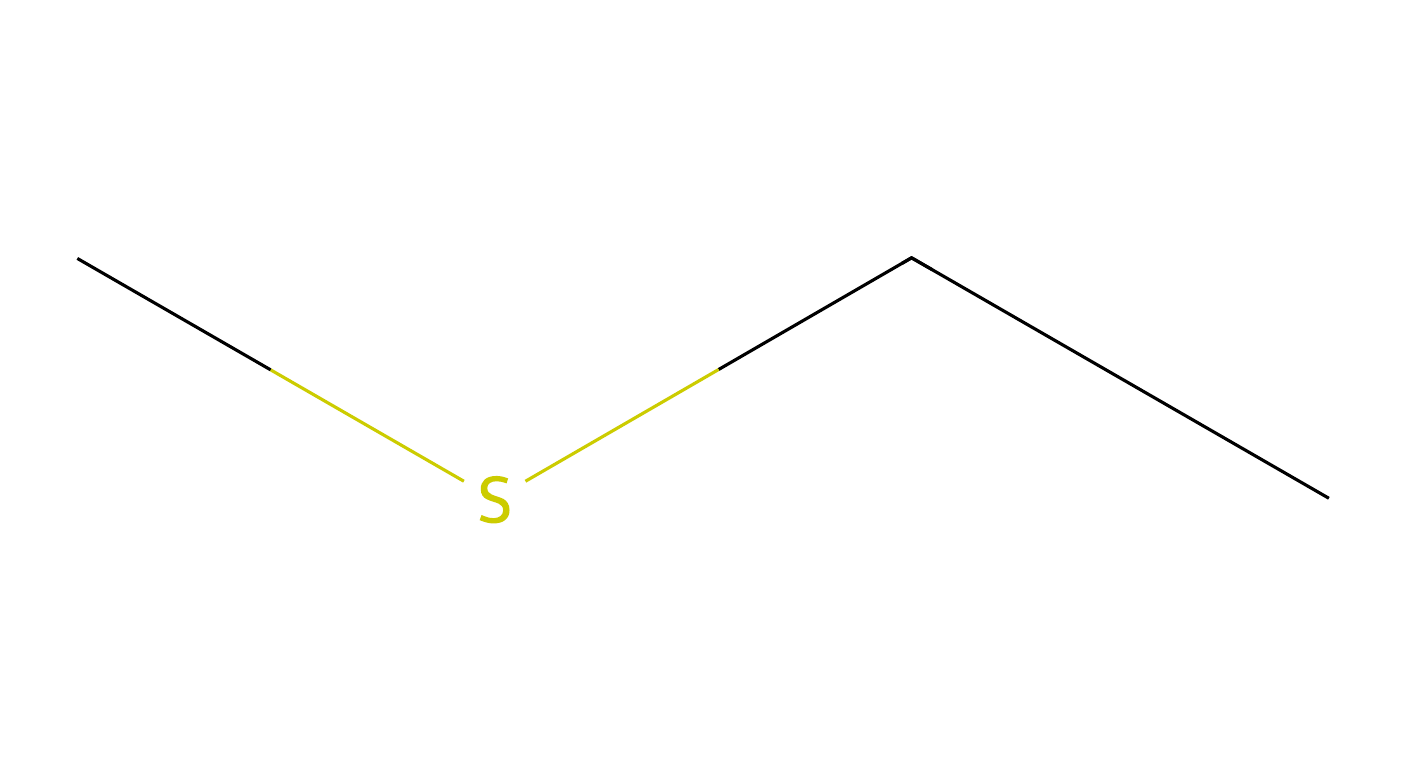}} \\
\parbox[c][1.48cm][c]{0.47\columnwidth}{\raggedright\scriptsize\ttfamily CN(\allowbreak{}C)\allowbreak{}C(\allowbreak{}=\allowbreak{}O)\allowbreak{}c1ccccc1}
 & \parbox[c][1.48cm][c]{0.27\columnwidth}{\raggedright\scriptsize\ttfamily mobley\_\allowbreak{}4188615}
 & \parbox[c][1.48cm][c]{0.18\columnwidth}{\centering\includegraphics[width=\linewidth,height=1.32cm,keepaspectratio]{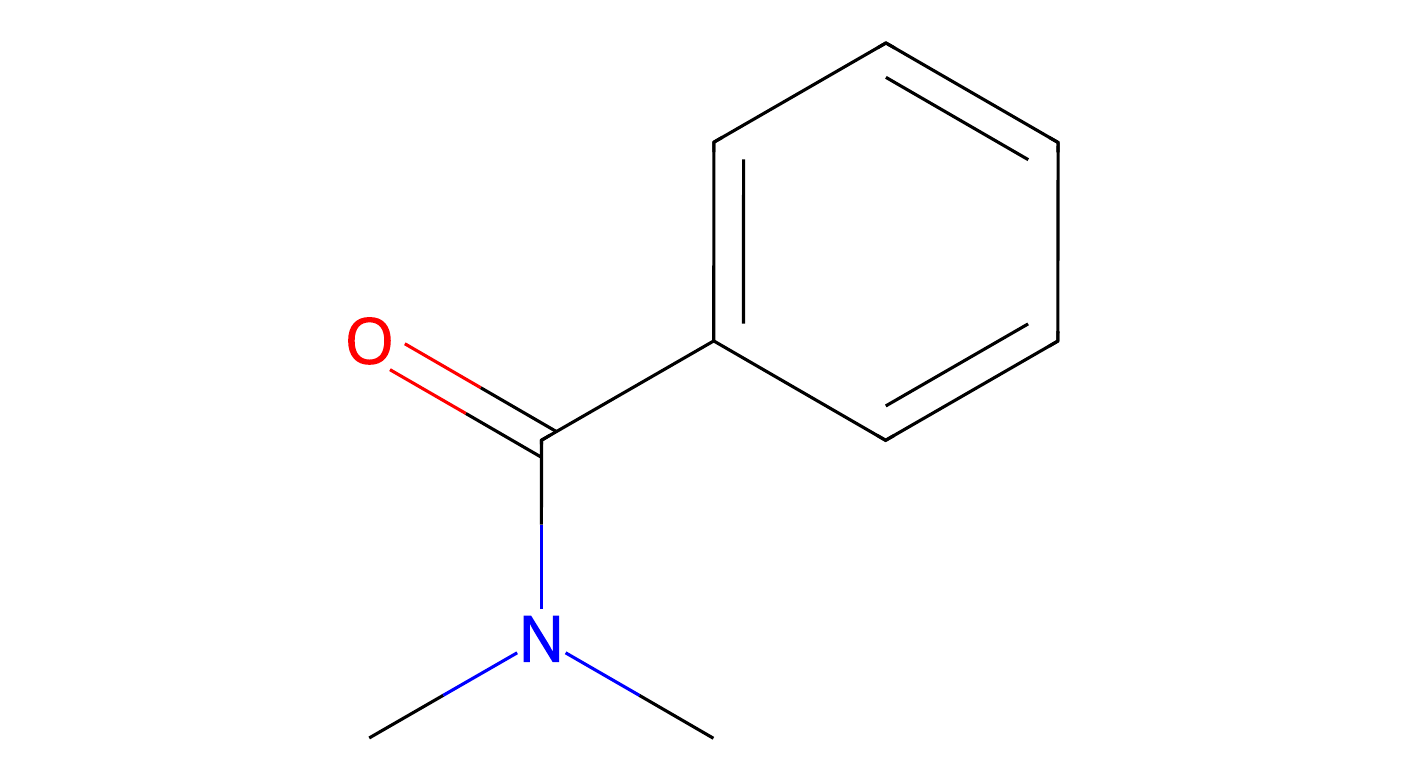}} \\
\parbox[c][1.48cm][c]{0.47\columnwidth}{\raggedright\scriptsize\ttfamily c1ccc2c(\allowbreak{}c1)\allowbreak{}cccc2O}
 & \parbox[c][1.48cm][c]{0.27\columnwidth}{\raggedright\scriptsize\ttfamily mobley\_\allowbreak{}7039935}
 & \parbox[c][1.48cm][c]{0.18\columnwidth}{\centering\includegraphics[width=\linewidth,height=1.32cm,keepaspectratio]{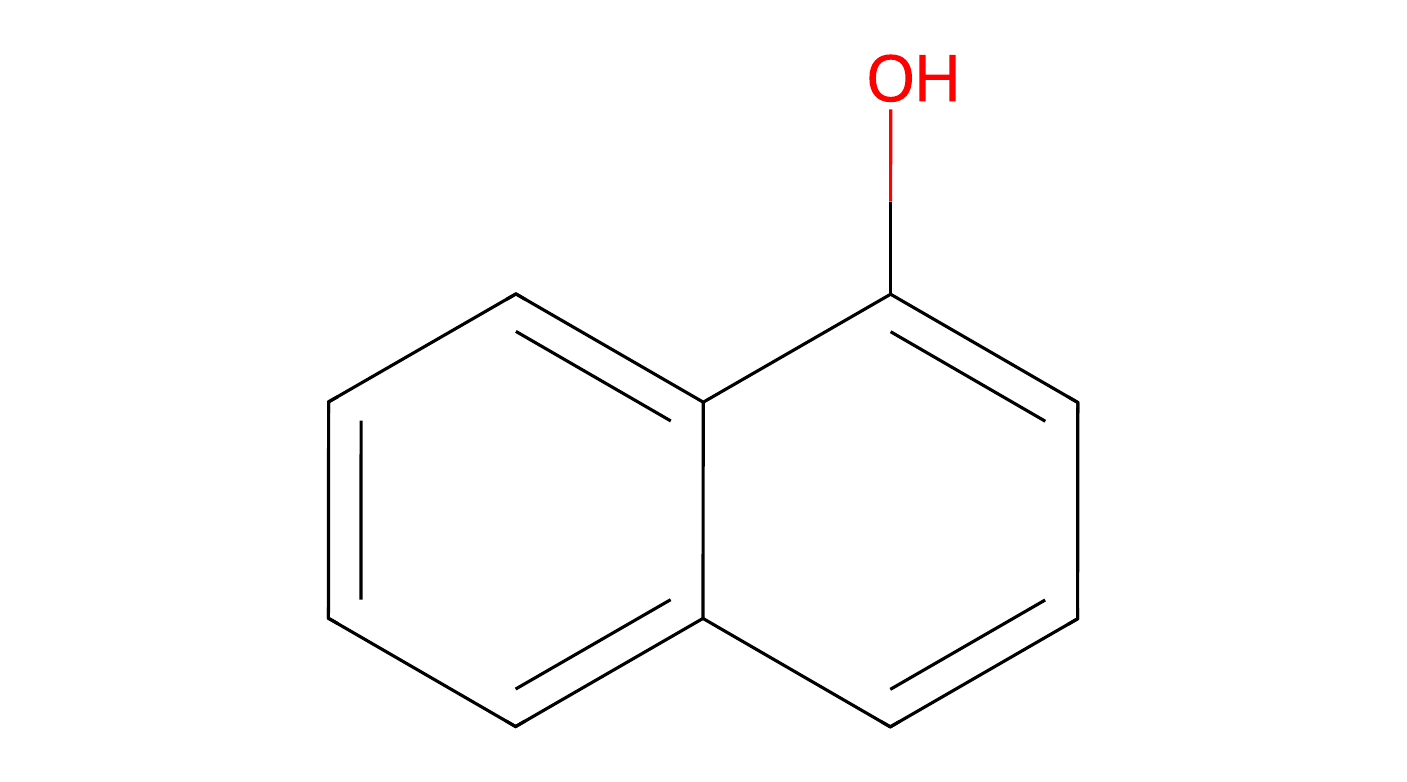}} \\
\parbox[c][1.48cm][c]{0.47\columnwidth}{\raggedright\scriptsize\ttfamily Cc1ccccc1O}
 & \parbox[c][1.48cm][c]{0.27\columnwidth}{\raggedright\scriptsize\ttfamily mobley\_\allowbreak{}4035953}
 & \parbox[c][1.48cm][c]{0.18\columnwidth}{\centering\includegraphics[width=\linewidth,height=1.32cm,keepaspectratio]{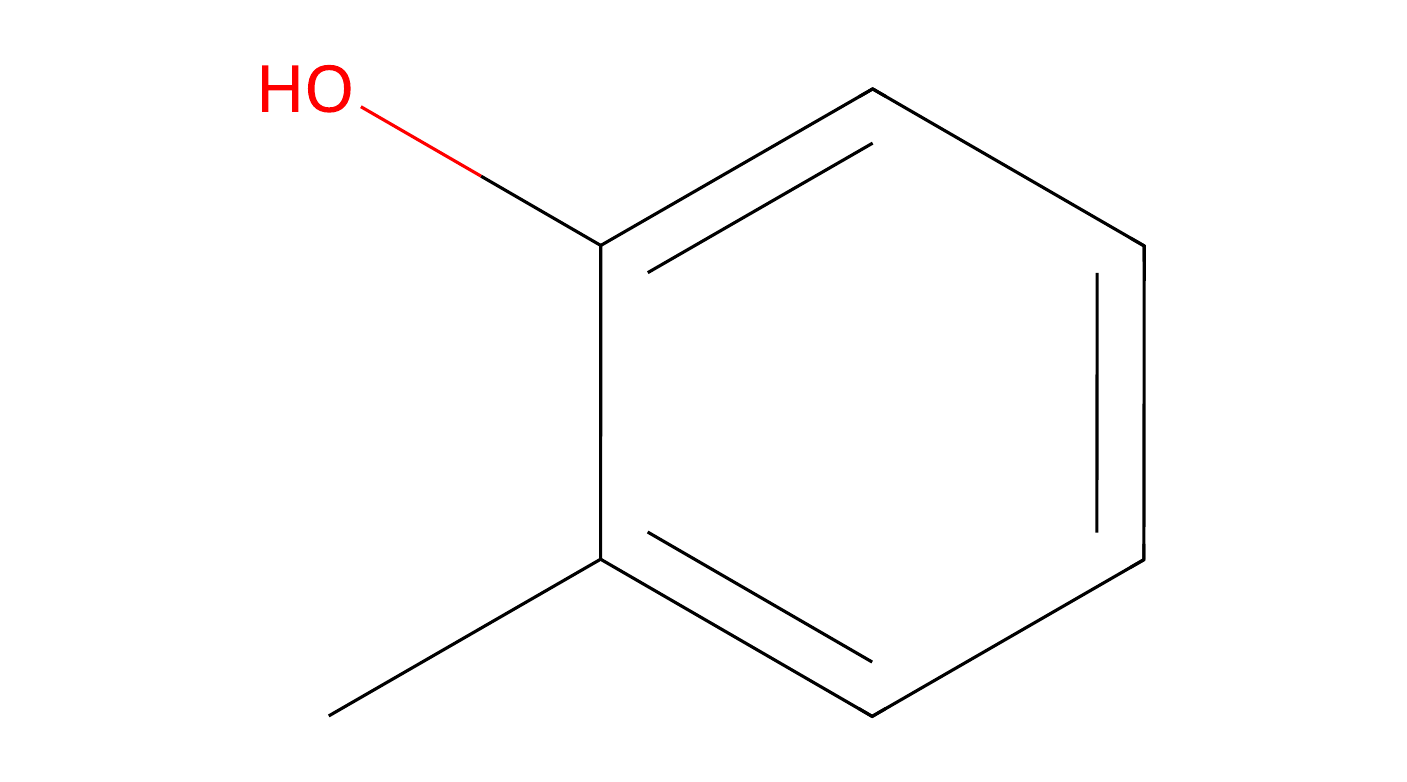}} \\
\parbox[c][1.48cm][c]{0.47\columnwidth}{\raggedright\scriptsize\ttfamily CCCC(\allowbreak{}=\allowbreak{}O)\allowbreak{}C}
 & \parbox[c][1.48cm][c]{0.27\columnwidth}{\raggedright\scriptsize\ttfamily mobley\_\allowbreak{}1881249}
 & \parbox[c][1.48cm][c]{0.18\columnwidth}{\centering\includegraphics[width=\linewidth,height=1.32cm,keepaspectratio]{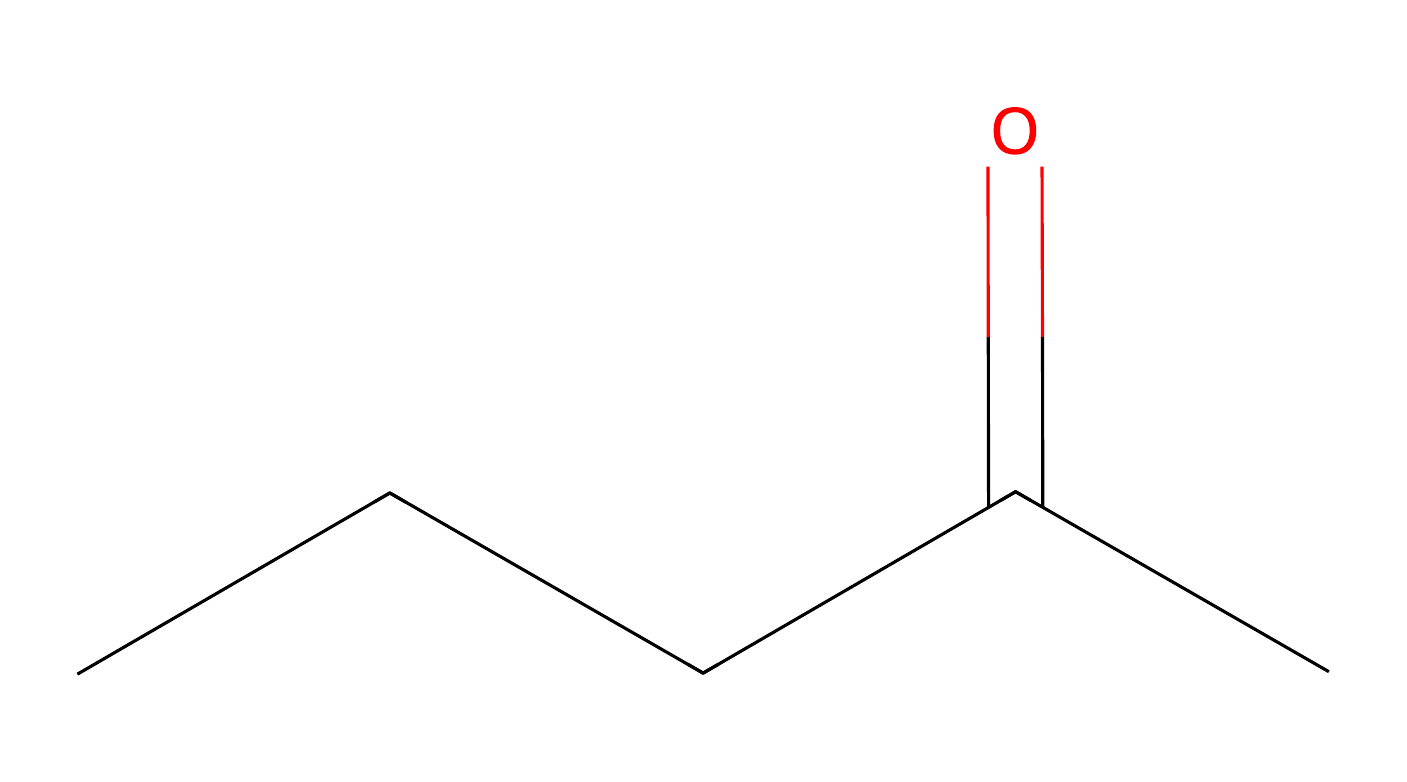}} \\
\parbox[c][1.48cm][c]{0.47\columnwidth}{\raggedright\scriptsize\ttfamily c1cc(\allowbreak{}cnc1)\allowbreak{}C\#\allowbreak{}N}
 & \parbox[c][1.48cm][c]{0.27\columnwidth}{\raggedright\scriptsize\ttfamily mobley\_\allowbreak{}8764620}
 & \parbox[c][1.48cm][c]{0.18\columnwidth}{\centering\includegraphics[width=\linewidth,height=1.32cm,keepaspectratio]{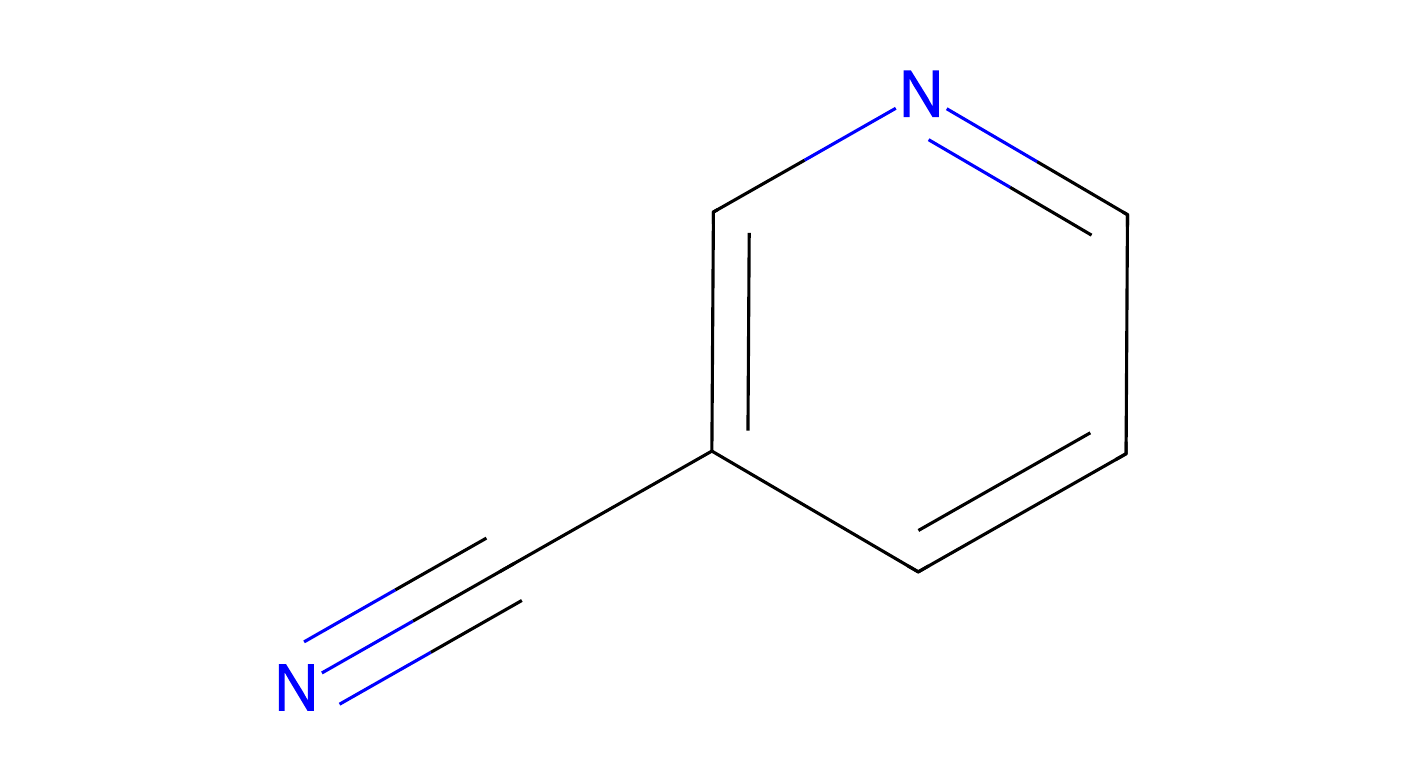}} \\
\parbox[c][1.48cm][c]{0.47\columnwidth}{\raggedright\scriptsize\ttfamily C1CCOC1}
 & \parbox[c][1.48cm][c]{0.27\columnwidth}{\raggedright\scriptsize\ttfamily mobley\_\allowbreak{}8882696}
 & \parbox[c][1.48cm][c]{0.18\columnwidth}{\centering\includegraphics[width=\linewidth,height=1.32cm,keepaspectratio]{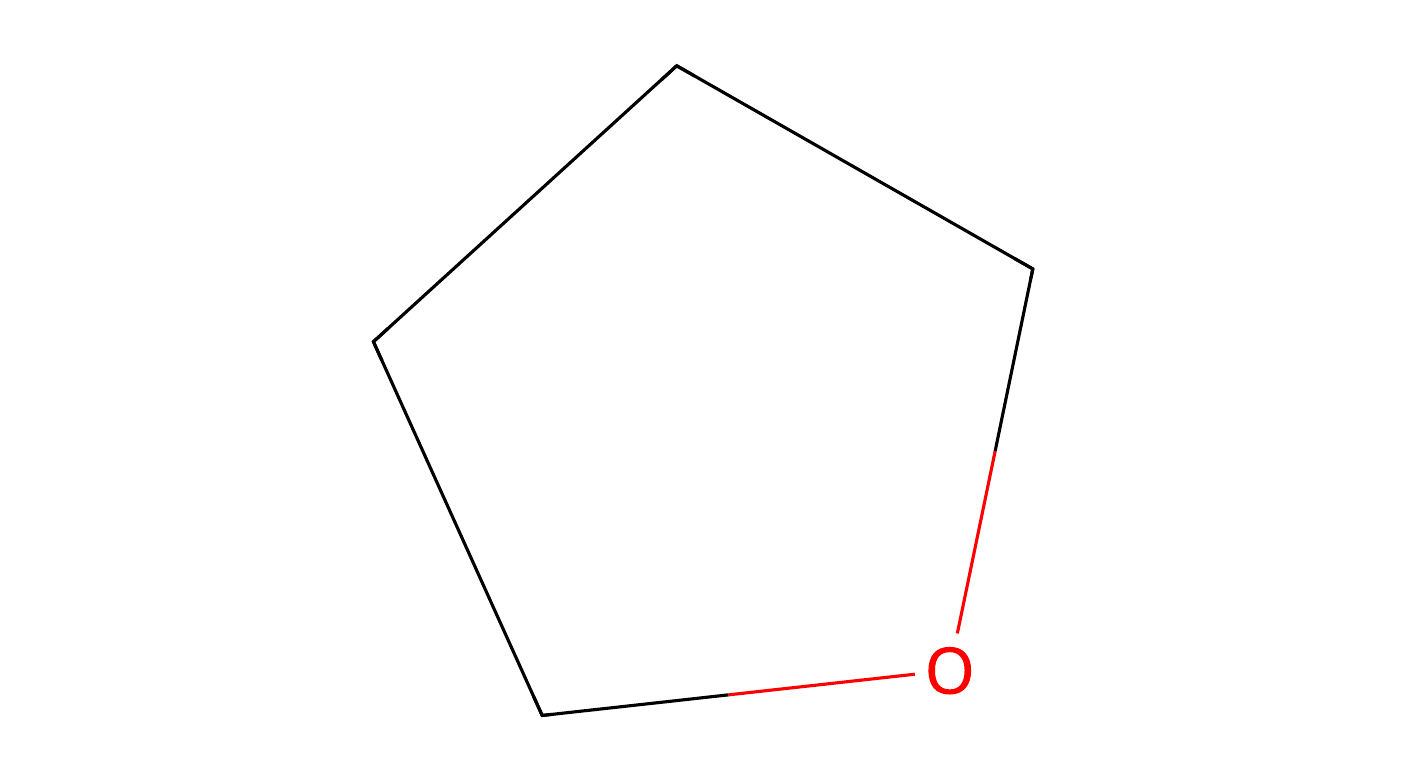}} \\
\bottomrule
\end{tabular}

\endgroup

\clearpage

\clearpage
\begingroup
\scriptsize
\setlength{\tabcolsep}{1.5pt}
\renewcommand{\arraystretch}{1.0}

\refstepcounter{table}
\label{tab:high-error-test-set-molecules}
\makeatletter
\@makecaption{\fnum@table}{Molecular structures for the compounds in the high-error test set.}
\makeatother
\vspace{2pt}
\noindent\begin{tabular}{@{}lll@{}}
\toprule
\parbox[t]{0.47\columnwidth}{\raggedright\bfseries SMILES} &
\parbox[t]{0.27\columnwidth}{\raggedright\bfseries freesolv\_id} &
\parbox[t]{0.18\columnwidth}{\centering\bfseries Molecule} \\
\midrule
\parbox[c][1.48cm][c]{0.47\columnwidth}{\raggedright\tiny\ttfamily C\allowbreak{}C\allowbreak{}O\allowbreak{}P\allowbreak{}(\allowbreak{}=\allowbreak{}S\allowbreak{})\allowbreak{}(\allowbreak{}O\allowbreak{}C\allowbreak{}C\allowbreak{})\allowbreak{}S\allowbreak{}[\allowbreak{}C\allowbreak{}@\allowbreak{}@\allowbreak{}H\allowbreak{}]\allowbreak{}(\allowbreak{}C\allowbreak{}C\allowbreak{}l\allowbreak{})\allowbreak{}N\allowbreak{}1\allowbreak{}C\allowbreak{}(\allowbreak{}=\allowbreak{}O\allowbreak{})\allowbreak{}c\allowbreak{}2\allowbreak{}c\allowbreak{}c\allowbreak{}c\allowbreak{}c\allowbreak{}c\allowbreak{}2\allowbreak{}C\allowbreak{}1\allowbreak{}=\allowbreak{}O\allowbreak{}}
 & \parbox[c][1.48cm][c]{0.27\columnwidth}{\raggedright\scriptsize\ttfamily m\allowbreak{}o\allowbreak{}b\allowbreak{}l\allowbreak{}e\allowbreak{}y\allowbreak{}\_\allowbreak{}2\allowbreak{}5\allowbreak{}1\allowbreak{}8\allowbreak{}9\allowbreak{}8\allowbreak{}9\allowbreak{}}
 & \parbox[c][1.48cm][c]{0.18\columnwidth}{\centering\includegraphics[width=\linewidth,height=1.32cm,keepaspectratio]{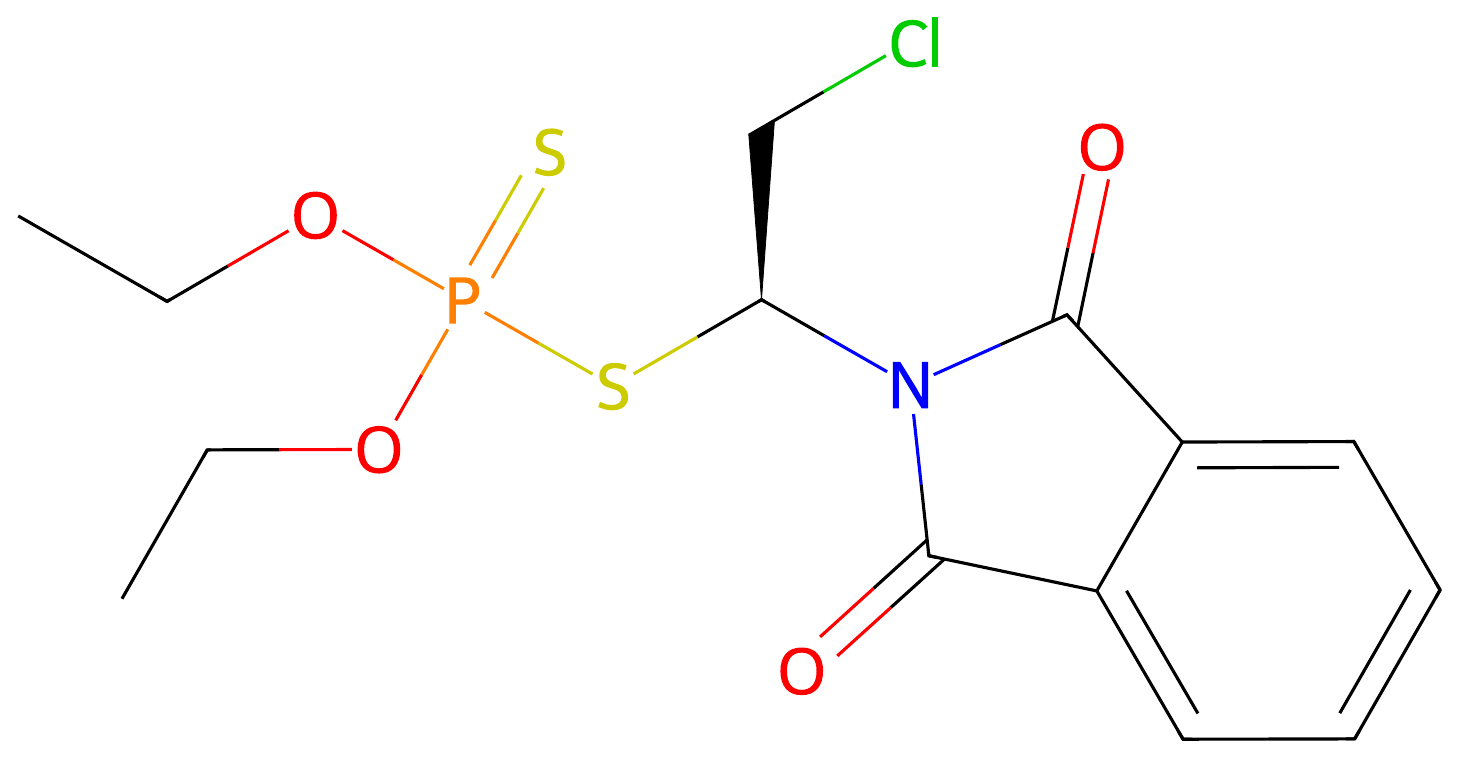}} \\
\parbox[c][1.48cm][c]{0.47\columnwidth}{\raggedright\tiny\ttfamily C\allowbreak{}1\allowbreak{}[\allowbreak{}C\allowbreak{}@\allowbreak{}@\allowbreak{}H\allowbreak{}]\allowbreak{}2\allowbreak{}[\allowbreak{}C\allowbreak{}@\allowbreak{}H\allowbreak{}]\allowbreak{}(\allowbreak{}C\allowbreak{}O\allowbreak{}S\allowbreak{}(\allowbreak{}=\allowbreak{}O\allowbreak{})\allowbreak{}O\allowbreak{}1\allowbreak{})\allowbreak{}[\allowbreak{}C\allowbreak{}@\allowbreak{}@\allowbreak{}]\allowbreak{}3\allowbreak{}(\allowbreak{}C\allowbreak{}(\allowbreak{}=\allowbreak{}C\allowbreak{}(\allowbreak{}[\allowbreak{}C\allowbreak{}@\allowbreak{}]\allowbreak{}2\allowbreak{}(\allowbreak{}C\allowbreak{}3\allowbreak{}(\allowbreak{}C\allowbreak{}l\allowbreak{})\allowbreak{}C\allowbreak{}l\allowbreak{})\allowbreak{}C\allowbreak{}l\allowbreak{})\allowbreak{}C\allowbreak{}l\allowbreak{})\allowbreak{}C\allowbreak{}l\allowbreak{})\allowbreak{}C\allowbreak{}l\allowbreak{}}
 & \parbox[c][1.48cm][c]{0.27\columnwidth}{\raggedright\scriptsize\ttfamily m\allowbreak{}o\allowbreak{}b\allowbreak{}l\allowbreak{}e\allowbreak{}y\allowbreak{}\_\allowbreak{}9\allowbreak{}5\allowbreak{}7\allowbreak{}1\allowbreak{}8\allowbreak{}8\allowbreak{}8\allowbreak{}}
 & \parbox[c][1.48cm][c]{0.18\columnwidth}{\centering\includegraphics[width=\linewidth,height=1.32cm,keepaspectratio]{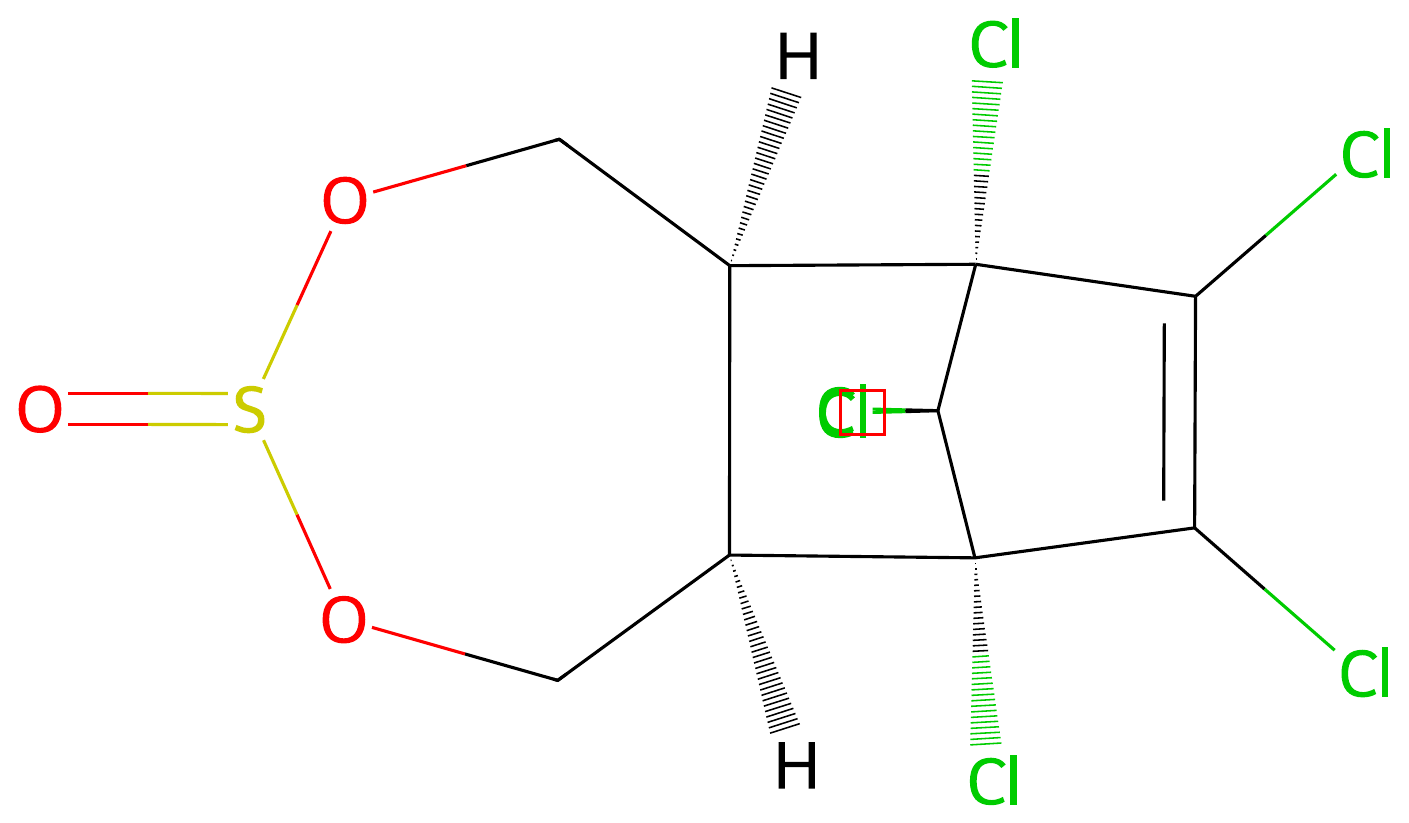}} \\
\parbox[c][1.48cm][c]{0.47\columnwidth}{\raggedright\scriptsize\ttfamily C\allowbreak{}n\allowbreak{}1\allowbreak{}c\allowbreak{}n\allowbreak{}c\allowbreak{}2\allowbreak{}c\allowbreak{}1\allowbreak{}c\allowbreak{}(\allowbreak{}=\allowbreak{}O\allowbreak{})\allowbreak{}n\allowbreak{}(\allowbreak{}c\allowbreak{}(\allowbreak{}=\allowbreak{}O\allowbreak{})\allowbreak{}n\allowbreak{}2\allowbreak{}C\allowbreak{})\allowbreak{}C\allowbreak{}}
 & \parbox[c][1.48cm][c]{0.27\columnwidth}{\raggedright\scriptsize\ttfamily m\allowbreak{}o\allowbreak{}b\allowbreak{}l\allowbreak{}e\allowbreak{}y\allowbreak{}\_\allowbreak{}7\allowbreak{}3\allowbreak{}7\allowbreak{}8\allowbreak{}9\allowbreak{}8\allowbreak{}7\allowbreak{}}
 & \parbox[c][1.48cm][c]{0.18\columnwidth}{\centering\includegraphics[width=\linewidth,height=1.32cm,keepaspectratio]{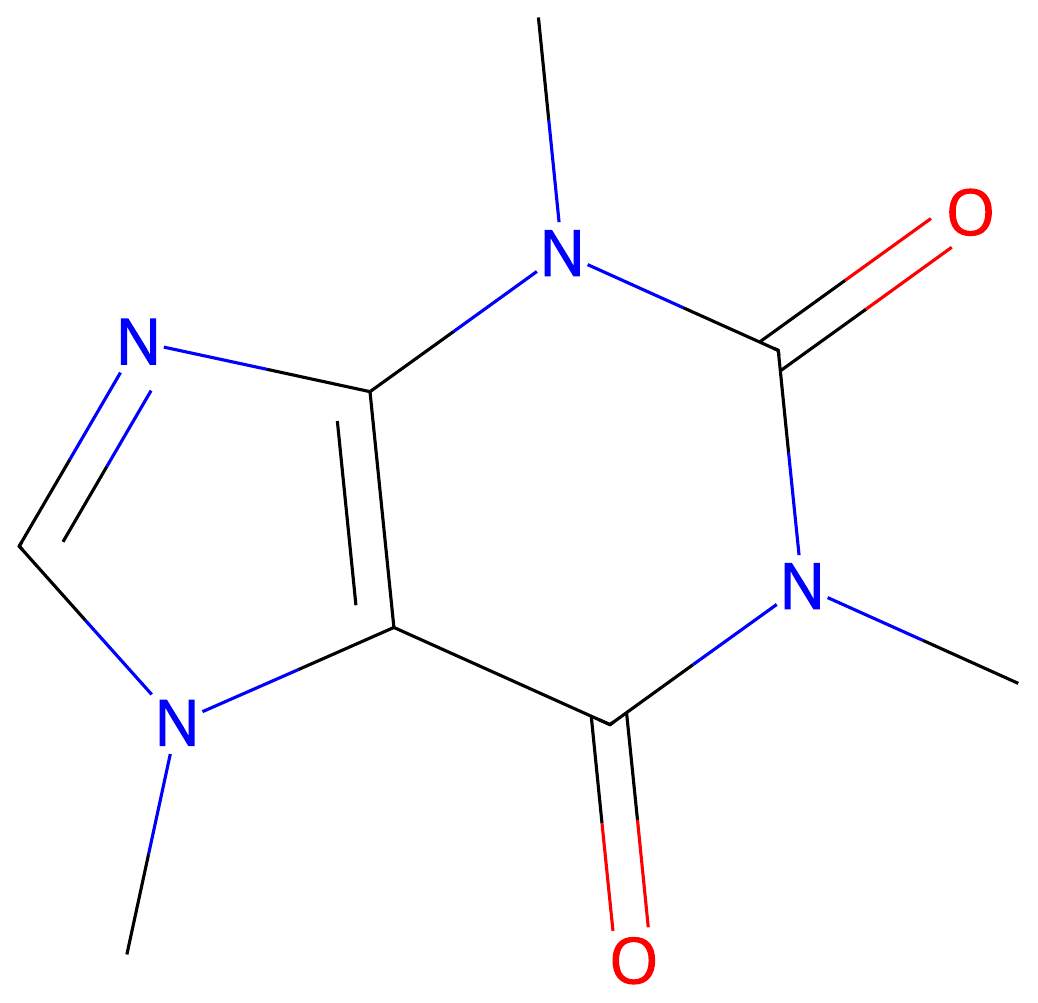}} \\
\parbox[c][1.48cm][c]{0.47\columnwidth}{\raggedright\scriptsize\ttfamily C\allowbreak{}c\allowbreak{}1\allowbreak{}c\allowbreak{}(\allowbreak{}n\allowbreak{}c\allowbreak{}(\allowbreak{}n\allowbreak{}c\allowbreak{}1\allowbreak{}O\allowbreak{}C\allowbreak{}(\allowbreak{}=\allowbreak{}O\allowbreak{})\allowbreak{}N\allowbreak{}(\allowbreak{}C\allowbreak{})\allowbreak{}C\allowbreak{})\allowbreak{}N\allowbreak{}(\allowbreak{}C\allowbreak{})\allowbreak{}C\allowbreak{})\allowbreak{}C\allowbreak{}}
 & \parbox[c][1.48cm][c]{0.27\columnwidth}{\raggedright\scriptsize\ttfamily m\allowbreak{}o\allowbreak{}b\allowbreak{}l\allowbreak{}e\allowbreak{}y\allowbreak{}\_\allowbreak{}3\allowbreak{}2\allowbreak{}0\allowbreak{}1\allowbreak{}7\allowbreak{}0\allowbreak{}1\allowbreak{}}
 & \parbox[c][1.48cm][c]{0.18\columnwidth}{\centering\includegraphics[width=\linewidth,height=1.32cm,keepaspectratio]{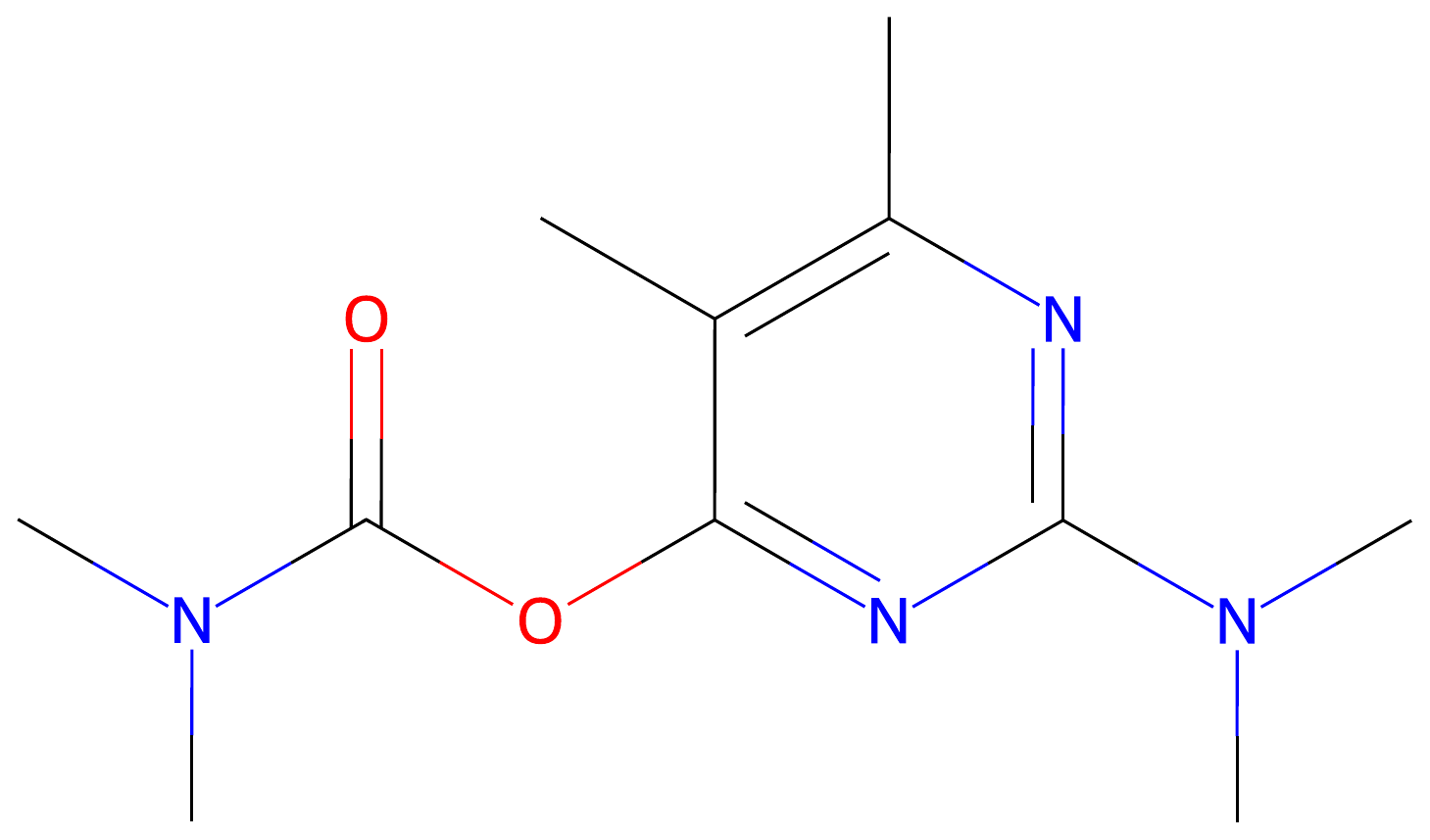}} \\
\parbox[c][1.48cm][c]{0.47\columnwidth}{\raggedright\scriptsize\ttfamily C\allowbreak{}O\allowbreak{}P\allowbreak{}(\allowbreak{}=\allowbreak{}S\allowbreak{})\allowbreak{}(\allowbreak{}O\allowbreak{}C\allowbreak{})\allowbreak{}S\allowbreak{}C\allowbreak{}n\allowbreak{}1\allowbreak{}c\allowbreak{}(\allowbreak{}=\allowbreak{}O\allowbreak{})\allowbreak{}c\allowbreak{}2\allowbreak{}c\allowbreak{}c\allowbreak{}c\allowbreak{}c\allowbreak{}c\allowbreak{}2\allowbreak{}n\allowbreak{}n\allowbreak{}1\allowbreak{}}
 & \parbox[c][1.48cm][c]{0.27\columnwidth}{\raggedright\scriptsize\ttfamily m\allowbreak{}o\allowbreak{}b\allowbreak{}l\allowbreak{}e\allowbreak{}y\allowbreak{}\_\allowbreak{}1\allowbreak{}7\allowbreak{}7\allowbreak{}0\allowbreak{}2\allowbreak{}0\allowbreak{}5\allowbreak{}}
 & \parbox[c][1.48cm][c]{0.18\columnwidth}{\centering\includegraphics[width=\linewidth,height=1.32cm,keepaspectratio]{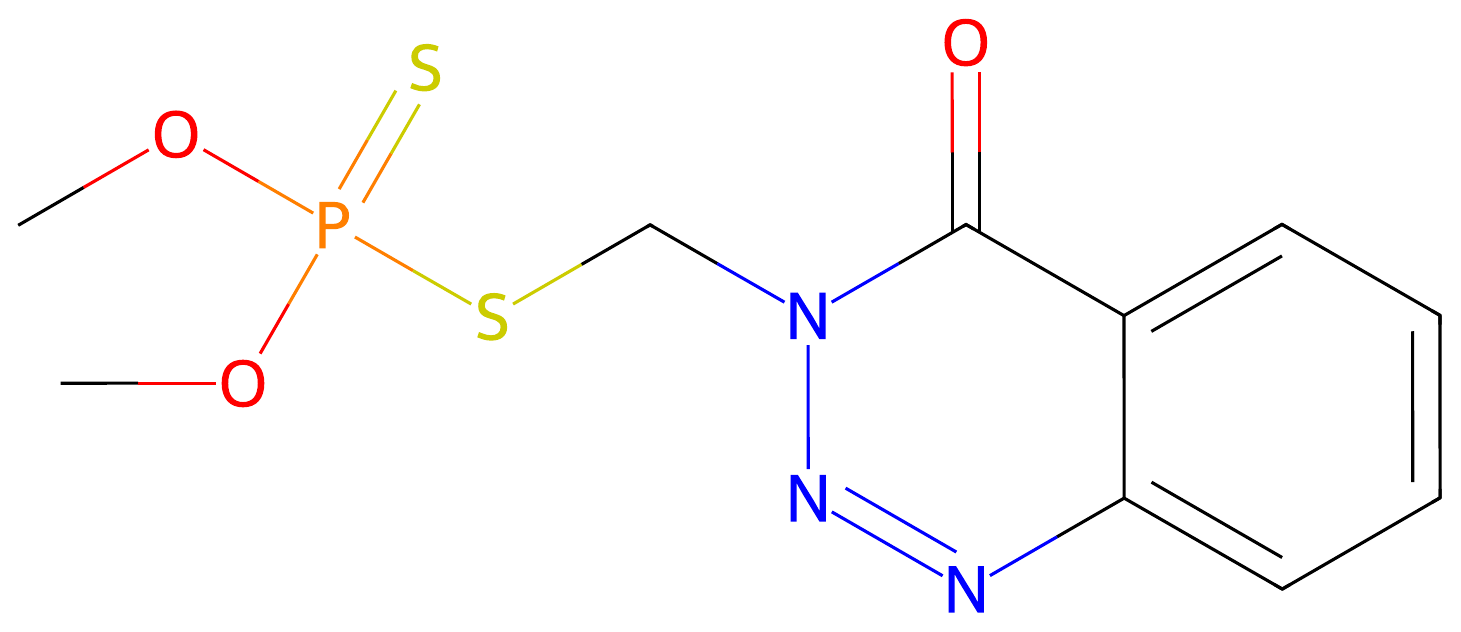}} \\
\parbox[c][1.48cm][c]{0.47\columnwidth}{\raggedright\scriptsize\ttfamily C\allowbreak{}O\allowbreak{}S\allowbreak{}(\allowbreak{}=\allowbreak{}O\allowbreak{})\allowbreak{}(\allowbreak{}=\allowbreak{}O\allowbreak{})\allowbreak{}C\allowbreak{}}
 & \parbox[c][1.48cm][c]{0.27\columnwidth}{\raggedright\scriptsize\ttfamily m\allowbreak{}o\allowbreak{}b\allowbreak{}l\allowbreak{}e\allowbreak{}y\allowbreak{}\_\allowbreak{}9\allowbreak{}2\allowbreak{}0\allowbreak{}1\allowbreak{}2\allowbreak{}6\allowbreak{}3\allowbreak{}}
 & \parbox[c][1.48cm][c]{0.18\columnwidth}{\centering\includegraphics[width=\linewidth,height=1.32cm,keepaspectratio]{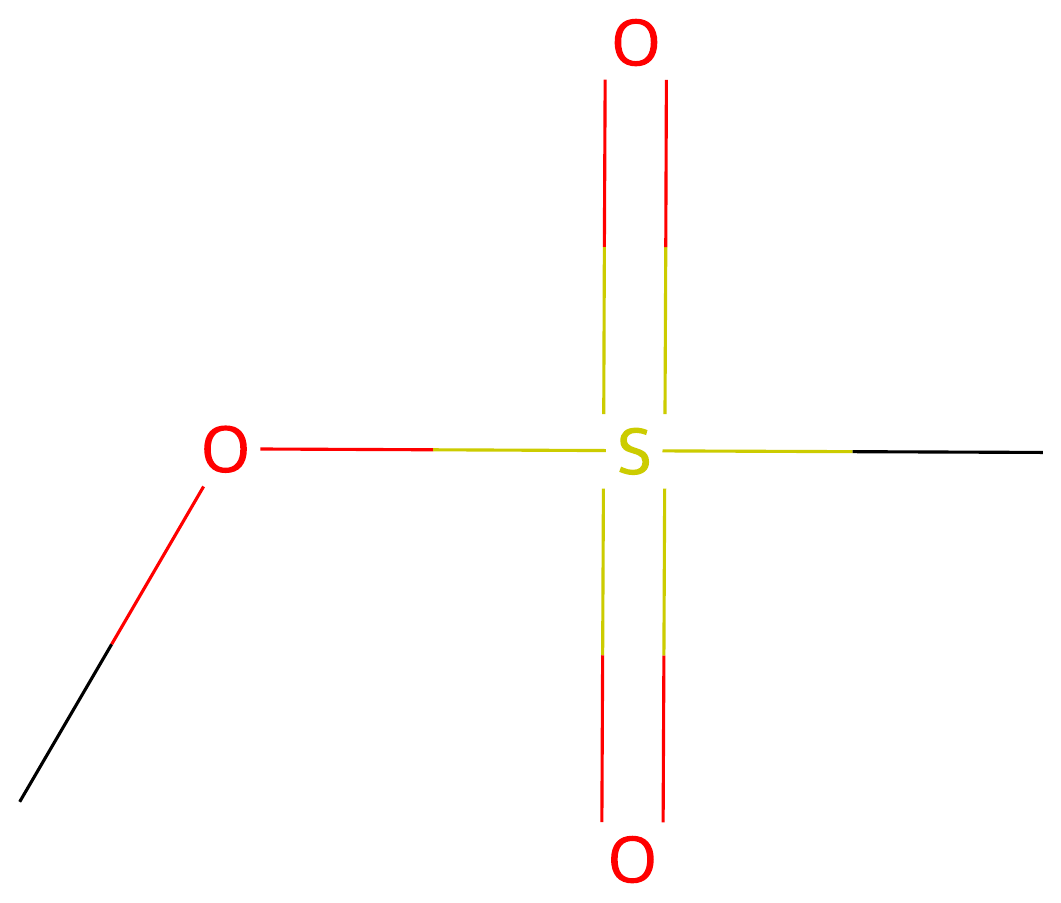}} \\
\parbox[c][1.48cm][c]{0.47\columnwidth}{\raggedright\scriptsize\ttfamily c\allowbreak{}1\allowbreak{}c\allowbreak{}c\allowbreak{}c\allowbreak{}2\allowbreak{}c\allowbreak{}(\allowbreak{}c\allowbreak{}1\allowbreak{})\allowbreak{}C\allowbreak{}(\allowbreak{}=\allowbreak{}O\allowbreak{})\allowbreak{}c\allowbreak{}3\allowbreak{}c\allowbreak{}(\allowbreak{}c\allowbreak{}c\allowbreak{}c\allowbreak{}(\allowbreak{}c\allowbreak{}3\allowbreak{}C\allowbreak{}2\allowbreak{}=\allowbreak{}O\allowbreak{})\allowbreak{}N\allowbreak{})\allowbreak{}N\allowbreak{}}
 & \parbox[c][1.48cm][c]{0.27\columnwidth}{\raggedright\scriptsize\ttfamily m\allowbreak{}o\allowbreak{}b\allowbreak{}l\allowbreak{}e\allowbreak{}y\allowbreak{}\_\allowbreak{}2\allowbreak{}4\allowbreak{}2\allowbreak{}4\allowbreak{}8\allowbreak{}0\allowbreak{}}
 & \parbox[c][1.48cm][c]{0.18\columnwidth}{\centering\includegraphics[width=\linewidth,height=1.32cm,keepaspectratio]{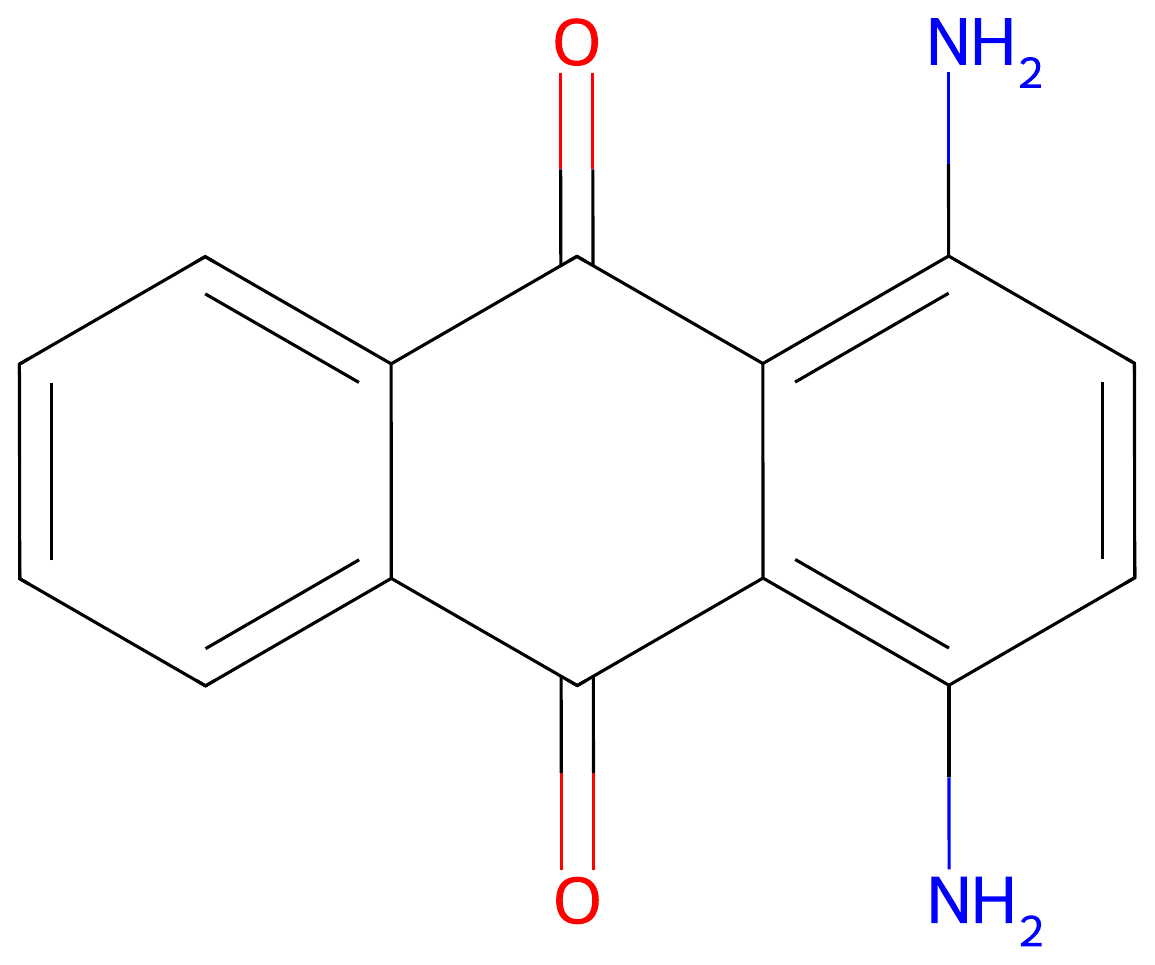}} \\
\bottomrule
\end{tabular}

\newpage
\noindent{\scriptsize\bfseries \tablename~\thetable{} -- continued}
\par\vspace{2pt}
\noindent\begin{tabular}{@{}lll@{}}
\toprule
\parbox[t]{0.47\columnwidth}{\raggedright\bfseries SMILES} &
\parbox[t]{0.27\columnwidth}{\raggedright\bfseries freesolv\_id} &
\parbox[t]{0.18\columnwidth}{\centering\bfseries Molecule} \\
\midrule
\parbox[c][1.48cm][c]{0.47\columnwidth}{\raggedright\scriptsize\ttfamily C\allowbreak{}O\allowbreak{}S\allowbreak{}(\allowbreak{}=\allowbreak{}O\allowbreak{})\allowbreak{}(\allowbreak{}=\allowbreak{}O\allowbreak{})\allowbreak{}O\allowbreak{}C\allowbreak{}}
 & \parbox[c][1.48cm][c]{0.27\columnwidth}{\raggedright\scriptsize\ttfamily m\allowbreak{}o\allowbreak{}b\allowbreak{}l\allowbreak{}e\allowbreak{}y\allowbreak{}\_\allowbreak{}5\allowbreak{}7\allowbreak{}6\allowbreak{}0\allowbreak{}5\allowbreak{}6\allowbreak{}3\allowbreak{}}
 & \parbox[c][1.48cm][c]{0.18\columnwidth}{\centering\includegraphics[width=\linewidth,height=1.32cm,keepaspectratio]{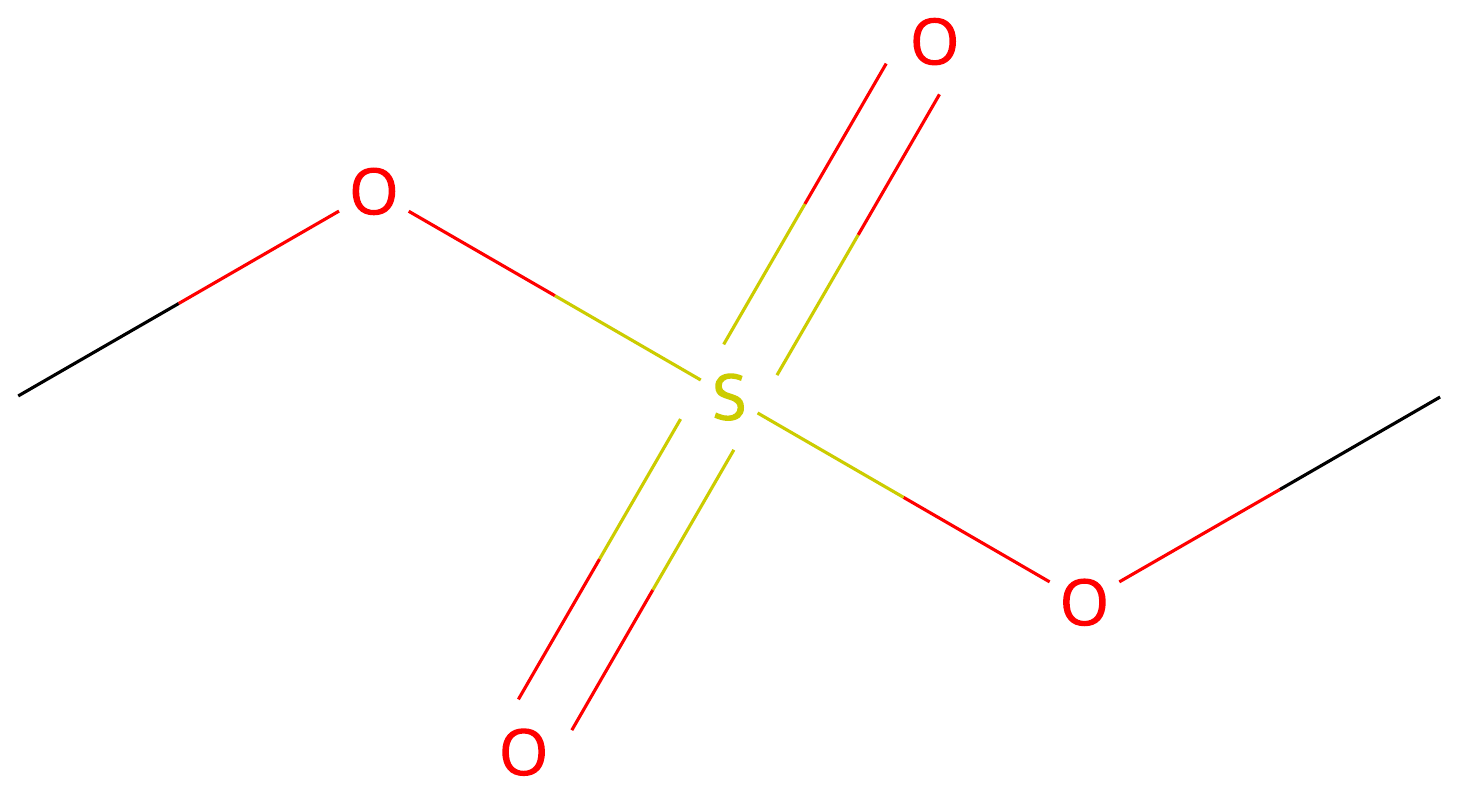}} \\
\parbox[c][1.48cm][c]{0.47\columnwidth}{\raggedright\scriptsize\ttfamily C\allowbreak{}(\allowbreak{}C\allowbreak{}(\allowbreak{}C\allowbreak{}O\allowbreak{})\allowbreak{}O\allowbreak{})\allowbreak{}O\allowbreak{}}
 & \parbox[c][1.48cm][c]{0.27\columnwidth}{\raggedright\scriptsize\ttfamily m\allowbreak{}o\allowbreak{}b\allowbreak{}l\allowbreak{}e\allowbreak{}y\allowbreak{}\_\allowbreak{}3\allowbreak{}2\allowbreak{}5\allowbreak{}9\allowbreak{}4\allowbreak{}1\allowbreak{}1\allowbreak{}}
 & \parbox[c][1.48cm][c]{0.18\columnwidth}{\centering\includegraphics[width=\linewidth,height=1.32cm,keepaspectratio]{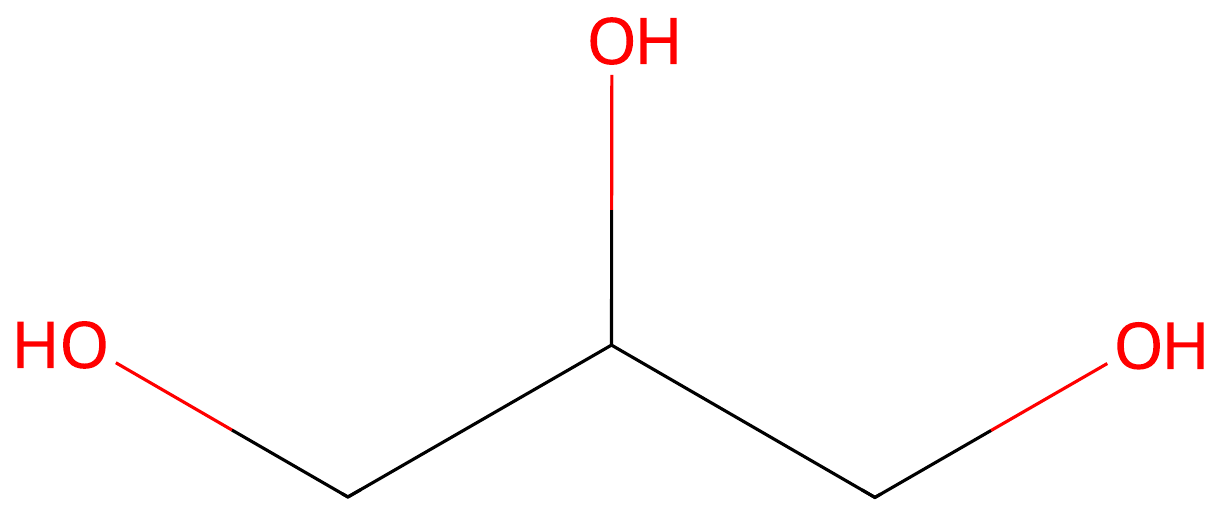}} \\
\parbox[c][1.48cm][c]{0.47\columnwidth}{\raggedright\scriptsize\ttfamily C\allowbreak{}C\allowbreak{}(\allowbreak{}O\allowbreak{}C\allowbreak{}(\allowbreak{}=\allowbreak{}O\allowbreak{})\allowbreak{}C\allowbreak{})\allowbreak{}O\allowbreak{}C\allowbreak{}(\allowbreak{}=\allowbreak{}O\allowbreak{})\allowbreak{}C\allowbreak{}}
 & \parbox[c][1.48cm][c]{0.27\columnwidth}{\raggedright\scriptsize\ttfamily m\allowbreak{}o\allowbreak{}b\allowbreak{}l\allowbreak{}e\allowbreak{}y\allowbreak{}\_\allowbreak{}4\allowbreak{}4\allowbreak{}6\allowbreak{}3\allowbreak{}9\allowbreak{}1\allowbreak{}3\allowbreak{}}
 & \parbox[c][1.48cm][c]{0.18\columnwidth}{\centering\includegraphics[width=\linewidth,height=1.32cm,keepaspectratio]{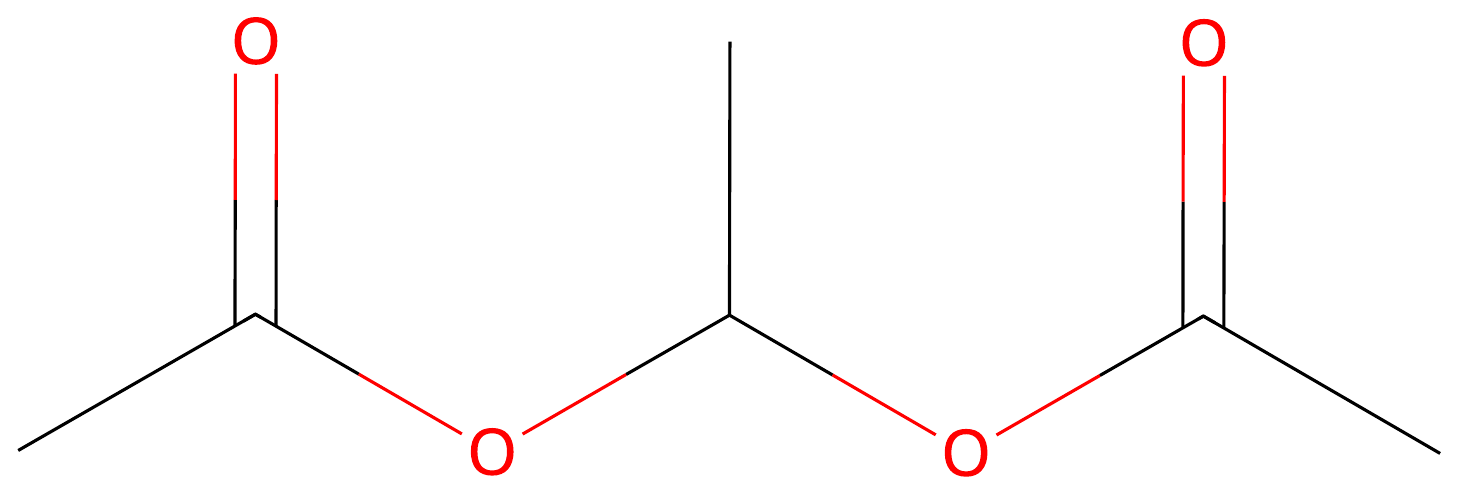}} \\
\parbox[c][1.48cm][c]{0.47\columnwidth}{\raggedright\scriptsize\ttfamily C\allowbreak{}C\allowbreak{}O\allowbreak{}C\allowbreak{}(\allowbreak{}=\allowbreak{}O\allowbreak{})\allowbreak{}C\allowbreak{}[\allowbreak{}C\allowbreak{}@\allowbreak{}H\allowbreak{}]\allowbreak{}(\allowbreak{}C\allowbreak{}(\allowbreak{}=\allowbreak{}O\allowbreak{})\allowbreak{}O\allowbreak{}C\allowbreak{}C\allowbreak{})\allowbreak{}S\allowbreak{}P\allowbreak{}(\allowbreak{}=\allowbreak{}S\allowbreak{})\allowbreak{}(\allowbreak{}O\allowbreak{}C\allowbreak{})\allowbreak{}O\allowbreak{}C\allowbreak{}}
 & \parbox[c][1.48cm][c]{0.27\columnwidth}{\raggedright\scriptsize\ttfamily m\allowbreak{}o\allowbreak{}b\allowbreak{}l\allowbreak{}e\allowbreak{}y\allowbreak{}\_\allowbreak{}8\allowbreak{}9\allowbreak{}1\allowbreak{}6\allowbreak{}4\allowbreak{}0\allowbreak{}9\allowbreak{}}
 & \parbox[c][1.48cm][c]{0.18\columnwidth}{\centering\includegraphics[width=\linewidth,height=1.32cm,keepaspectratio]{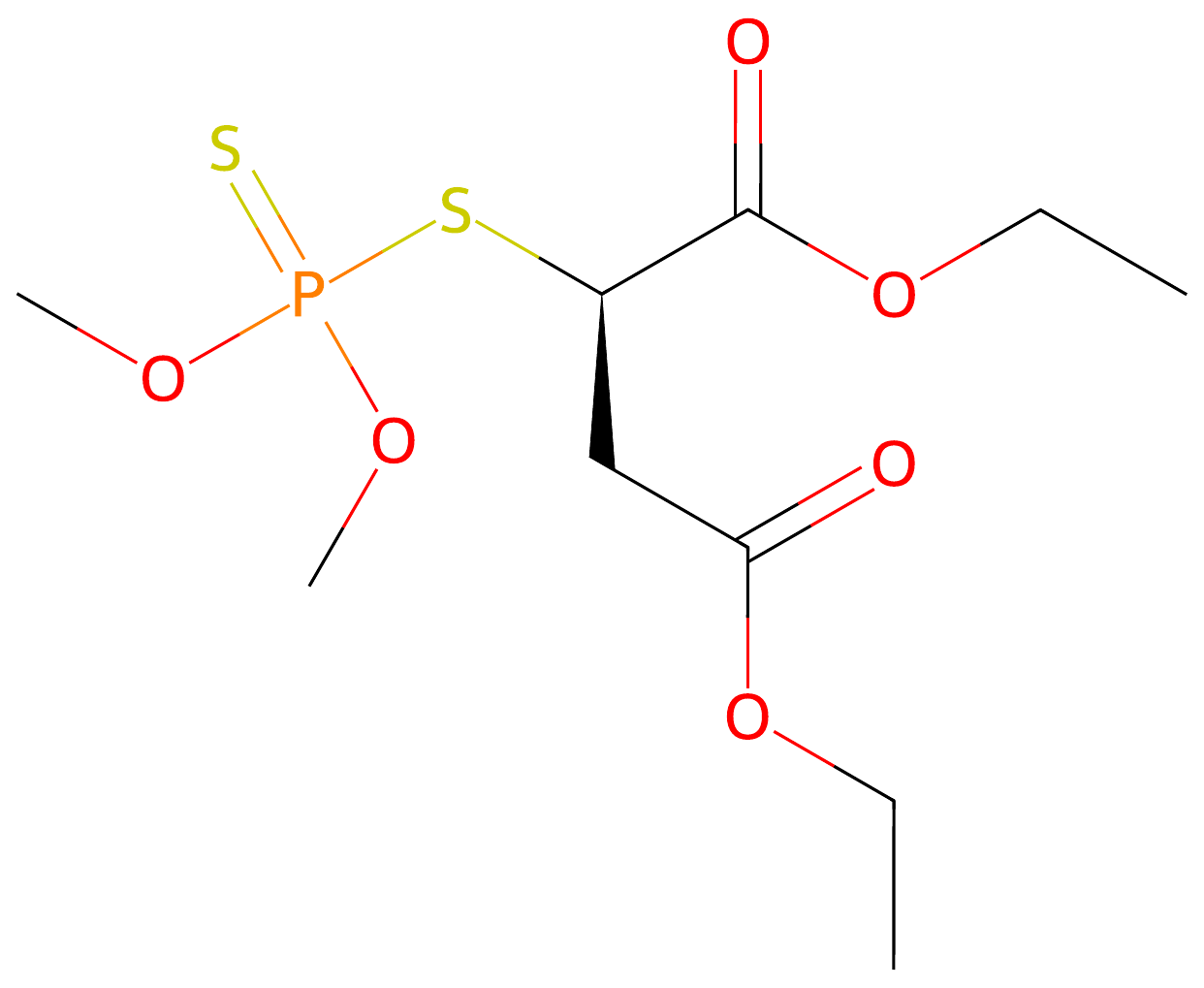}} \\
\parbox[c][1.48cm][c]{0.47\columnwidth}{\raggedright\scriptsize\ttfamily c\allowbreak{}1\allowbreak{}c\allowbreak{}c\allowbreak{}c\allowbreak{}(\allowbreak{}c\allowbreak{}(\allowbreak{}c\allowbreak{}1\allowbreak{})\allowbreak{}O\allowbreak{})\allowbreak{}I\allowbreak{}}
 & \parbox[c][1.48cm][c]{0.27\columnwidth}{\raggedright\scriptsize\ttfamily m\allowbreak{}o\allowbreak{}b\allowbreak{}l\allowbreak{}e\allowbreak{}y\allowbreak{}\_\allowbreak{}8\allowbreak{}7\allowbreak{}8\allowbreak{}9\allowbreak{}4\allowbreak{}6\allowbreak{}5\allowbreak{}}
 & \parbox[c][1.48cm][c]{0.18\columnwidth}{\centering\includegraphics[width=\linewidth,height=1.32cm,keepaspectratio]{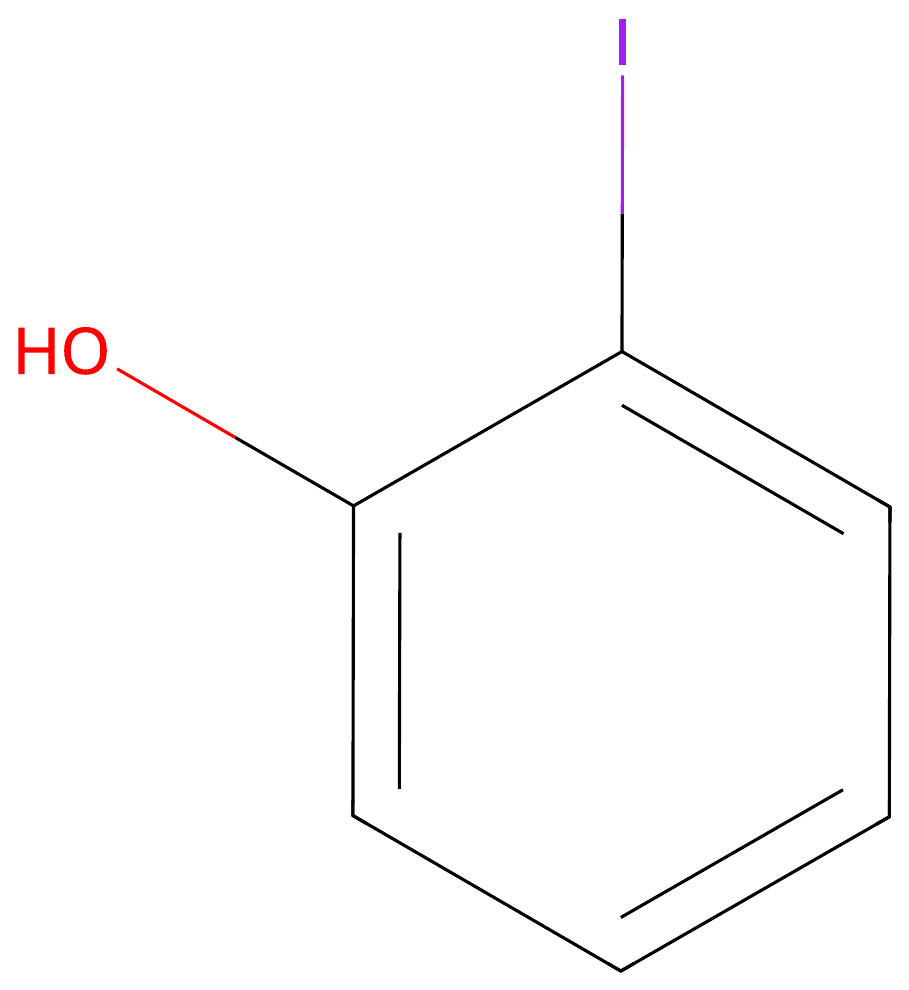}} \\
\parbox[c][1.48cm][c]{0.47\columnwidth}{\raggedright\scriptsize\ttfamily C\allowbreak{}C\allowbreak{}N\allowbreak{}c\allowbreak{}1\allowbreak{}n\allowbreak{}c\allowbreak{}(\allowbreak{}n\allowbreak{}c\allowbreak{}(\allowbreak{}n\allowbreak{}1\allowbreak{})\allowbreak{}S\allowbreak{}C\allowbreak{})\allowbreak{}N\allowbreak{}C\allowbreak{}(\allowbreak{}C\allowbreak{})\allowbreak{}C\allowbreak{}}
 & \parbox[c][1.48cm][c]{0.27\columnwidth}{\raggedright\scriptsize\ttfamily m\allowbreak{}o\allowbreak{}b\allowbreak{}l\allowbreak{}e\allowbreak{}y\allowbreak{}\_\allowbreak{}3\allowbreak{}0\allowbreak{}4\allowbreak{}7\allowbreak{}3\allowbreak{}6\allowbreak{}4\allowbreak{}}
 & \parbox[c][1.48cm][c]{0.18\columnwidth}{\centering\includegraphics[width=\linewidth,height=1.32cm,keepaspectratio]{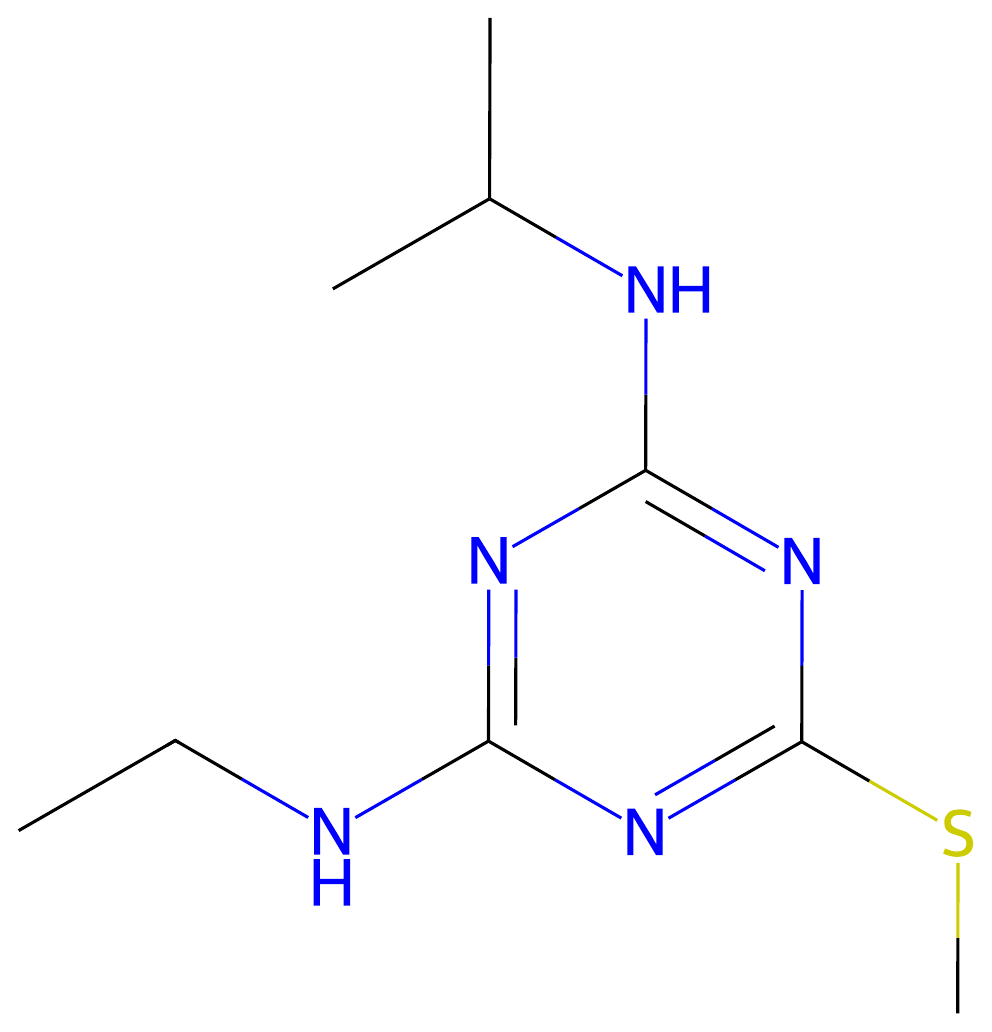}} \\
\parbox[c][1.48cm][c]{0.47\columnwidth}{\raggedright\scriptsize\ttfamily c\allowbreak{}1\allowbreak{}c\allowbreak{}c\allowbreak{}c\allowbreak{}2\allowbreak{}c\allowbreak{}(\allowbreak{}c\allowbreak{}1\allowbreak{})\allowbreak{}C\allowbreak{}(\allowbreak{}=\allowbreak{}O\allowbreak{})\allowbreak{}c\allowbreak{}3\allowbreak{}c\allowbreak{}c\allowbreak{}c\allowbreak{}c\allowbreak{}(\allowbreak{}c\allowbreak{}3\allowbreak{}C\allowbreak{}2\allowbreak{}=\allowbreak{}O\allowbreak{})\allowbreak{}N\allowbreak{}}
 & \parbox[c][1.48cm][c]{0.27\columnwidth}{\raggedright\scriptsize\ttfamily m\allowbreak{}o\allowbreak{}b\allowbreak{}l\allowbreak{}e\allowbreak{}y\allowbreak{}\_\allowbreak{}8\allowbreak{}0\allowbreak{}5\allowbreak{}2\allowbreak{}2\allowbreak{}4\allowbreak{}0\allowbreak{}}
 & \parbox[c][1.48cm][c]{0.18\columnwidth}{\centering\includegraphics[width=\linewidth,height=1.32cm,keepaspectratio]{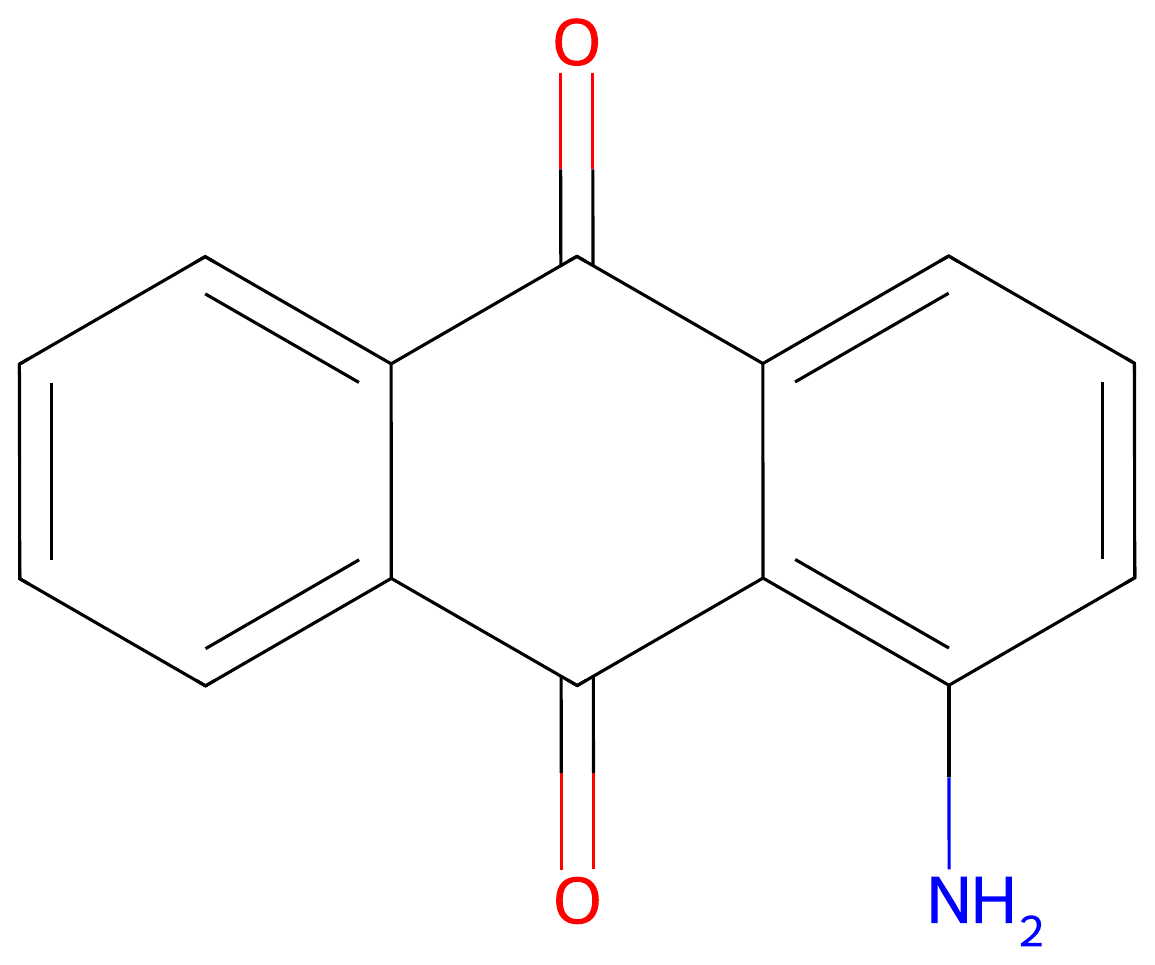}} \\
\bottomrule
\end{tabular}

\endgroup

\end{document}